\documentclass[
11pt, 
oneside, 
english, 
singlespacing, 
headsepline, 
]{MastersDoctoralThesis} 

\let\ltverbatim\verbatim
\let\ltendverbatim\endverbatim
\usepackage{tcolorbox}
\let\verbatim\ltverbatim
\let\endverbatim\ltendverbatim

\usepackage{tabularx}
\usepackage{array}
\usepackage{colortbl}
\tcbuselibrary{skins}

\newcolumntype{Y}{>{\raggedleft\arraybackslash}X}

\tcbset{
    tablestyle_main/.style={
        enhanced,
        fonttitle=\bfseries,
        fontupper=\normalsize\sffamily,
        colback=gray!20!white,
        colframe=gray,
        colbacktitle=gray,
        coltitle=black,center title,
        sharp corners,
        frame hidden
    }
}

\usepackage[utf8]{inputenc} 
\usepackage[T1]{fontenc} 

\usepackage{mathpazo} 
\usepackage{amsmath} 
\usepackage{amssymb}
\usepackage{graphicx} 
\usepackage{url}
\usepackage{hyperref}
\usepackage{mdframed}
\usepackage{algorithm}
\usepackage{algpseudocode}
\usepackage{spverbatim}

\usepackage{listings} 
\usepackage[autostyle=true]{csquotes} 

\usepackage[
backend=biber,
style=numeric,
sorting=none,
defernumbers=true,
sortcites=true,
]{biblatex}
\makeatletter
\let\blx@rerun@biber\relax
\makeatother

\usepackage{titlesec}

\newcommand{\harpoon}{\overset{\rightharpoonup}} 
\newcommand{\footnoteref}[1]{\textsuperscript{\ref{#1}}} 

\definecolor{x11gray}{rgb}{0.9, 0.9, 0.9}

\thesistitle{Hardware Acceleration of Progressive Refinement Radiosity using Nvidia RTX} 
\supervisor{Prof. Dr. Günter \textsc{Rote}} 
\degree{Master of Science} 
\author{Benjamin \textsc{Kahl}} 

\university{\href{https://www.fu-berlin.de/}{Freie Universität Berlin}} 
\department{\href{https://www.mi.fu-berlin.de/en/index.html}{Department of Mathematics and Computer Science}} 
\group{{\href{https://www.mi.fu-berlin.de/en/index.html}}{Research Group}} 
\faculty{\href{https://www.mi.fu-berlin.de/en/inf/index.html}{Institute of Computer Science}} 

\AtBeginDocument{
\hypersetup{pdftitle=\ttitle} 
\hypersetup{pdfauthor=\authorname} 
}

\begin{document}


\frontmatter 

\pagestyle{plain} 


\begin{titlepage}
\begin{center}

\vspace*{.06\textheight}
{\scshape\LARGE \univname\par}\vspace{1.5cm} 
\textsc{\Large Master's Thesis}\\[0.5cm] 

\HRule \\[0.4cm] 
{\huge \bfseries \ttitle\par}\vspace{0.4cm} 
\HRule \\[1.5cm] 
 
\begin{minipage}[t]{0.4\textwidth}
\begin{flushleft} \large
\emph{Author:}\\
\href{https://benjamin.kahl.fi}{\authorname} 
\end{flushleft}
\end{minipage}
\begin{minipage}[t]{0.4\textwidth}
\begin{flushright} \large
\emph{Supervisor:} \\
\href{http://www.mi.fu-berlin.de/inf/groups/ag-ti/members/professoren/Rote_Guenter.html}{\supname} 
\end{flushright}
\end{minipage}\\[3cm]
 
\vfill

\large \textit{Revised and corrected version of a thesis submitted in fulfillment of the requirements\\ for the degree of \degreename\\ on}\\[0.3cm] 
October 13, 2022\\[0.4cm]
\textit{in the}\\[0.4cm]
\deptname\\[2cm] 
 
\vfill

{\large \today}\\[0cm] 
 
\vfill
\end{center}
\end{titlepage}

\begin{abstract}
\addchaptertocentry{\abstractname} 
A vital component of photo-realistic image synthesis is the simulation of indirect diffuse reflections, which still remain a quintessential hurdle that modern rendering engines struggle to overcome.
Real-time applications typically pre-generate diffuse lighting information offline using \textit{radiosity} to avoid performing costly computations at run-time.

In this thesis we present a variant of \textit{progressive refinement radiosity} that utilizes Nvidia's novel RTX technology to accelerate the process of form-factor computation without compromising on visual fidelity.
Through a modern implementation built on DirectX 12 we demonstrate that offloading radiosity's visibility component to \textit{RT cores} significantly improves the lightmap generation process and potentially propels it into the domain of real-time.
\end{abstract}


\tableofcontents 


\mainmatter 

\pagestyle{thesis} 



\chapter{Preface} 

\label{Chapter1} 


\newcommand{\keyword}[1]{\textbf{#1}}
\newcommand{\tabhead}[1]{\textbf{#1}}
\newcommand{\code}[1]{\texttt{#1}}
\newcommand{\file}[1]{\texttt{\bfseries#1}}
\newcommand{\option}[1]{\texttt{\itshape#1}}


The computational synthesis of photorealistic images has been a quintessential challenge in the computer graphics domain since its inception. Growing industries such as video games, virtual reality and visual effects have induced a veritable explosion in demand for increased realism over the last few decades. A major step towards this goal was taken by James Kajiya in 1989 when he formulated the \textit{rendering equation} \cite{Kajiya}, which provides a general mathematical description of how light propagates through a 3D environment.

Unfortunately, the rendering equation proved far too complex to solve linearly. Every surface can receive light from infinitely many directions and then scatter it diffusely, effectively qualifying the surface as a separate light-source itself.

Computer graphics researchers have spent a large part of their endeavour grappling with the conundrum that is finding an ideal, numerically solvable model to this infinitely recursive complexity. Indeed, rendering algorithms we see employed today can all be regarded as approximations, shortcuts or simplifications of the rendering equation.

With regard to \textit{global illumination}, two of these have stood the test of time: \textit{ray tracing} for the generation of individual, highly realistic images and \textit{radiosity} for real-time use cases that continuously render the same, static geometry from a large set of camera angles.

\section{Nvidia RTX}

Over the last decades raytracing has generally found its place as a crude and expensive approach that nevertheless provides a very high degree of photorealism, albeit at a proportionally high cost in required computation time.

Yet in 2018, fifty years after the first computer-based ray-tracer was created \cite{first_raytracer}, the American tech company \textit{Nvidia} unveiled their \textit{GeForce RTX} series of graphics cards. Uniquely, these contain specialized computation units that can speed up raytracing-related operations to such a degree that it propels this blunt, brute-force approach into the domain of real-time \cite{turing_whitepaper}. 

The mathematical challenges faced by raytracing and radiosity are fundamentally identical and thus inextricably linked. In this thesis we argue that the considerable performance increase enabled by the RTX platform ought to be reflected in radiosity to the same degree it is seen in raytracing.

\section{Motivation}

Global illumination solutions based on radiosity typically generate lighting information and then export it into a texture, which can be rendered a-posteriori within consumer applications at virtually no cost at all. Despite great rendering performance, the process of generating these textures remains a computationally expensive process that can severely hamstring the development and design process of complex 3D environments.

Radiosity's performance bottleneck unequivocally lies with the vast amount of visibility calculations required \cite{radiosity}.
Although this problem is, in theory, highly parallelizable \cite{RadiosityParallelization}, implementations of the radiosity model seem to generally favour multi-core CPUs (approx. 4-16 high performance cores) over GPUs (approx. 1-10 thousand low performance cores) \cite{ray_engine}, because GPU variants rely on hemicuboid z-Buffering for visibility determination \cite{RadiosityOnGPUs_Coombe}.

In this thesis we investigate if an RTX-based visibility solution provides a performance improvement sufficient enough to fully advance radiosity into the realm of GPUs and parallel computing.

Not only is raytracing a highly adequate solution for visibility, but RTX GPUs also perform their operations on dedicated hardware in the form of a moderate amount (30 to 80) of highly specialized \textit{RT cores} \cite{turing_whitepaper}. This intermediate solution between the parallelization levels of a CPU and a GPU may prove ideal for the acceleration of the radiosity algorithm.

\section{Objective}

The intended goal behind this thesis is to further the acceleration of radiosity computations for developers and designers of 3D environments working on machines compatible with RTX. Once lighting textures have been generated, they can in turn be rendered on almost any graphics hardware, regardless of RTX compatibility.

To accomplish this, we grapple a common variant of radiosity found on GPUs, known as \textit{progressive refinement radiosity}, and substitute its z-Buffering components with an RTX-based approach. We will also investigate and examine potential performance improvements in addition to how well this approach compares to already existing solutions.

\section{Thesis Structure}

The next chapter will cover the theoretical knowledge required for the remainder of the thesis by deriving the rendering equation and providing mathematical models for several global illumination solutions.

Afterwards, chapter \ref{Chapter3} takes a deep dive into RTX technology by examining and reviewing the underlying \textit{Turing architecture} and \textit{DirectX} raytracing pipeline.

Chapters \ref{Chapter4} and \ref{Chapter5} will present \textit{RTRad}, our RTX-accelerated progressive refinement radiosity implementation as well as any related tweaks and potential performance improvements.

Lastly, in chapter \ref{Chapter6}, we compare and analyze the performance of this implementation upon which we draw our conclusions in chapter \ref{Chapter7}.


\chapter{Introduction} 

\label{Chapter2} 


This chapter outlines the background knowledge and core concepts that are required in the subsequent chapters. The first section commences by deriving the core problem of computer graphics starting at the root. Afterwards, we show how the global illumination problem is tackled specifically by radiosity and raytracing.

\section{The Speed - Realism Dichotomy}

A classical image-synthetization process computes how light scattered into an environment translates into pixel colors on a retina.
The color an object should adopt on a virtual sensor can be traced back to the wavelengths absorbed by its surface in relation to the light incident on it, which is turn affected by the light that is reflected, refracted or emitted by other surfaces around it.

Combine this endless recursion with the vast amount of intrinsics this process is subject to, such as physical properties or geometric arrangements, and it quickly becomes clear that a complete, physically accurate light simulation is an unfeasible computational task that needs to be approximated.

Indeed, even highly photorealistic, computationally heavy methods employ a significant amount of approximation and reductionism. The question therefore becomes which simplifications one is willing to make and what their payoff is in computational expense.

The balance of speed vs. realism that underpins this challenge divides it into two distinct problem domains: Whilst some industries, like CG film-making and SFX, are more geared towards realism, other areas have driven an increased demand of faster, more responsive graphics, known under the umbrella term of \textit{real-time rendering}.

\section{Rendering Optics}

The essence of generating images from abstract descriptions can be narrowed down to the simulation of a real-world camera in a virtual environment. As such, we commence by examining the characteristics of virtual cameras as well as the related concepts from \textit{radiometry} that help us model light propagation.

\subsection{Camera Optics}

Most genuine cameras have a series of common denominators arranged in a similar construction:

An aperture allows light to enter through a convex lens, which casts an image onto a light-capturing sensor.
The convexity of the lens ensures that the direction in which light hits the sensor is restricted, thus focusing the image with a limited depth of field, determined by the focal length.

\begin{figure}[th]
\centering
\includegraphics[scale=0.3]{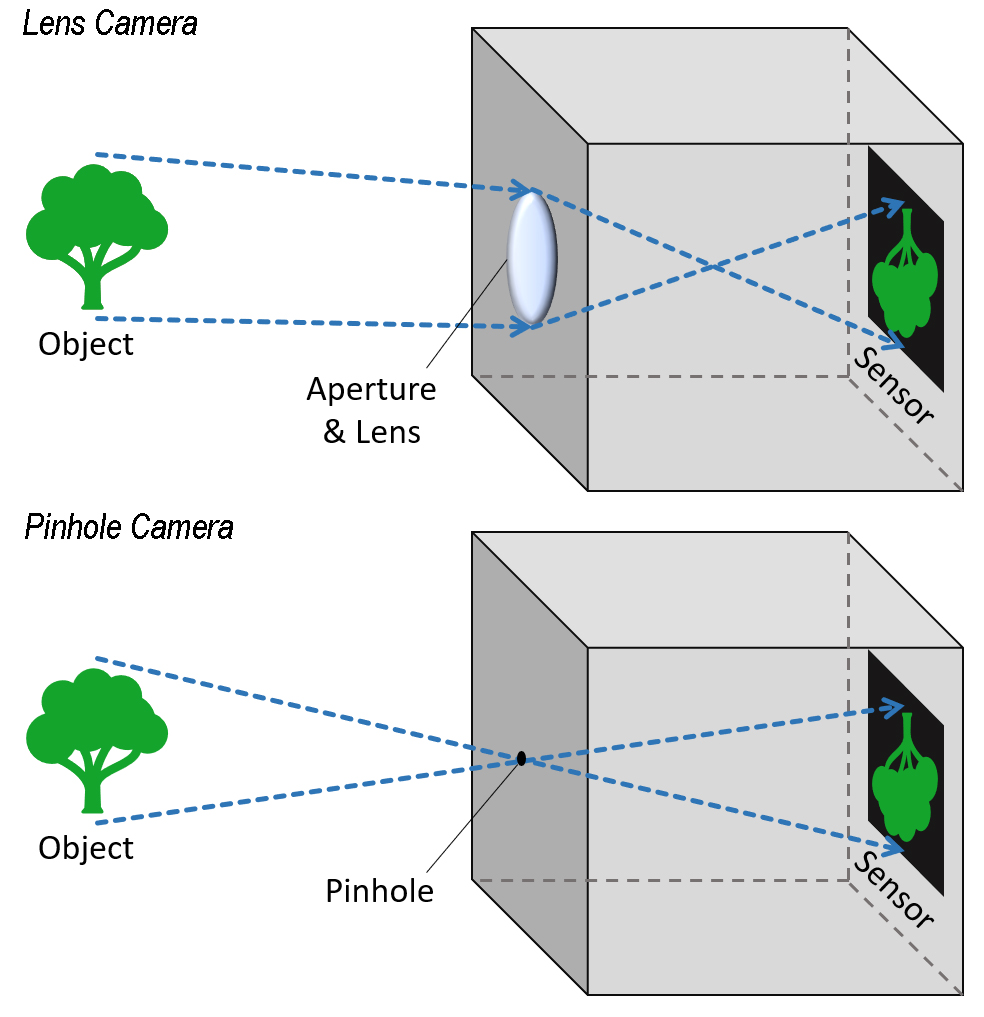}
\decoRule
\caption[]{Comparison of the optics behind a lens-based and pinhole camera.}
\label{lens_vs_pinhole}
\end{figure}

This directional limitation of incident light can also be accomplished without a lens, by severely limiting the size of the camera's aperture, which is the principle of the \textit{pinhole camera} \cite{pinhole_virtual_camera} (as seen in fig. \ref{lens_vs_pinhole}).

\begin{figure}[h]
\centering
\includegraphics[scale=1.15]{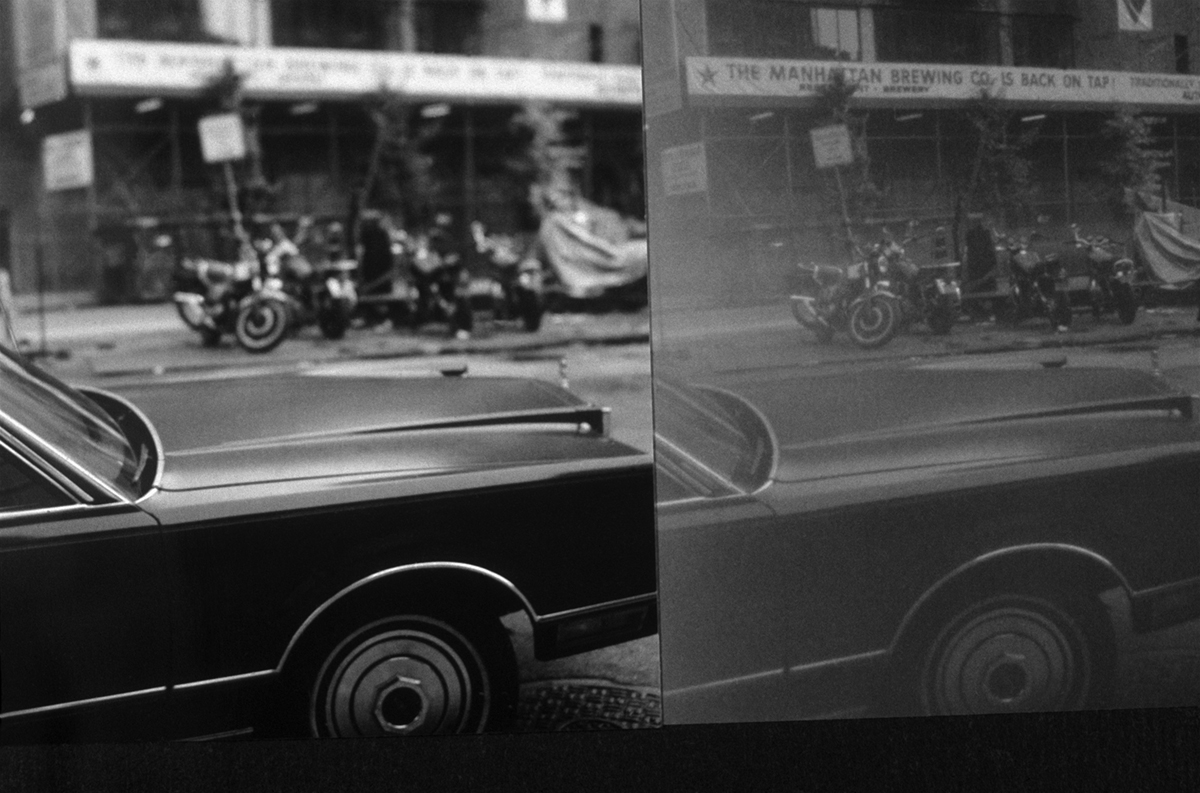}
\decoRule
\caption[]{Normal lens (left) vs. Pinhole lens (right). The pinhole has greater depth of field, but the image sharpness decreases with pinhole size. Image credited to Leonard Lessin/Science Source \cite{ScienceSource}.}
\label{pinhole_vs_lens_photo}
\end{figure}

Pinhole cameras have nearly infinite depth of field and, unlike lens-based cameras, do not suffer from lens distortion (see fig. \ref{pinhole_vs_lens_photo}). However, their minuscule apertures require proportionally lengthy exposure times to produce serviceable photographs \cite{pinhole_virtual_camera}, for which reason pinhole cameras tend to find little to no use in real-life photography \cite{PinholeOptics}.


However, virtual environments are not subject to the same physical constraints, as any numerical value for light can arbitrarily be multiplied by some factor to control for brightness or exposure. As such, the notion of an \textit{exposure time} is not a valid one within a virtual context.

\subsection{Virtual Camera}

A common observation one can make in computer-generated images is that they tend to have unlimited depth of field. This is, indeed, because virtual cameras strongly mimic the simplified optics of a pinhole camera \cite{pinhole_virtual_camera}. 

\begin{figure}[th]
\centering
\includegraphics[scale=0.33]{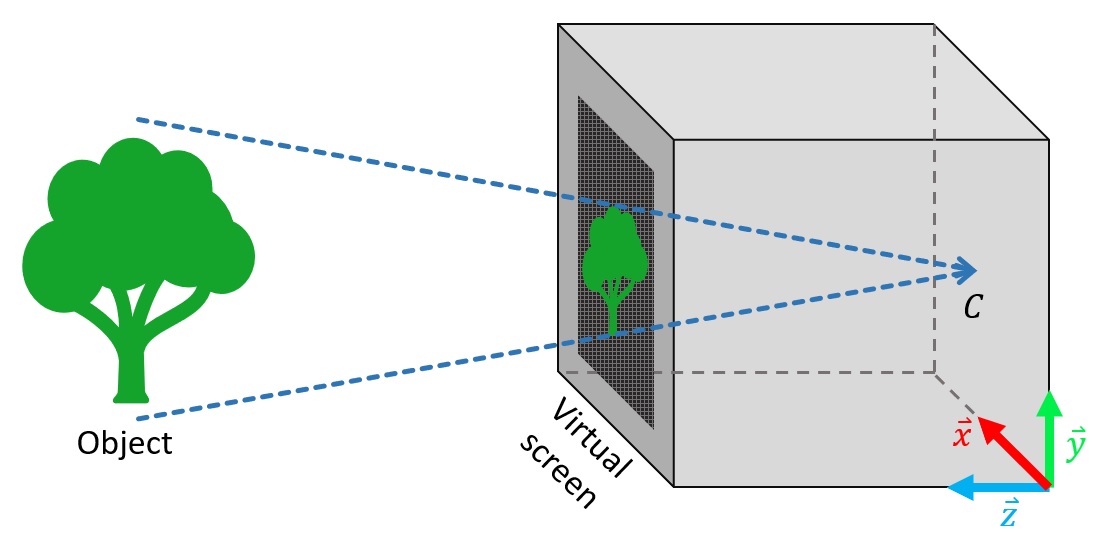}
\decoRule
\caption[]{Optics of a virtual camera. The dimensions of the virtual screen are directly tied to the angles $\measuredangle_{x}$, $\measuredangle_{y}$ making them independent of their distance from $C$.}
\label{virtual_camera}
\end{figure}

In reality, increasing the distance between a pinhole and its sensor would produce a weaker image due to inverse-square attenuation. But since brightness factors and exposure times are irrelevant in a virtual context, the pinhole-sensor distance can be entirely discarded.
As such, the sensor can be regarded as being a virtual screen \textit{in front} of the camera \cite{pinhole_virtual_camera}, where light enters through grates corresponding to pixels on the final image. This arrangement is depicted in fig. \ref{virtual_camera}.

By designating the location of the pinhole as the camera's position, we are left with a location vector, a view direction and two \textit{field-of-view} (FOV) angles that are proportional to the height and width of the resulting image respectively \footnote{Most computer graphics domains expand this definition by also including a near and far \textit{clipping plane}, thus forming a \textit{view frustum} \cite{pinhole_virtual_camera}.}.

\begin{equation}
Camera = \{C, \harpoon x, \harpoon y, \harpoon z, \measuredangle_{x}, \measuredangle_{y}\}
\end{equation}

where $C$ is the location of the camera, $\harpoon x$, $\harpoon y$ and $\harpoon z$ are the camera's right, upward and forward directions respectively and $\measuredangle_{x}$, $\measuredangle_{y}$ are the FOV angles for the $x$ and $y$ directions.

In most practical cases, the given directions form an orthogonal coordinate system with $\hat{z} = \hat{x} \times \hat{y}$.

\section{Radiometric Quantities}

The process of rendering a virtual object from the perspective of a virtual camera now comes down to measuring the light that the object emits or reflects towards the location of the virtual pinhole $C$.

The direction of incident photons from a given surface cannot be described by a simple 3D vector, as surfaces subtend an infinite amount of directions towards a single point. Thus, it is useful to define a measurement of \textit{solid angle} for the total \textit{field of view} an object occupies towards an observer.

\subsection{Solid Angle}

Regular 2D angles can be adequately represented by the length of the arc they cover on a unit circle. Analogously, solid angles are proportional to the area a surface projects onto a unit sphere around the point of origin \cite{radiosity, fu_cg}.
Solid angles are measured in \textit{steradians} and limited by the total surface area $4\pi$ of a unit sphere.

\begin{figure}[H]
\centering
\includegraphics[scale=0.355]{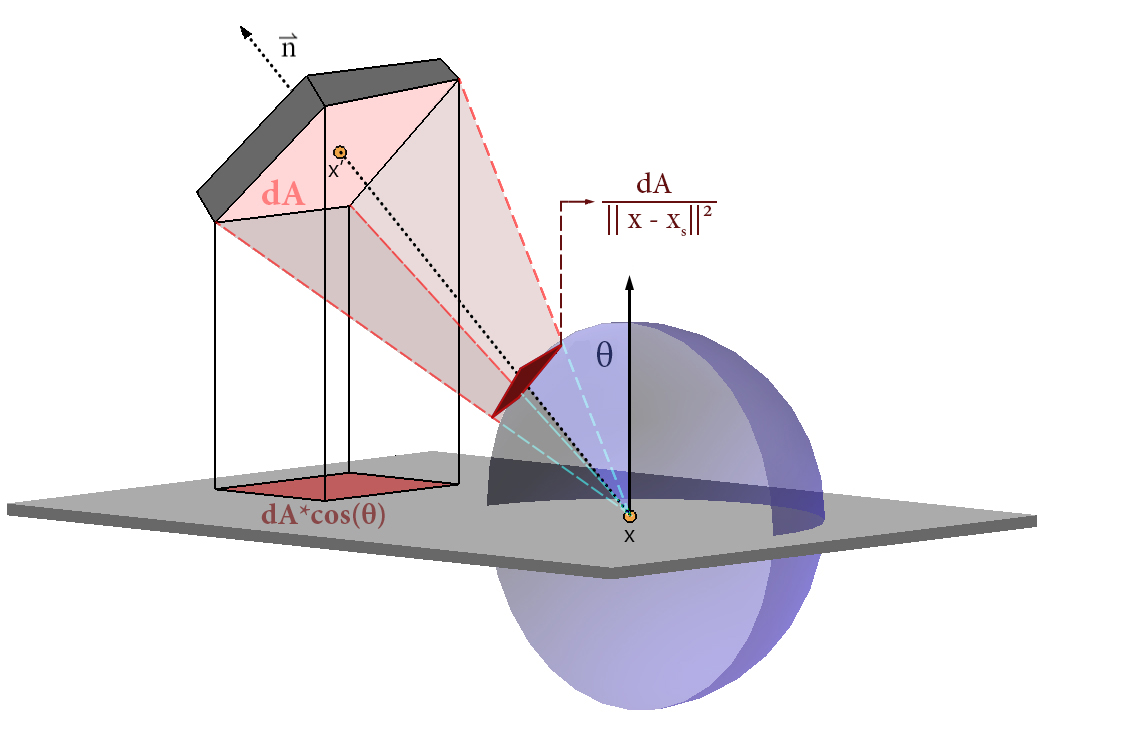}
\decoRule
\caption[Solid Angle]{A surface projected onto a plane results in a surface area of $dA\cos \theta$. Projecting a perpendicular surface towards the center of a sphere results in a projected surface area on the sphere of $dA / \lVert x-x'\rVert^2$. Image from work by Benjamin Kahl \cite{VXCT}.}
\label{solid_angle_adv}
\end{figure}

Let $dA$ be a differential, arbitrarily rotated surface area at point $x'$ with the normal vector $\harpoon n$. The solid angle that $dA$ occupies at another point $x$ can be calculated using two separate operations \cite{radiosity}, both of which are illustrated in fig. \ref{solid_angle_adv}.

\begin{itemize}
    \item The surface area $dA$ projects onto a plane perpendicular to $x'-x$ is equal to $dA \cos(\theta)$, where $\theta$ is the angle between $\harpoon n$ and $x'-x$ \cite{radiosity}. This operation accounts for the \textit{rotation} of $dA$ and gives us the surface area of $dA$ that is perpendicular to $x$.
    \item Projecting a perpendicular, differential surface onto a unit sphere around $x$ is simply given by the inverse square of their distance \cite{radiosity}. This operation is given by the \textit{inverse square law} and accounts for the distance between $x$ and $x'$.
\end{itemize}

Combining both these operations into a single formula, will take both rotation and distance into consideration \cite{radiosity, c3}:

\begin{equation}\label{SolidAngleProjectionRelation}
\omega = \frac{\cos(\theta) dA}{\lVert  x'-x \rVert ^2}
\end{equation}

The field-of-view that an object occupies on the sensor of a virtual camera can be quantified through a solid angle. What color the corresponding pixels should adopt now depends on the light that the object emits, reflects or refracts towards the camera, which is given by its \textit{radiance}.

\subsection{Radiance}

\textit{Radiance} describes the radiant energy propagating in a given direction $\harpoon\omega$ at a given location $x$.

Let $p(x, \harpoon\omega, \lambda)$ be the volume density of photons of wavelength $\lambda$ at position $x$ that are travelling in direction $\harpoon\omega$, then the corresponding radiance $L(x, \harpoon \omega)$ equates to the product of said photon density and the energy of a single photon $\frac{hc}{\lambda}$, integrated over all wavelengths \cite{radiosity}:

\begin{equation}
L(x, \harpoon \omega) = \int_{\lambda } p(x, \harpoon \omega, \lambda) \frac{hc}{\lambda}
\label{eqn:RadiancePhotons}
\end{equation}

Since the retinas in human eyes consist of three different types of photoreceptor cones (red, green and blue respectively), pixel colors are usually modelled as 3D vectors with each dimension corresponding to a respective color component. A commonly employed format is the RGB 24-bit color depth format, which assigns each component 8 bits of depth, with another optional 8 bits for transparency in an \textit{alpha channel}.

The overall magnitude of a color-vector is a measure of its overall energy, which corresponds to the radiometric quantity of \textit{flux}.

\subsection{Radiant Flux}

In computer graphics the quantum nature of light (photon density) tends to be discarded in favour of \textit{radiant flux} $\Phi$, which amounts to a general measure of radiant energy per unit time \cite{radiosity, c3}:

\begin{equation}
\Phi = \frac{\partial Q}{\partial t} [W]
\end{equation}

where $Q$ is the energy emitted, transmitted or reflected.

The total flux a surface $A$ emanates is equal to the total radiance all the points on this surface emit in all directions \cite{radiosity}:

\begin{equation}
\Phi_o = \int_{x\in A} \int_{\omega} L(x, \harpoon \omega)
\end{equation}

Respectively, the radiance exiting a surface in a particular direction is the total flux the surface emits per unit solid angle per unit projected surface area \cite{VXCT}:

\begin{equation}
L = \frac{d\Phi}{d\omega dA_{\perp}} = \frac{d\Phi}{d\omega dA\cos \theta}
\label{eqn:Radiance}
\end{equation}

This formulation of radiance is of particular importance, as it acts as a measure of how bright a surface would appear to a camera in direction $\harpoon \omega$ \cite{radiosity}.

Henceforth we will refer to radiance \textit{exiting} a surface point in a certain direction as $L_o(x, \harpoon \omega)$ and radiance \textit{incident} on that point from $\harpoon \omega$ as $L_i(x, \harpoon \omega)$.

Under the assumption that a surface does not emit any light of its own, we can discern that the radiance $L_o$ emitted towards a camera would have to be less or equal to the flux incident on the surface, as per conservation of energy.
The total flux incident on a given surface can be quantified by the value of \textit{irradiance}.

\subsection{Irradiance and Radiant Exitance}

The total flux incident on a surface per unit surface area is termed the \textit{irradiance} $E$ of that surface (sometimes known as \textit{illuminance}) \cite{c3, radiosity}:

\begin{equation}
E = \frac{d\Phi_i}{dA} [W/m^2]
\label{eqn:Irradiance}
\end{equation}

It is equivalent to the radiance incident from all directions in a hemisphere $\Omega$ above the surface \cite{c3, radiosity}:

\begin{equation}
E = \frac{d\Phi_i}{dA} = \int_{\Omega} L_i(\omega) \cos \theta d\omega
\label{eqn:IrradianceExpanded}
\end{equation}

\begin{figure}[h]
\centering
\includegraphics[scale=1.0]{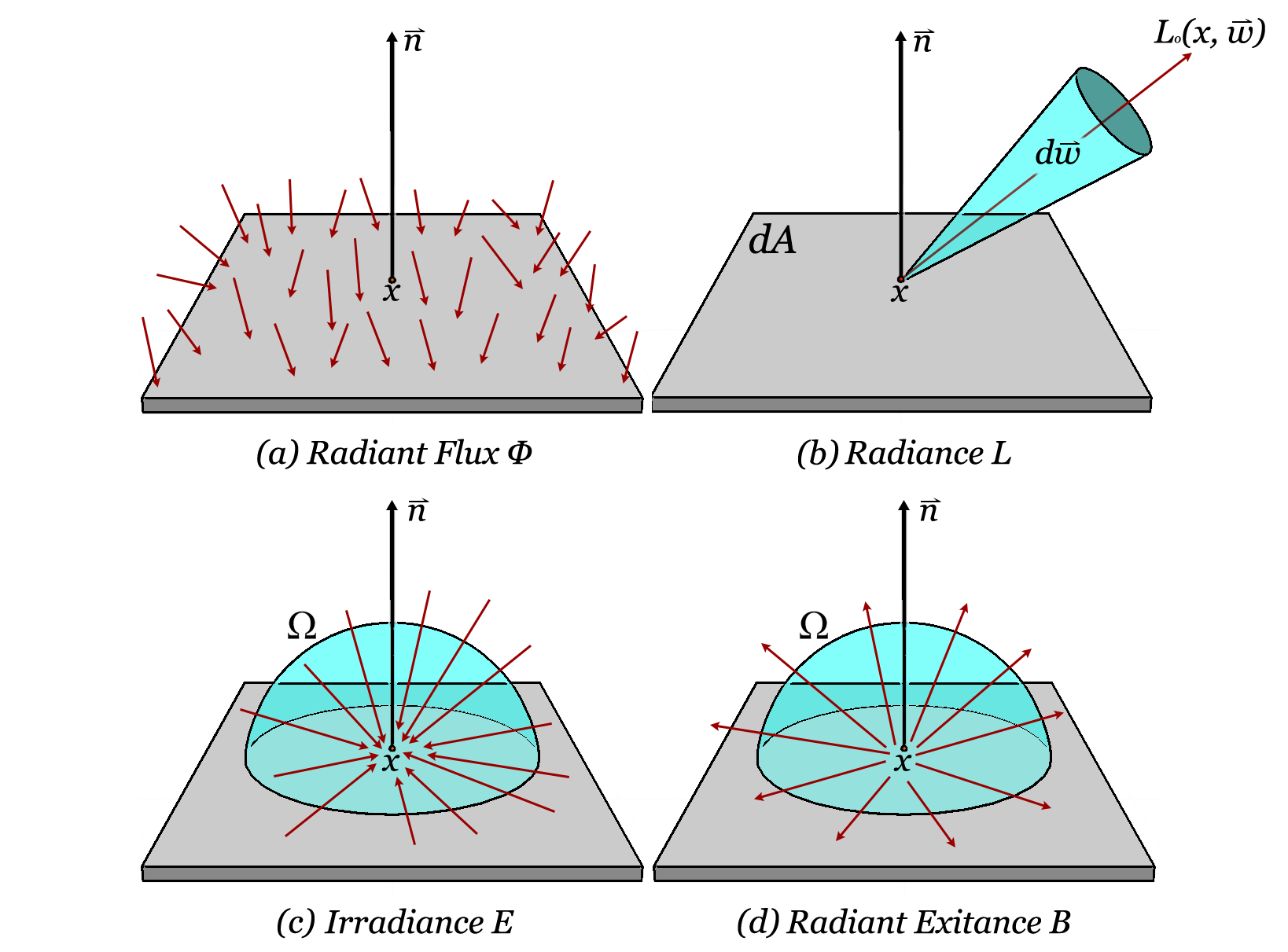}
\decoRule
\caption[]{Visualization of flux, radiance, irradiance and radiant exitance (a-d). Images based on depictions by Jarosz et al. \cite{jarosz12theory}.}
\end{figure}

The opposite of irradiance is the \textit{radiant exitance} $B$, which is defined as the flux per surface
area leaving or being emanated from a surface \cite{Radiometry_Book}:

\begin{equation}
B = \frac{d\Phi_o}{dA} = \int_{\Omega} L_o(\omega) \cos \theta d\omega
\end{equation}

The first law of thermodynamics dictates that energy is neither created nor destroyed. In the context of computer graphics it implies that the radiance reflected by a non-emissive surface must be less or equal to the radiance it receives.

Put differently, the radiance of a point $x$ must be proportional to its irradiance under a coefficient of one or less \cite{radiosity}:

\begin{equation}
dL_o(\harpoon \omega_o) \propto dE(\harpoon \omega_i)
\end{equation}

The coefficient of proportionality that ordains this relationship is given by the \textit{bidirectional reflectance distribution function} of $x$.

\subsection{BRDF and the Reflectance Equation}

For any given pair of differential solid angles (i.e. directions) $\harpoon \omega_i$ and $\harpoon \omega_o$, a material's {\it{bidirectional reflectance distribution function}} (BRDF) defines the ratio of flux concentration per steradian incident from $\harpoon \omega_i$ that is reflected into $\harpoon \omega_o$ \cite{radiosity, fu_cg}:

\begin{equation}
f_r(\harpoon\omega_i \rightarrow \harpoon\omega_o) = \frac{L_o(\harpoon\omega_o)}{E(\harpoon\omega_i)} = \frac{L_o(\harpoon\omega_o)}{L_i(\harpoon\omega_i) \cos \theta d\omega_i}
\end{equation}

If we solve this equation for $L_o$ and perform the same hemispherical integral over the set of all incident directions $\Omega$ as in (\ref{eqn:IrradianceExpanded}), we arrive at the total amount of light reflected by a surface in a specified direction, also known as the {\it{reflectance equation}} \cite{radiosity}:

\begin{equation}\label{eqn:ReflectanceEquation}
\begin{aligned}
& f_r(\harpoon \omega_i \rightarrow \harpoon \omega_o) = \frac{L_o(\harpoon\omega_o)}{E(\harpoon\omega_i)} \\
\Leftrightarrow   & L_o(\harpoon\omega_o) = \int_{\Omega} f_r(\harpoon\omega_i \rightarrow \harpoon\omega_o) L_i(\harpoon \omega_i) \cos \theta_i d\omega_i
\end{aligned}
\end{equation}

Put simply, the reflectance equation prescribes that the radiance $L_o$ a surface reflects in a particular direction $\harpoon\omega_o$ equates to its BRDF weighted irradiance.

\section{The Rendering Equation}

\begin{figure}[H]
\centering
\includegraphics[scale=0.35]{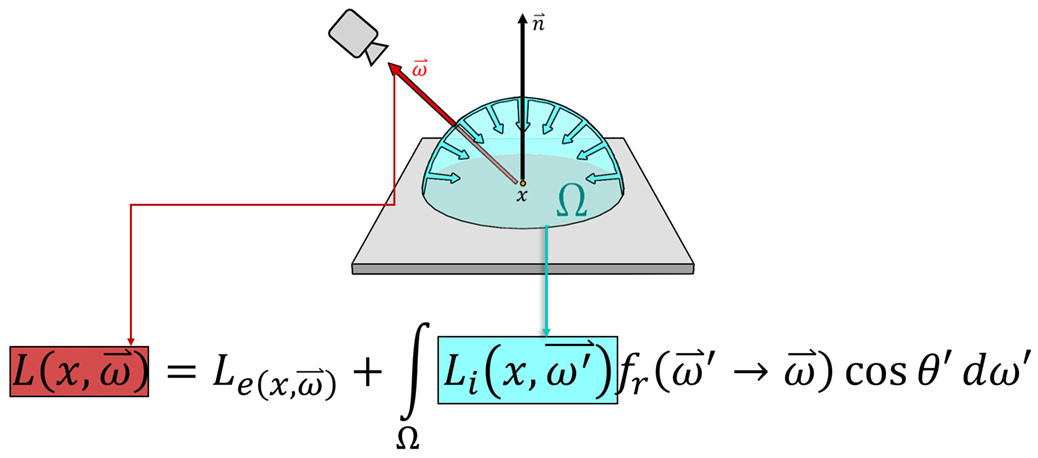}
\decoRule
\caption[]{The fundamentals of the rendering equation: $L$ is a weighted integral over all incident directions.}
\label{RendEquBase}
\end{figure}

The reflectance equation yields the light a surface point reflects towards a camera. Including a term $L_e(x, \harpoon\omega)$ for emission (light the point $x$ emits itself) provides the total radiance the surface emanates in a direction $\harpoon \omega$ \cite{radiosity} (see fig. \ref{RendEquBase}).
This sum constitutes the \textit{Rendering Equation}, as originally formulated by Kayija et al. in 1986 \cite{Kajiya, radiosity}:

\begin{equation}\label{RenderingEquation_general}
\begin{aligned}
L_o(x, \harpoon\omega) & = L_e(x, \harpoon\omega) + \int_{\Omega} f_r(\harpoon\omega_i, \harpoon\omega, x) L_i(x, \harpoon \omega_i) (\harpoon \omega_i \cdot \harpoon {n_x}) d\omega_i
\end{aligned}
\end{equation}

The rendering equation states that "the transport intensity of light from one surface point to another is simply the sum of emitted light and the total light intensity which is scattered toward $x$ form all other surface points" \cite{Kajiya}.

An alternative formulation of this equation can be derived by replacing the hemispheric integral of directions by an integral over all other surface points \cite{Kajiya, radiosity}.

Let $S$ be the set of all surfaces in the scene, then the irradiance $E$ incident on a given surface point $x$ is the integral over all light leaving any other surface point towards $x$, so long as $x$ and the other point $x'$ are mutually visible. The term of visibility - or occlusion - is given by a function $V(x, x')$ which is equal to 1 if $x$ and $x'$ are mutually visible, 0 if not:

\begin{equation}
E = \int_{S} L_o(x', \harpoon \omega') V(x, x') \frac{\cos \theta \cos \theta'}{\lVert x-x' \rVert^2}dA
\end{equation}

The multiplication of the two cosines $\cos{\theta}$ and $\cos{\theta'}$ accounts for the mutually projected surface area of the two locations and the division by the square of their distance stems from the inverse-square law.

Applying this equation for irradiance to the reflectance equation in (\ref{eqn:ReflectanceEquation}) and adding the emissive component $L_e$, leaves us with the following variant of the rendering equation \cite{Kajiya, radiosity}:

\begin{equation}\label{RenderingEquation_surface}
L_o(x, \harpoon\omega) = L_e(x, \harpoon\omega) + \int_{S} f_r(\harpoon\omega', \harpoon\omega, x) L_o(x', \harpoon \omega') V(x, x') \frac{\cos \theta \cos \theta'}{\lVert x-x' \rVert ^2}dA
\end{equation}

For the sake of simplicity, we will henceforth refer to these individual variants as the \textit{hemispheric-} and \textit{surface-based} rendering equation respectively.

\section{Specular and Diffuse BRDFs}

Materials we encounter in reality tend to be highly granular and possess immense detail on a microscopic scale, which leads to light being reflected in complex distributions of outgoing directions that our simple, computational models cannot fully replicate.

\begin{figure}[th]
\centering
\includegraphics[scale=0.7]{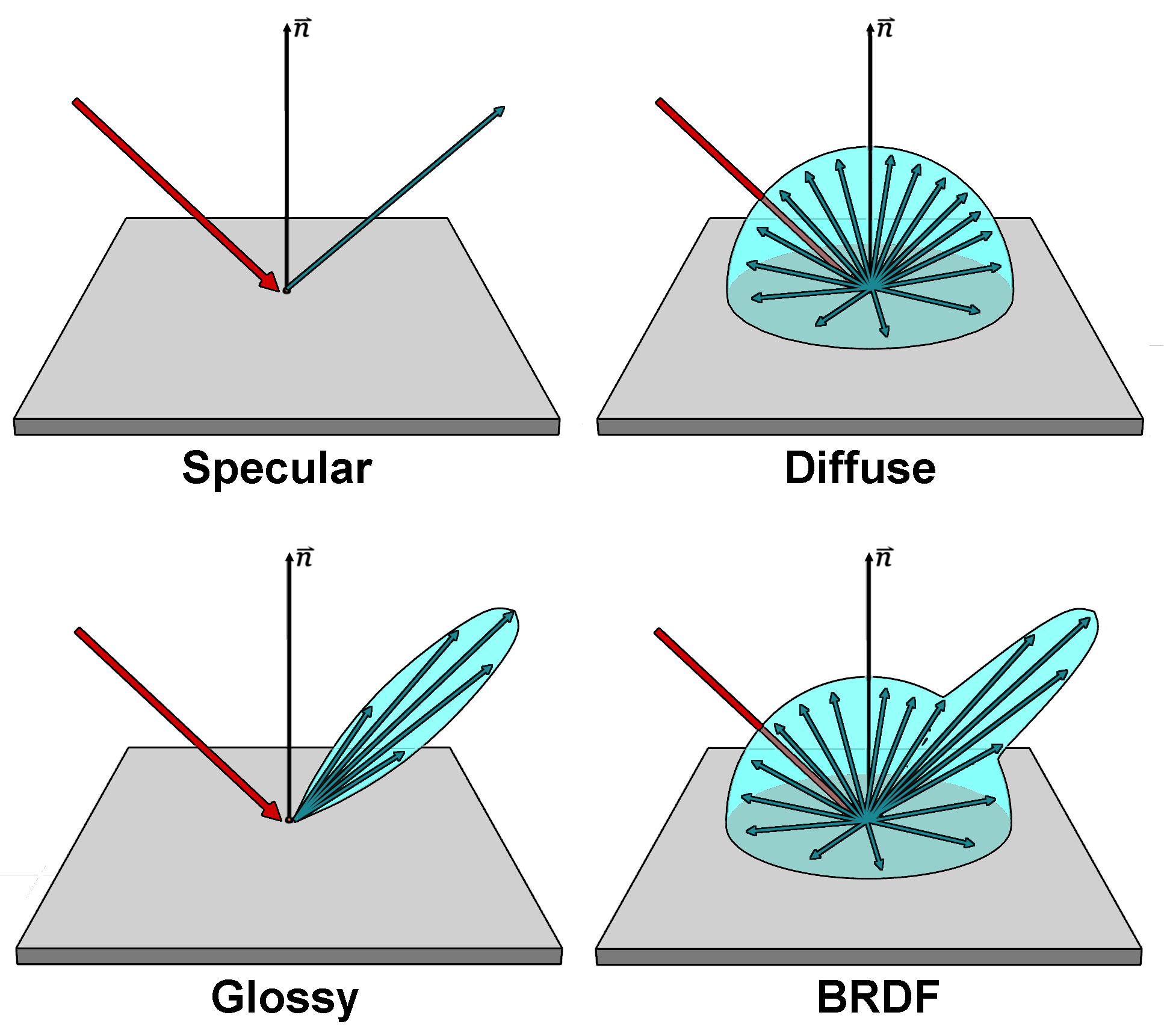}
\decoRule
\caption[]{Reflectance components as part of a BRDF for a given angle of incidence (in red).}
\label{ReflectanceComponents}
\end{figure}

\begin{figure}[th]
\centering
\includegraphics[scale=1.27]{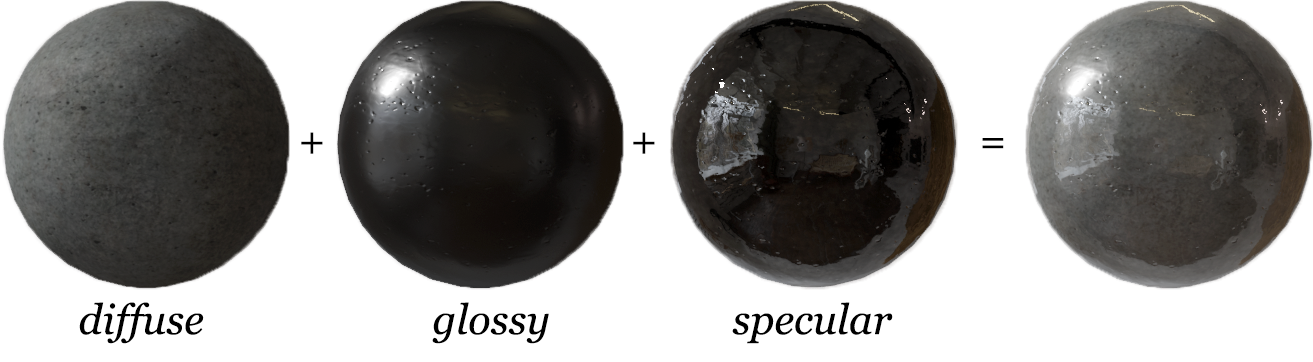}
\decoRule
\caption[]{Reflectance components on a sphere rendered in the Blender 2.8 EEVEE engine.}
\label{ReflectanceComponent_Render}
\end{figure}

The tendency in computer graphics is to differentiate between three distinct reflectance components: diffuse, specular and glossy \cite{radiosity, VXCT}, as portrayed in fig. \ref{ReflectanceComponents} and fig. \ref{ReflectanceComponent_Render}. These components are weighed in various proportions to one another, depending on the underlying material's physical properties and parameters.

\subsection{Specular Reflections}\label{SpecularReflection}

In a \textit{specular} reflection the incident lights' trajectory is perfectly mirrored across the surface's normal vector.
We can model this behaviour by applying a \textit{dirac delta function} to the angle around the normal \cite{radiosity}:

\begin{equation}
f_r(\harpoon \omega_i \rightarrow \harpoon \omega_o) = \frac{\delta(\cos \theta_i - \cos \theta_o)}{\cos \theta_i} \delta(\phi_i - (\phi_o \pm \pi))
\end{equation}

The values of $\theta$ and $\phi$ correspond to the angles with and around the upward axis respectively. The delta function $\delta$ yields zero for any non-zero parameter.

\subsection{Glossy Reflections}\label{GlossyReflection}

Glossy reflections typically describe specular reflections with a small to moderate amount of scattering and variation. A material's \textit{glossiness} determines the degree of diffusion that occurs.

Mathematically, a broader version of the delta function can be modelled by taking the dot product of two normalized vectors and raising it to some high exponent \cite{learnopengl, VXCT}.

Let $\harpoon R$ be the vector of a perfectly specular reflection and ${\harpoon\omega_o}$ be a vector that points from a surface towards the camera, then the glossy radiance equates to the following \cite{learnopengl, phong, VXCT}:

\begin{equation}
L_o(\harpoon\omega_o) = L_{i}d\omega_{i}(\max(\harpoon R\cdot\harpoon\omega_o), 0)^{\alpha}
\end{equation}

The value of $\max(\harpoon R\cdot\harpoon\omega_o), 0)$ is zero for angles larger than 90° and increases if $\harpoon R$ and ${\harpoon\omega_o}$ are of similar angle to the surface normal. The glossiness $\alpha$ dictates the narrowness of the underlying pseudo-delta function.

\subsection{Diffuse Reflections}\label{DiffuseReflection}

As depicted in fig. \ref{ReflectanceComponents}, \textit{labertian diffuse} reflections are defined as completely isotropic, where the apparent brightness and surface color remain equal for all angles of observation \cite{radiosity}:

\begin{equation}
\forall \alpha, \beta \in \Omega: \quad f_r(x \rightarrow \alpha) = f_r(x \rightarrow \beta)
\end{equation}

The corresponding BRDF $f_r$ thus acts as a constant, since it retains the same value for all parameters.

Replacing the BRDF in the reflectance equation (\ref{eqn:ReflectanceEquation}) by a constant allows for it to be separated from the integrand, providing us with a constant fraction of the irradiance $E$ \cite{radiosity}:

\begin{equation}
\begin{aligned}
L_o(\harpoon\omega_o) & = \int_{\Omega} f_r L_i(\harpoon \omega_i) \cos \theta_i d\omega_i \\
& = f_r \int_{\Omega} L_i(\harpoon \omega_i) \cos \theta_i d\omega_i \\
& = f_r E
\end{aligned}
\end{equation}

Intuitively, this equation describes that if the light incident on a surface gets scattered evenly, then the same fraction of the irradiance is reflected in all directions.

Since genuine surfaces typically absorb a portion of the light incident on them, it is useful to define a measure of \textit{reflectivity} $\rho$ that defines what percentage of irradiance is reflected into radiant exitance \cite{radiosity}:

\begin{equation}
\rho = \frac{B}{E} = \frac{\int_{\Omega_o} L_o(\harpoon\omega_o)\cos \theta_o d\omega_o}{E}
\end{equation}

$L_o$ is constant for all directions and $\Omega$ constitutes a hemisphere, as such, $\rho$ constitutes the BRDF constant multiplied by $\pi$ \cite{radiosity}: 

\begin{equation}
\begin{aligned}
\rho & = \frac{\int_{\Omega_o} L_o(\harpoon\omega_o)\cos \theta_o d\omega_o}{E} \\
& = \frac{L_o \int_{\Omega_o} \cos \theta_o d\omega_o}{E} \\
& = \frac{L_o \pi }{E} \\
& = \pi f_r
\end{aligned}
\end{equation}

Which, in turn, implies that the BRDF for a lambertian diffuse reflection is equal to a multiplication by $\frac{\rho}{\pi}$ \cite{radiosity}:

\begin{equation}\label{equn:diffuse_brdf}
    f_r = \frac{\rho}{\pi}
\end{equation}

Diffuse reflections are the most difficult type of reflections to model accurately, as they require knowing the total light incident for \textit{all} directions.
 
\section{Rasterization}

The rendering equation provides a mathematical model of what brightness and color a given surface adopts for a given observer. To compile this information into an actual image, we require the set of \textit{pixels} on the image that a surface occupies.

In practice, three-dimensional scenes are usually described by a series of simple polygons - known as \textit{primitives} - that span areas between 3D points known as \textit{vertices}.
The conversion process of a primitive into a corresponding set of pixels that it occupies on a screen is known as \textit{rasterization} and forms a vital step in the majority of rendering pipelines. A more detailed account of this process is given in section \ref{rast_pipeline}.

\begin{figure}[H]
\centering
\includegraphics[scale=0.19]{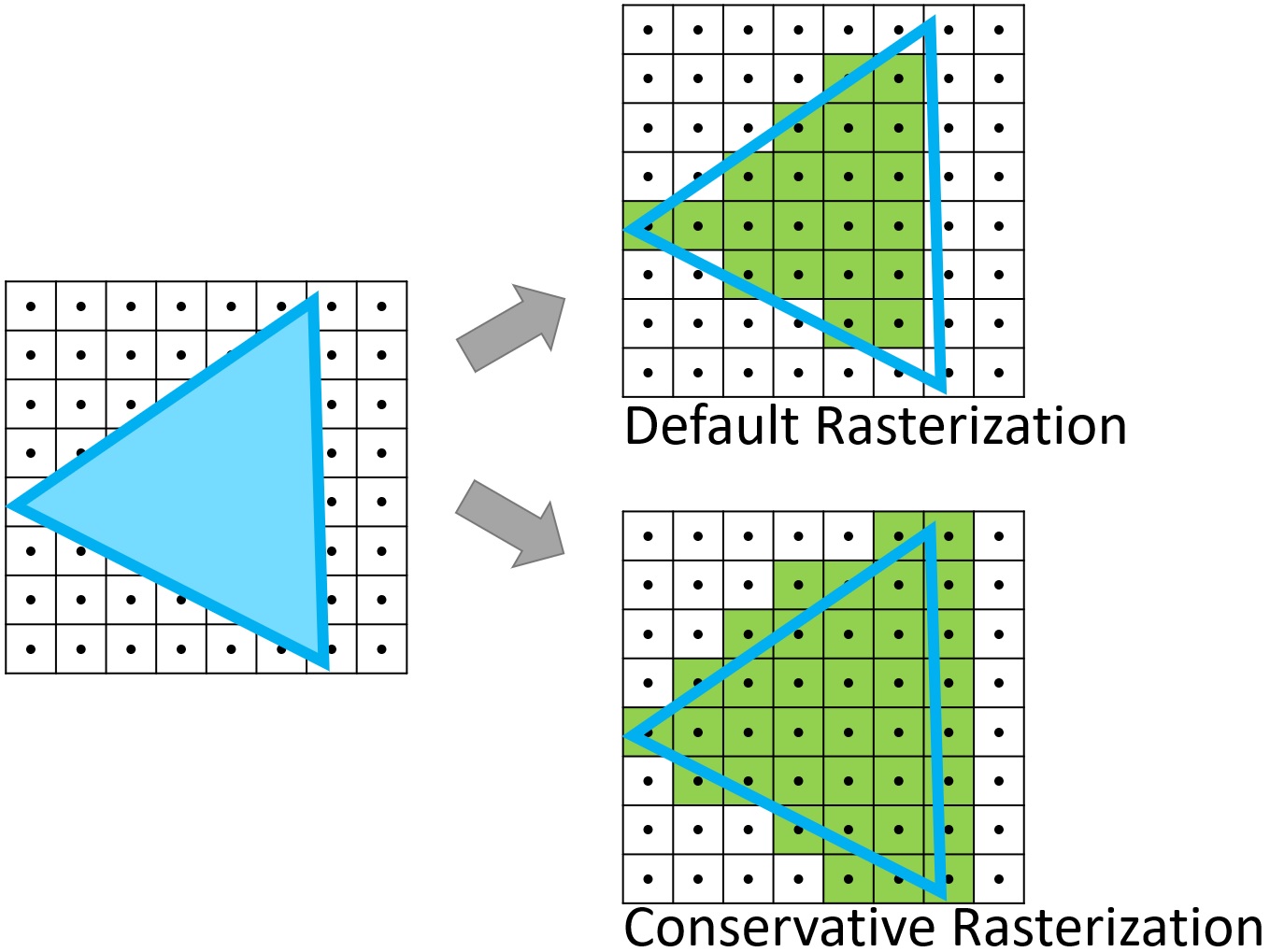}
\decoRule
\caption[Rasterization]{Default and conservative rasterization in comparison: Default rasterization only rasterizes fragments if their center is covered by the triangle. Conservative rasterization additionally includes all fragments partially covered by the triangle.}
\label{def_vs_const_rast}
\end{figure}

The most common rasterization methods triangulate higher degree polygons and proceed to only rasterize pixels if their center lies completely inside a triangle. \textit{Conservative rasterization} can add some certainty to pixel rendering, as all pixels that are at least partially covered by a rendered primitive are rasterized \cite{directx12_docs} (see fig. \ref{def_vs_const_rast}).

\begin{figure}[H]
\centering
\includegraphics[scale=0.48]{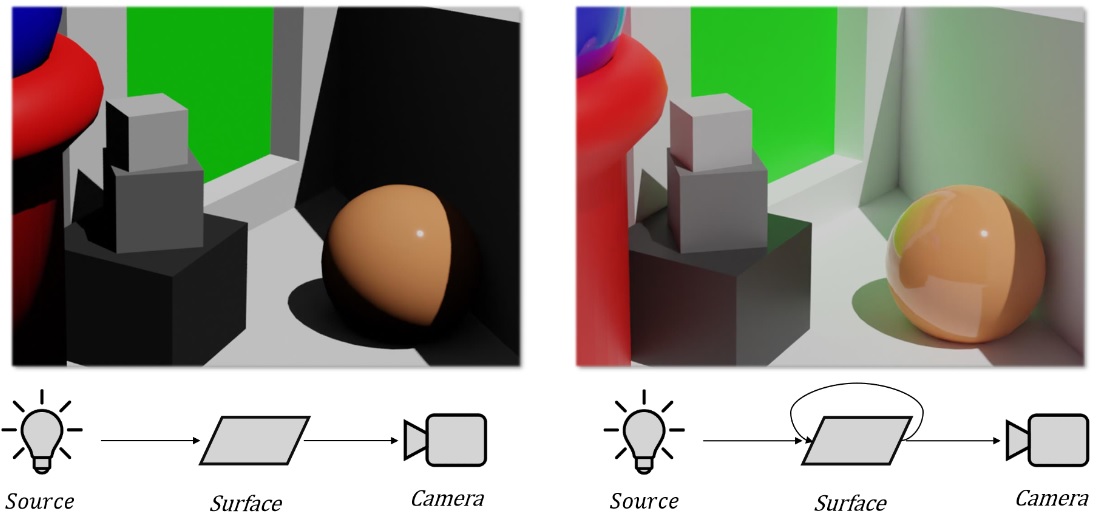}
\decoRule
\caption[Rasterization]{Local illumination (left) vs. global illumination (right) in the Unity 2020.2 Engine. Note in particular the indirect, green light on the sphere and angled surfaces.}
\label{gi_vs_li}
\end{figure}

Once the pixels a surface occupies have been determined, a plethora of different lighting models can be applied to compute the color for each pixel individually, such as the \textit{Phong local illumination model} \cite{phong}.
So called \textit{global illumination} models will include indirect light, i.e. light that bounces more than once before reaching the camera (see fig. \ref{gi_vs_li}).

All illumination models are, in essence, an exercise in solving the rendering equation through approximation.

\section{Raytracing}\label{Raytracing}

The infinite recursion and integration of the rendering equation might appear to have one simple solution: Apply a maximum recursion-depth limit and approximate the hemispheric integral as a finite sum of directions. This exact thought-process lies behind \textit{raytracing}, which has long stood as one the simplest and most intuitive approaches of numerically solving the rendering equation.

The core concept is simple and consists of tracing rays starting from the camera position through each pixel of the virtual screen and then setting the pixel colors based on the surfaces the respective rays collide with \cite{rtx_gems, fu_cg}.
Shooting rays directly from the camera makes a discrete rasterization unnecessary.

\begin{figure}[H]
\centering
\includegraphics[scale=0.5]{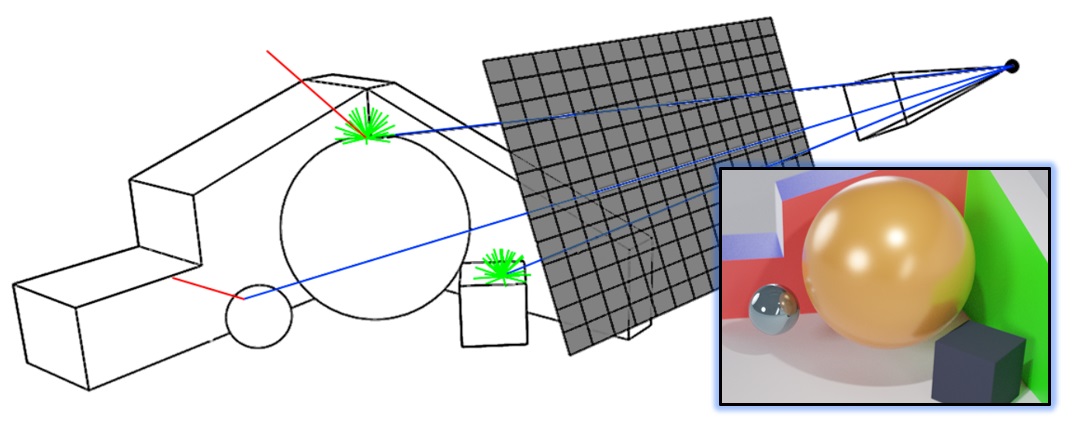}
\decoRule
\caption[Raytracing]{Illustration of the raytracing process with the resulting image. Specular rays are marked in red, diffuse rays in green and initial rays in blue.}
\label{raytracing_example}
\end{figure}

Computing a surface's lit color usually involves launching further sets of rays, which depend on the surface's material properties as well as the specific raytracing variant that is being utilized \cite{rtx_gems}.
For instance, if the intersection surface is smooth, a specular ray is cast in the reflection direction and whichever surface this \textit{specular ray} collides with will be mirror-reflected by the former surface. Rough surfaces, on the other hand, require sampling a myriad of rays to account for diffuse light (see fig. \ref{raytracing_example}).

This process occurs recursively, with each intersection shooting its own set of rays until either a light source is hit or a maximum amount of bounces is reached. Whereupon the tree of rays is evaluated bottom-up until, at the root, the pixel's final color can be calculated \cite{rtx_gems}.

This process is, in essence, a brute-force approach at computing the Riemann sum of the hemispheric rendering equation, as defined in (\ref{RenderingEquation_general}):

\begin{equation}
L_o(x, \harpoon\omega) = L_e(x, \harpoon\omega) + \int_{\Omega} f_r(\harpoon\omega_i, \harpoon\omega, x) L_i(\harpoon \omega_i, x) (\harpoon \omega_i \cdot \harpoon {n_x}) d\omega_i
\end{equation}

Let $I(x, \harpoon \omega)$ be a function that yields the closest surface intersected by a ray along $\harpoon\omega$ starting from $x$. The radiance incident at a surface point $x$ from a given direction $\harpoon\omega$ must be equal to the light exiting the closest surface visible from $\harpoon\omega$ in the reverse direction taken to the inverse square of their distance:

\begin{equation}
    L_i(\harpoon \omega, x) = \frac{L_o(I(x, \harpoon \omega), -\harpoon \omega)}{\lVert x - I(x, \harpoon \omega) \rVert^2}
\end{equation}

Based on this concept, raytracing collapses the hemispherical integral of the rendering equation into a finite sum of directions that are randomly and isotropically selected from the hemisphere:

\begin{equation}\label{RenderingEquation_Raytracing}
\begin{aligned}
L_o(x, \harpoon\omega)
= L_e(x, \harpoon\omega) + \frac{1}{|\Omega|}\sum_{\harpoon\omega_i \in \Omega} f_r(\harpoon\omega_i, \harpoon\omega, x)  L_o(I(x, \harpoon \omega_i), -\harpoon \omega) \frac{\harpoon \omega_i \cdot \harpoon {n_x}}{\lVert x - I(x, \harpoon \omega_i) \rVert^2}
\end{aligned}
\end{equation}

where $\Omega$ is such a finite set of directions. 

Despite being fundamentally simple, raytracing can produce highly photorealistic images because physical effects like refractions or caustics can easily be replicated. However, it does come at an equally high computational cost, as the recursive process can create a colossal amount of rays, all of which have to undergo a vast amount of intersection tests with the scene's geometry.

\subsection{Ray Definition}
\label{RayDefinition}

Rays follow the mathematical description of a three-dimensional half-line constituted by an origin $o$ and a direction $\harpoon d$ \cite{rtx_gems}. In its parametric form, a ray can be described as:

\begin{equation}\label{eqn:ray_common}
    R(t) = o + t\harpoon d \;\;\;\;\text{for } 0\leq t < \infty
\end{equation}

In practice it can be useful to compute cosines between vectors via dot-products, as is done in glossy reflections (see section \ref{GlossyReflection}). To facilitate this process, direction vectors are often restricted to be normalized unit vectors $\hat{d}$ \cite{rtx_gems}.

This restriction also implies that the distance from the origin is directly represented by $t$.
More generally, the difference in $t$ value is equal to the distance between the respective points \cite{rtx_gems}:

\begin{equation}
    \lVert R(t_1) - R(t_2) \rVert = | t_1 - t_2 |
\end{equation}

\begin{figure}[h]
\centering
\includegraphics[scale=0.5]{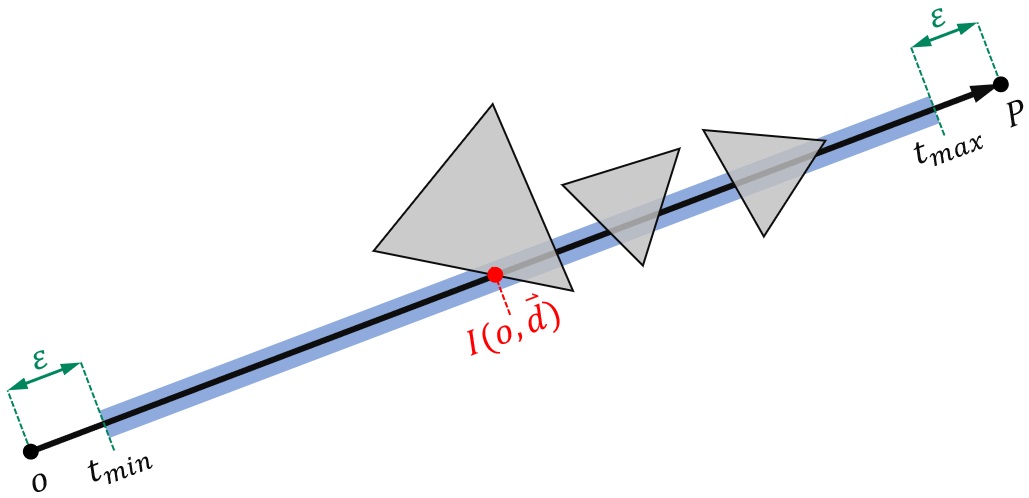}
\decoRule
\caption[Raytracing]{A ray launched from $o$ to $P$. To avoid precision problems the interval is offset by $\epsilon$. Of the three intersections the first is reported as the \textit{closest hit}. Image based on a depiction from \textit{Raytracing Gems} \cite{rtx_gems}.}
\label{ray_def}
\end{figure}

In raytracing, rays are intersected with a finite amount of geometry to determine collision points. As such, the semi-infinite description of a ray is not practical. Instead, rays frequently are defined with an additional interval that constitutes the range of t-values for which an intersection is useful: $t \in [t_{min}, t_{max}]$. Intersections that lie outside this interval are not reported (see fig. \ref{ray_def}).

The utility of this interval comes in the form of several advantages: Firstly, the minimum value can help prevent self-intersections with the geometry itself that can arise from floating-point inaccuracies \cite{rtx_gems}. Secondly, the maximum value can speed up intersection calculations when hits beyond a certain point do not matter, like with a shadow ray that is shot towards a light-source \cite{rtx_gems}.

\subsection{Ray-triangle Intersection}

Ray-triangle intersection tests are at the heart of the raytracing algorithm. Given a triangle $T$ with the three vertices $V_1$, $V_2$ and $V_3$ as well as a ray $R$ with origin $o$ and direction $\harpoon d$, we seek the intersection point where $R$ is equal to $T$.

One of the most commonly employed intersection algorithms in computer graphics is the \textit{Möller-Trumbore} algorithm \cite{moeller_trumbore}, which functions with minimal memory requirements by defining triangles in their parametric form of their \textit{barycentric coordinates}:

\begin{equation}\label{eqn:Triangle_bary}
    T(u,v) = (1-u-v)V_1 + uV_2 + vV_3
\end{equation}

where the barycentric coordinates $u$ and $v$ must fulfill the conditions $u\geq0$, $v\geq0$ and $u+v\leq1$.

\begin{figure}[th]
\centering
\includegraphics[scale=0.7]{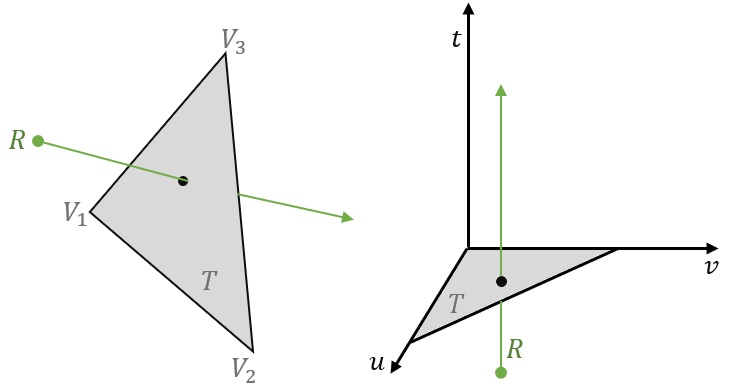}
\decoRule
\caption[Raytracing]{The system of linear equations produced in Möller-Trumbore algorithm effectively expresses the point of intersection in $t$, $u$, $v$ space. The unit triangle spanned by $u$ and $v$ corresponds to $T$. Image based on depiction by Scratchapixel \cite{moeller_trumbore_image}.}
\label{moeller-trumbore}
\end{figure}

Equating $T$ with the definition for a ray put forth in (\ref{eqn:ray_common}) yields the following expression \cite{moeller_trumbore}:

\begin{equation}
\begin{aligned}
R(t) & = T(u,v) \\
\Longleftrightarrow o + t\harpoon d & = (1-u-v)V_1 + uV_2 + vV_3 \\
\Longleftrightarrow o + t\harpoon d & = V_1 + u(V_2 - V_1) + v(V_3 - V_1) \\
\Longleftrightarrow o - V_1 & = -t\harpoon d + u(V_2 - V_1) + v(V_3 - V_1)
\end{aligned}
\end{equation}

which can be written as a linear system of equations \cite{moeller_trumbore}:

\begin{equation}
    o - V_1 = \begin{bmatrix}-D & (V_2 - V_1) & (V_3 - V_1)\end{bmatrix} \begin{bmatrix}t\\u\\v\end{bmatrix}
\end{equation}

This process is equivalent with expressing the point of intersection in a coordinate system spanned by the $t$, $u$ and $v$ axes, as depicted in fig. \ref{moeller-trumbore}.

There are several methods for solving linear equation systems, but the one employed by Möller and Trumbore is \textit{Cramer's Rule} \cite{moeller_trumbore}.

The resulting $t$, $u$ and $v$ values give both the distance from the ray origin as well as the barycentric triangle coordinates of the intersection. If the conditions from (\ref{eqn:Triangle_bary}) are not met or $t$ lies outside of its defined interval, then no intersection exists \cite{moeller_trumbore}.

Running this algorithm for every triangle in a scene whilst minimizing for $t$ gives us the first - or closest - point that a ray encounters, which corresponds to the function $I$ defined above:

\begin{equation}
\begin{aligned}
I(o,\harpoon d) = o + t\harpoon d \;\;\; & \text{for $t$ being the minimal value in $[t_{min}, t_{max}]$ that satisfies} \\
& \text{$R(o, \harpoon d) = T_i(u,v)$ with} \\
& \text{$u\geq0$, $v\geq0$ and $u+v\leq1$} \\
& \text{for any triangle $T_i$ in the scene}
\end{aligned}
\end{equation}

\begin{figure}[th]
\centering
\includegraphics[scale=0.33]{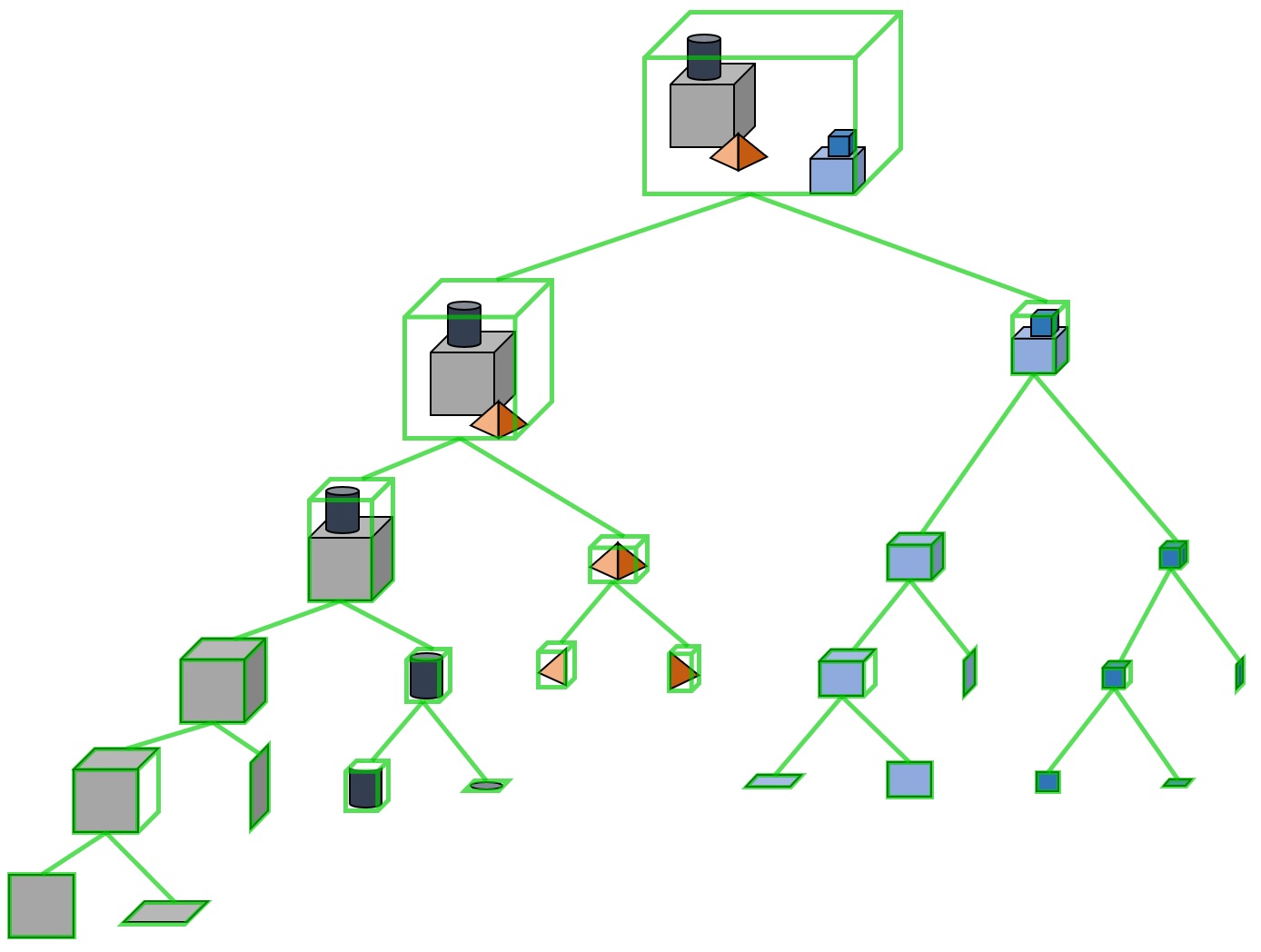}
\decoRule
\caption[BVH]{Simplified illustration of a BVH of a simple 3D scene.}
\label{bvh_example}
\end{figure}

\subsection{Bounding Volume Hierarchies}\label{BVH_theory}

The intersection tests required for tracing a ray can constitute a huge computational endeavour that \textit{acceleration data structures} aim to ameliorate.

Primitives can be grouped into bounding boxes that can in turn be grouped as well, thereby forming a hierarchy \cite{rtx_gems}, as depicted in fig. \ref{bvh_example}. Naturally, if a ray does not intersect a bounding volume it cannot intersect any of the contained primitives.

A \textit{bounding volume hierarchy} (BVH) stores the triangles in the leaves of a tree structure with each node corresponding to a bounding volume. Ray traversal can then commence at the root and progresses into the child nodes, in practice, reducing the time complexity to logarithmic in the number of primitives \cite{rtx_gems, RTX_visibility_koch}.

Other acceleration structures like \textit{binary space partitions} can also be used, but today's consensus is that BVHs are generally the best suited for raytracing, as they guarantee maximum memory usage threshold \cite{rtx_gems}.

\section{Radiosity}

In computer graphics, the term \textit{radiosity} refers to a finite-element approximation of the rendering equation for diffuse reflections. This method has its roots in heat transfer models used in thermal engineering \cite{radiosity_og} and, in contrast to raytracing, is far more adequate for real-time rendering purposes.

In (\ref{equn:diffuse_brdf}) we defined diffuse BRDFs as being constant and irrespective of the angle of observation. This means that, within a static environment, every surface retains its diffuse color no matter what perspective it is being rendered from.
Radiosity makes use of this fact by offloading the computation costs required for global illumination to a phase of pre-computation. Pre-computed values are stored in a so-called \textit{lightmap}, then simply retrieved on demand at minimal cost.

This method can only account for diffuse reflections on static (non-moving) geometry. Any displaced object will require the entire generation process to be repeated, while specular reflections have to be deferred to a different method, such as raytracing \cite{wallace}.

The mathematical basis behind radiosity can be derived from the surface-based rendering equation defined in (\ref{RenderingEquation_surface}):

\begin{equation}
L_o(x, \harpoon\omega) = L_e(x, \harpoon\omega) + \int_{S} f_r(\harpoon\omega', \harpoon\omega, x) L_o(x, \harpoon \omega') V(x, x') \frac{\cos \theta \cos \theta'}{\lVert x-x' \rVert ^2}dA
\end{equation}

Since only diffuse reflections are accounted for, the BRDF $f_r$ can entirely be replaced by the constant $\frac{\rho}{\pi}$ defined in section \ref{DiffuseReflection}. Furthermore, any directional parameters such as $\harpoon\omega$ can be dropped, as lambertian diffuse reflections are equal in all directions \cite{radiosity}:

\begin{equation}
L_o(x) = L_e(x) + \frac{\rho(x)}{\pi} \int_{S} L_o(x) V(x, x') \frac{\cos \theta \cos \theta'}{\lVert x-x' \rVert^2}dA
\end{equation}

\begin{figure}[H]
\centering
\includegraphics[scale=0.9]{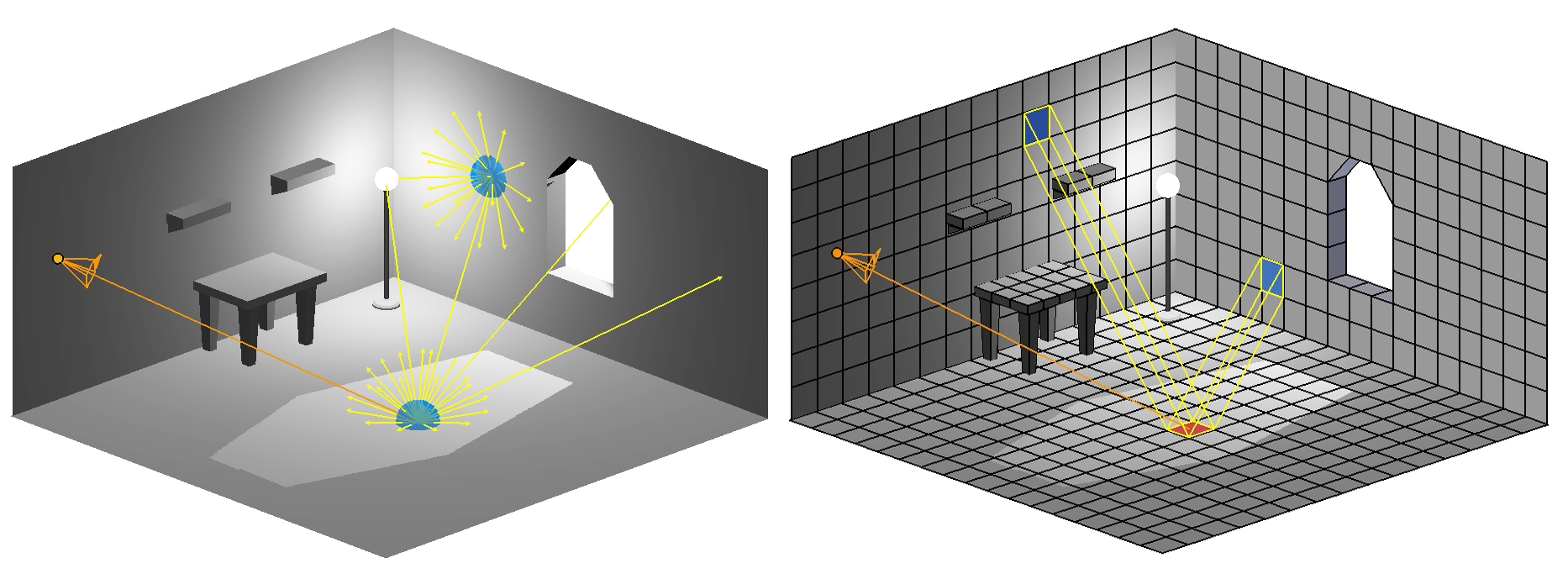}
\decoRule
\caption[]{Illustration of raytracing (left) and radiosity (right), as depicted by Benjamin Kahl \cite{VXCT}.}
\label{raytracing_vs_radiosity}
\end{figure}

Similarly to raytracing, radiosity makes use of the {\it{finite element method}} by subdividing the environment's geometry into a series of small patches, thus collapsing the area-integral into a finite sum \cite{radiosity}. Fig. \ref{raytracing_vs_radiosity} illustrates the two Riemann sums these approaches correspond to.

\begin{figure}[h]
\centering
\includegraphics[scale=0.44]{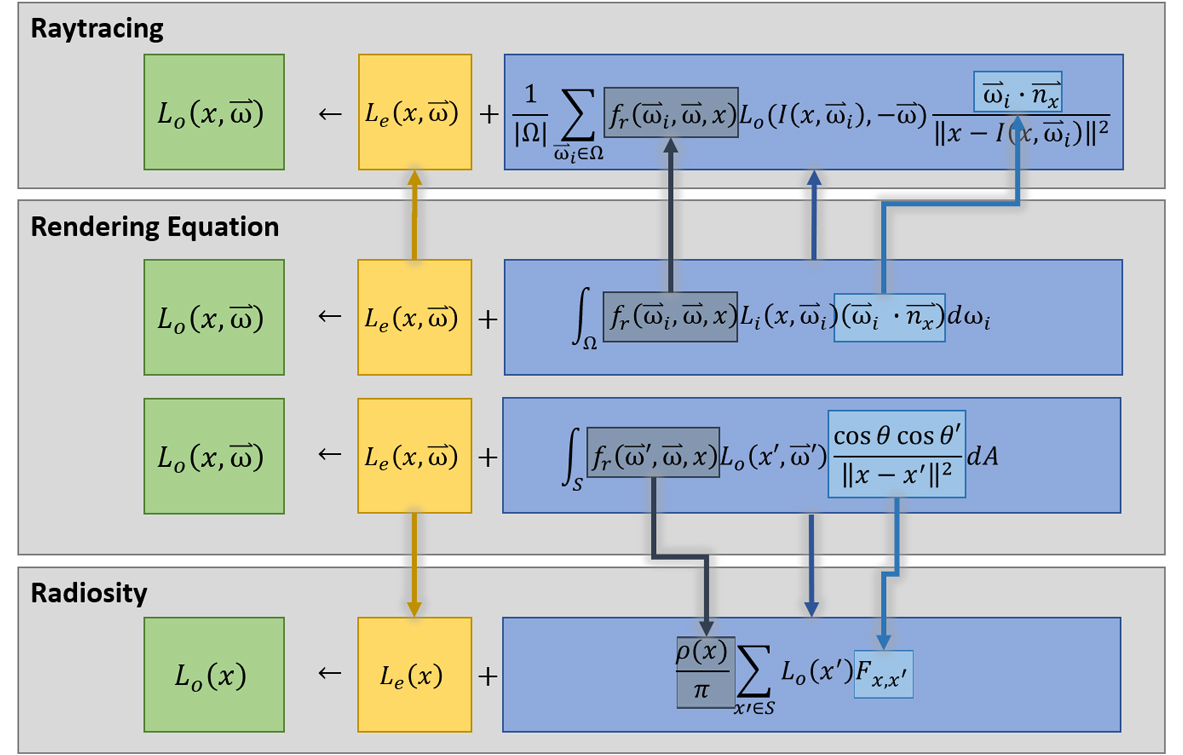}
\decoRule
\caption[]{Approximating the rendering equation through radiosity and raytracing. The inverse square factor is contained within the definition of $L_i$. Image based on a depiction by Benjamin Kahl \cite{VXCT}.}
\label{raytracing_radiosity_eqn_derivation}
\end{figure}

The term $V(x, x') \frac{\cos \theta \cos \theta'}{\lVert x-x' \rVert^2}dA$ is separated into a \textit{view factor} $F_{i,j}$ that describes the fraction of the energy leaving patch $i$ that arrives at $j$. Let $n$ be the amount of surfaces - e.g. patches - in our scene. Then the diffuse radiance of a patch $i$ can be approximated by the following equation \cite{radiosity_og, radiosity}:

\begin{equation}\label{eqn:RadiosityEquation}
L_o(i) = L_e(i) + \frac{\rho(i)}{\pi} \sum_{j=1}^n L_o(j) F_{i,j}
\end{equation}

which constitutes the \textit{Radiosity Equation} as it was formulated in 1984 by Goral et al. \cite{radiosity_og}. The individual components behind this mathematical transformation is illustrated in fig. \ref{raytracing_radiosity_eqn_derivation}.

\subsection{View Factor}

Also known as \textit{form factor}, a view factor $F_{i,j}$ describes how well two surfaces $i$ and $j$ are visible to one another and consists purely of geometric parameters.

In the engineering field of \textit{heat transfer}, view factors are calculated through the same geometric term we abstracted from the rendering equation above, integrated for all point-pairs on the two surfaces \cite{FormFactorLecture}:

\begin{equation}\label{eqn:ViewFactorHeatTransfer}
    F_{i,j} = \frac{1}{A_i} \int_{A_i}\int_{A_j} \frac{\cos \theta_i \cos \theta_j}{\lVert x_i-x_j \rVert^2}dA_idA_j
\end{equation}

The division over the surface area $A_i$ results in the \textit{reciprocity rule}, which states that if $A_i$ and $A_j$ are of equal size, then $F_{i,j}$ is equal to $F_{j,i}$ \cite{FormFactorLecture}:

\begin{equation}\label{eqn:ReciprocityRule}
    F_{i,j}A_i = F_{j,i}A_j
\end{equation}

This implies that if both $A_i$ and $A_j$ are known quantities, only half of the form factors need to computed or stored, as each respective mirror pair follows from $F_{i,j} = F_{j,i}\frac{A_j}{A_i}$.

\subsection{The Nusselt Analog}

The numerical computation of form factors is not a wholly simple task, as differential surfaces are difficult to establish. An analog to differential form factors developed by Wilhem Nusselt can provide some useful intuition for the algorithms that follow \cite{radiosity}.

The Nusselt analog corresponds to the same procedures outlined in the section on solid angles (see fig. \ref{solid_angle_adv}), where a patch $A_j$ is projected onto an imaginary unit hemisphere centered at $A_i$ and then orthogonally down onto the base of the hemisphere \cite{radiosity, hemicube}. 
Thus, the view factor equates to the area projected onto the base divided by the area of the base itself.

\subsubsection{The Hemicube Approximation}\label{ViewFactor_Hemicube}

\begin{figure}[H]
\centering
\includegraphics[scale=0.28]{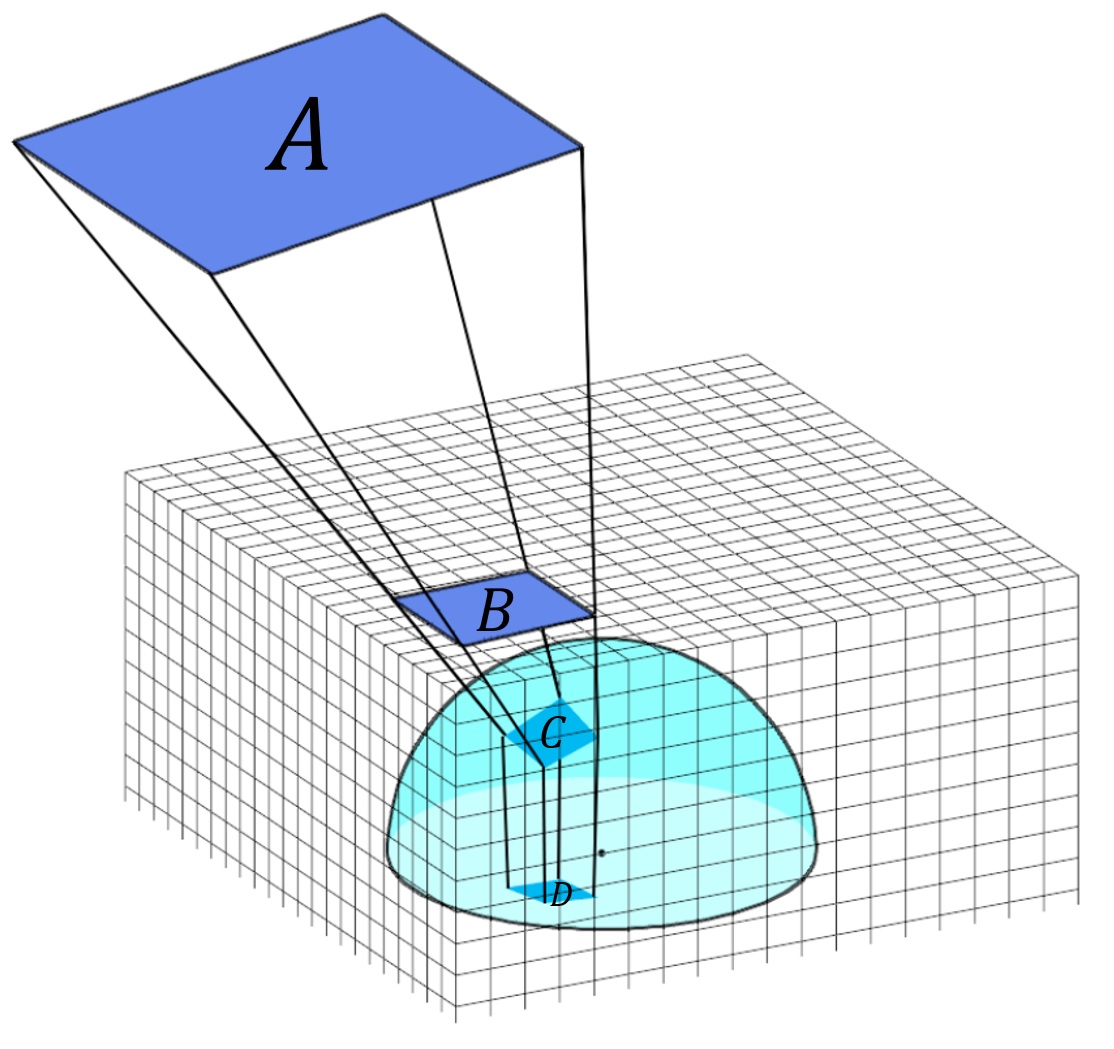}
\decoRule
\caption[]{The justification behind using a hemicube: Patches $A$, $B$ and $C$ have the same view factor, with $D$ corresponding to the Nusselt analog. The size of $B$ can be approximated by the amount of cells it occupies on the hemicube. Depiction based on an image by Watt et al. \cite{wattwatt}.}
\label{hemicube_approx}
\label{hemicube}
\end{figure}

Nusselt's analog illustrates how two differential surfaces that occupy the same solid angle must have the same form factor. Likewise, if a surface is projected radially onto an intermediate surface, such as a hemicube, the form factor of the projection will be the same as that of the original element \cite{radiosity} (see fig. \ref{hemicube_approx}).

This is the justification behind the \textit{hemicube approximation}, devised by Cohen et al. in 1985 \cite{hemicube}.
It approximates the hemisphere with a hemicube, the faces of which are subdivided into small cells.
Once one establishes how many of these cells are occupied by a patch projected onto the hemicube, this results in an amount proportional to the patches' view factor.

\begin{figure}[ht]
\centering
\includegraphics[scale=0.39]{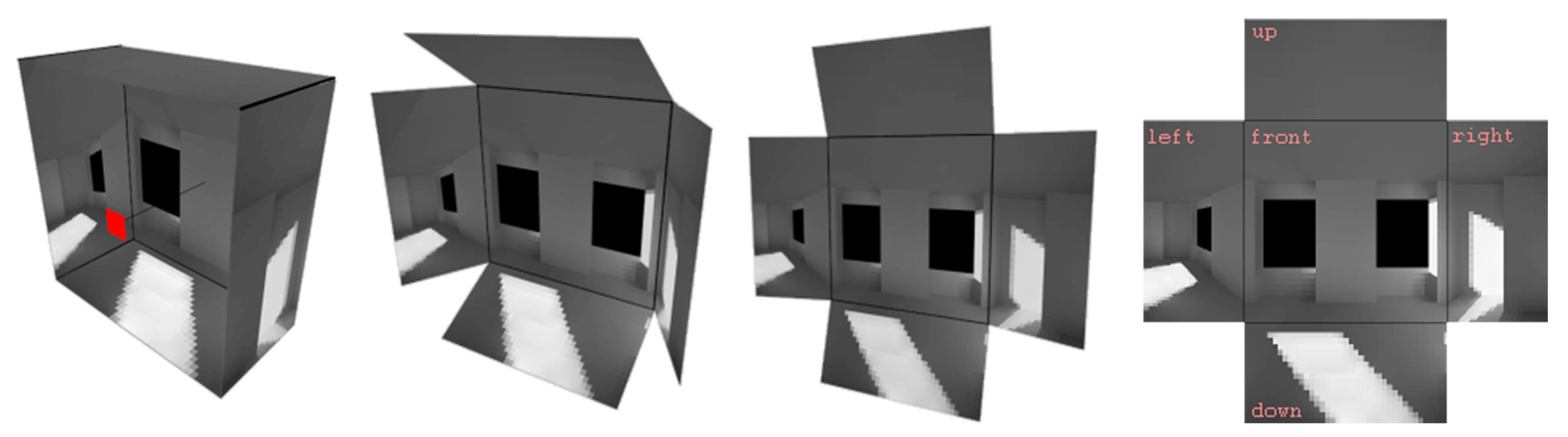}
\decoRule
\caption[]{Rendering a scene through a hemicube to determine visible patches and their view factors, giving an overall approximation of irradiance. Image adapted from work by Hugo Elias \cite{HugoElias_Radiosity}.}
\label{hemicube_elias}
\end{figure}

In graphics card programs the cells of a hemicube can easily be encoded as pixels on a \textit{cubemap}. Figuring out which cells a surface occupies then simply amounts to rasterizing said surface into a cubemap from the perspective of the patch (see fig. \ref{hemicube_elias}). Rasterizing all other surfaces alongside it lets us make a full accounting of which surface has what contribution (view-factor) from the current patch, including the visibility term $V$. This process is closely related to \textit{Z-Buffering}, which is described in further detail in section \ref{zbuffering}.

\subsubsection{Monte-Carlo Integration}\label{ViewFactor_MonteCarlo}

\begin{figure}[H]
\centering
\includegraphics[scale=0.23]{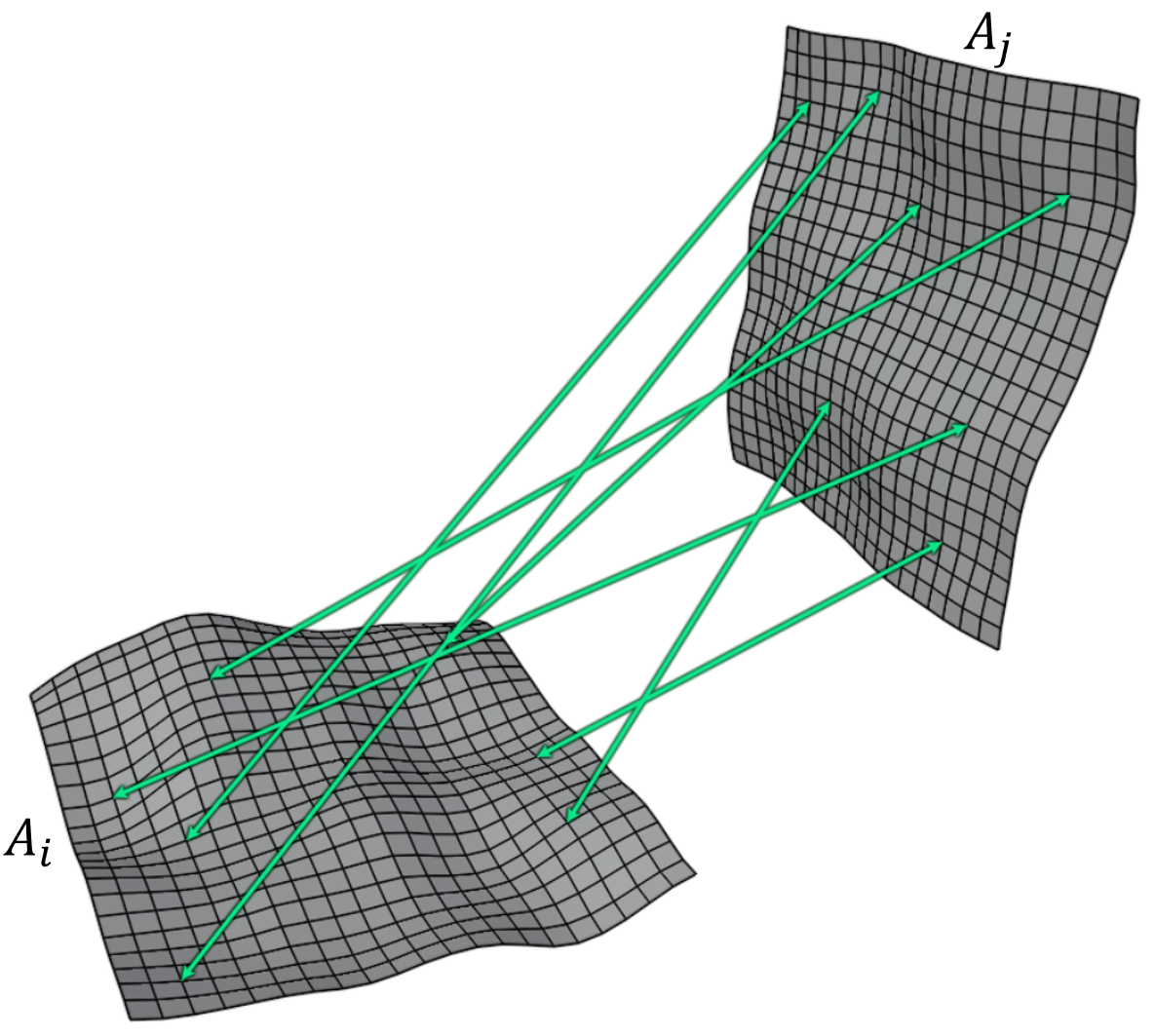}
\decoRule
\caption[]{Form factor calculation requires solving a double integral over surfaces of patches. We can instead randomly sample the 4D space that each patch-pair lies in to get an accurate estimate.}
\label{MC_fiewfactor}
\end{figure}

An alternative method of determining a view factor is to simply use a randomized sample set of pairs that are uniformly distributed points from each surface. Let $x_{ki}$ and $x_{kj}$ be the $k$'th random pair of  points for the two surfaces $A_i$ and $A_j$ respectively. Then the form factor between $A_i$ and $A_j$ can be computed through a Monte-Carlo approximation of $K$ samples \cite{HierarchalMonteCarloRadiosity} (see fig. \ref{MC_fiewfactor}): 

\begin{equation}\label{eqn:ViewFactor_MonteCarlo}
    F_{i,j} = \frac{1}{K} \sum_{k=1}^{K} A_j \frac{\cos \theta_{ki} \cos \theta_{kj}}{\lVert x_{ki}-x_{kj} \rVert^2} V(x_{ki},x_{kj})
\end{equation}

\subsection{Classical Radiosity}

Recall the radiosity equation as defined in (\ref{eqn:RadiosityEquation}):

\begin{equation}
L_o(i) = L_e(i) + \frac{\rho(i)}{\pi} \sum_{j=1}^N L_o(j) F_{i,j}
\end{equation}

Now, where the view factors $F_{i,j}$ have been defined, the radiance values $L_o$ can be formulated as a solution vector, which allows the problem to be entirely expressed as a matrix equation \cite{radiosity, FormFactorLecture}:

\begin{equation}
    \begin{bmatrix}L_o(1) \\ L_o(2) \\ ... \\ L_o(n) \end{bmatrix} = 
    \begin{bmatrix}L_e(1) \\ L_e(2) \\ ... \\ L_e(n) \end{bmatrix} +
    \begin{bmatrix}\rho_1 & 0 & ... & 0 \\ 0 & \rho_2 & ... & 0 \\ ... & ... & ... & ... \\ 0 & 0 & ... & \rho_n \end{bmatrix}
    \begin{bmatrix}F_{11} & F_{12} & ... & F_{1n} \\ F_{21} & F_{22} & ... & F_{2n} \\ ... & ... & ... & ... \\ F_{n1} & F_{n2} & ... & F_{nn} \end{bmatrix}
    \begin{bmatrix}L_o(1) \\ L_o(2) \\ ... \\ L_o(n) \end{bmatrix}
\end{equation}

Through some algebraic transformation (see Cohen et al. \cite{radiosity}), this can formally be written as:

\begin{equation}
    L_o = (I - \rho F)^{-1} L_e
\end{equation}

where $I$ is an identity matrix of size $n\times n$.

Solving this system yields the complete solution to the radiosity equation directly, but requires the entire computational cost to be paid upfront, which becomes prohibitive for larger values of $n$. Instead, it is common practice to solve the equation progressively, with each bounce of light performed separately \cite{Weimar_radiosity}.

\subsection{Progressive Radiosity}\label{ProgressiveRadiosity}

The nature of diffuse reflections combined with the inverse square law implies that lighting values converge rather quickly, which can be leveraged in iterative solutions where each iteration applies the calculations for a single bounce of light. The amount of iterations - e.g. passes - will determine the brightness and fidelity of the scene but also linearly impact the required computation time. Since each subsequent bounce has a lesser impact on the resulting image, fewer than 16 iterations tend to be sufficient in most cases, after which the difference in radiance tends to become non-tangible.

\begin{figure}[h]
\centering
\includegraphics[scale=0.3]{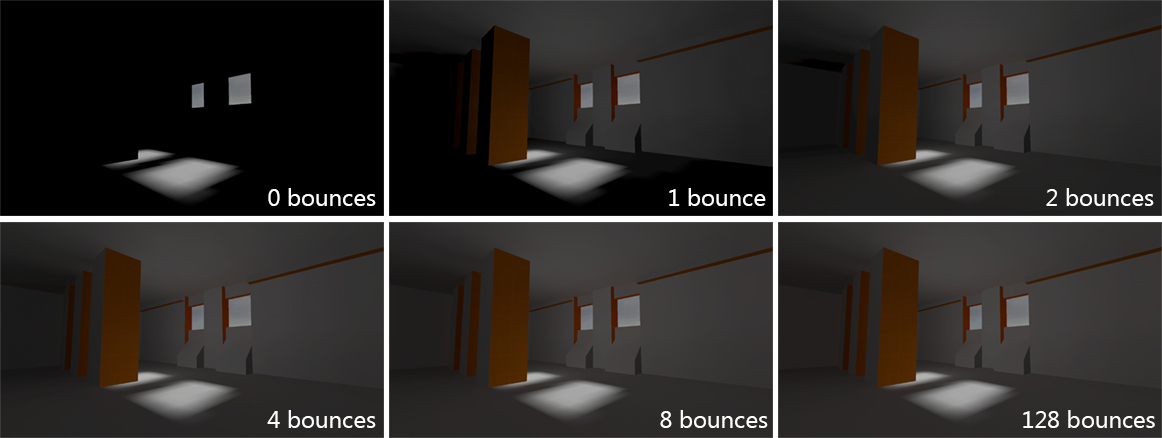}
\decoRule
\caption[]{Progressive radiosity after a specified amount of additional light bounces, as done by the 2014 VRAD tool and rendered by the 2009 Source Engine. Image from work by Benjamin Kahl \cite{VXCT}.}
\label{radiosity_bounces}
\end{figure}

The prevalence of algorithms such as \textit{instant radiosity} \cite{instant_radiosity} and \textit{voxel cone tracing} \cite{Crassin} imply that, frequently, a mere two bounces tend to be sufficient for an adequate approximation of indirect light (see fig. \ref{radiosity_bounces}). Progressive radiosity can be halted after any iteration, once a desired solution has been reached. 

Standard iterative methods for solving matrix equations include the \textit{Jacobi iteration} and the \textit{Gauss-Seidel method} \cite{FormFactorLecture}. The solution can also be configured as a \textit{shooting} or \textit{gathering} variant, depending on which patches are processed in the algorithms outermost for-each loop \cite{gathering_shooting}.

A generalized pseudo-code of progressive radiosity (adapted from Wüthrich \cite{Weimar_radiosity}) is listed below:

\begin{algorithm}[H]
    \caption{Progressive Radiosity}
    \begin{algorithmic}[1]
      \For{each iteration}
        \For{$A_i \in S$}
            \For{$A_j \in S$}
                \State Calculate or retrieve $F_{i,j}$
                \State Update radiosity of $A_j$
                \State Update emission of $A_j$
            \EndFor
            \State Set emission of $A_i$ to 0
        \EndFor
    \EndFor
    \end{algorithmic}
\end{algorithm}

Henceforth, we will refer to this algorithm as \textit{pure} progressive radiosity, to differentiate it from its variants that employ additional enhancements to improve performance.

\subsection{Progressive Refinement Radiosity}
\label{ProgressiveRefinementRadiosity}

An extension to pure radiosity, \textit{progressive refinement radiosity}, was first introduced by Cohen et al. in 1988 \cite{progressive_refinement}. This reformulation of the original algorithm eliminates the memory requirements for view factors entirely by computing them on-the-fly. Patches are processed in sorted order according to their energy contribution to the environment and then updated simultaneously after each pass.

\begin{figure}[th]
\centering
\includegraphics[scale=1.3]{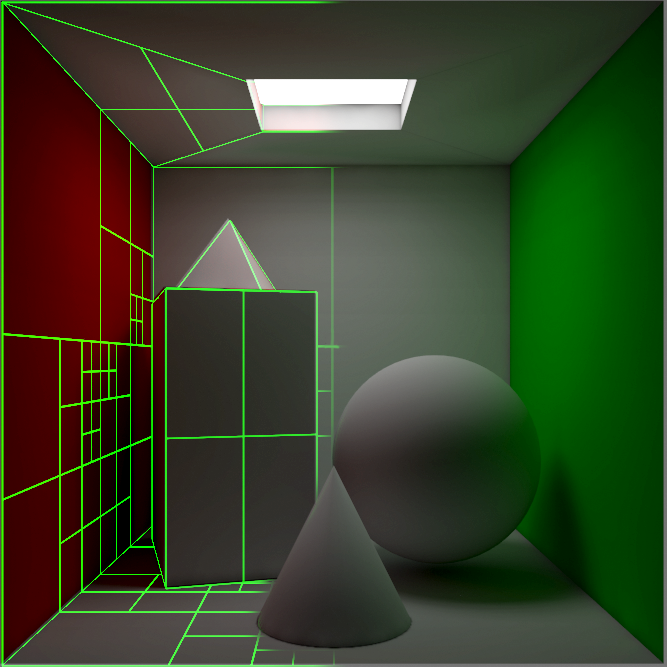}
\decoRule
\caption[]{Radiosity with adaptive subdivision. More patches are allocated to areas with a high gradient, such as shadow boundaries. Image based on a concept by Coombe et al. \cite{RadiosityOnGPUs_Coombe}.}
\label{adaptive_subdivision_scene}
\end{figure}

More importantly is the use of \textit{refinement} through adaptive subdivision, which had already been introduced by Cohen et al. in 1986 \cite{adaptive_subdivision}. This process dynamically sub-divides or merges individual radiosity patches depending on the gradient across them. The resulting \textit{quad-tree} will have more leaves in places of relevance, such as the boundary of a shadow, whilst treating, flat mono-colored surfaces as single patches (as depicted in fig. \ref{adaptive_subdivision_scene}).

Refinement in general is not exclusive to progressive radiosity and can be performed in a number of varieties. Most techniques operate \textit{a posteriori}, meaning they adjust the amount of patches based on the output of each iteration. These are commonly categorized as follows (as done by Slusallek et al. \cite{FormFactorLecture}):

\begin{itemize}
    \item \textit{r-refinement:} Repositions vertices of a mesh based on the lighting gradient.
    \item \textit{h-refinement:} Stores lighting data in a quad-tree and subdivide each node depending on a gradient threshold.
    \item \textit{p-refinement:} Increases polynomial order of patches depending on a gradient threshold.
    \item \textit{remeshing:} Re-computes an entirely new mesh, with edges and vertices aligned along shadow boundaries.
\end{itemize}

The primary goal behind refinement is to drive down the need for large patch amounts, thus improving performance.

Some solutions (see Coombe et al. \cite{gpu_gems_2005}) model the patches that are sampled (i.e shot towards) as separate sub-set of the set of all patches in scene. This allows one to maintain a large set of evenly distributed patches, whilst only sampling the most important ones, in accordance to a subdivided quad-tree.

\subsection{Instant Radiosity and Sampling Approaches}\label{instant_radiosity_section}

To complement radiosity's slow rate of convergence and static geometry constraints, \textit{Instant Radiosity} was introduced by Keller in 1997 \cite{instant_radiosity}. It approximates global illumination effects by creating additional, \textit{virtual light-sources} on inter-reflecting surfaces, making it perfectly fit to be used within real-time requirements \cite{instant_radiosity_2}.

\begin{figure}[th]
\centering
\includegraphics[scale=0.25]{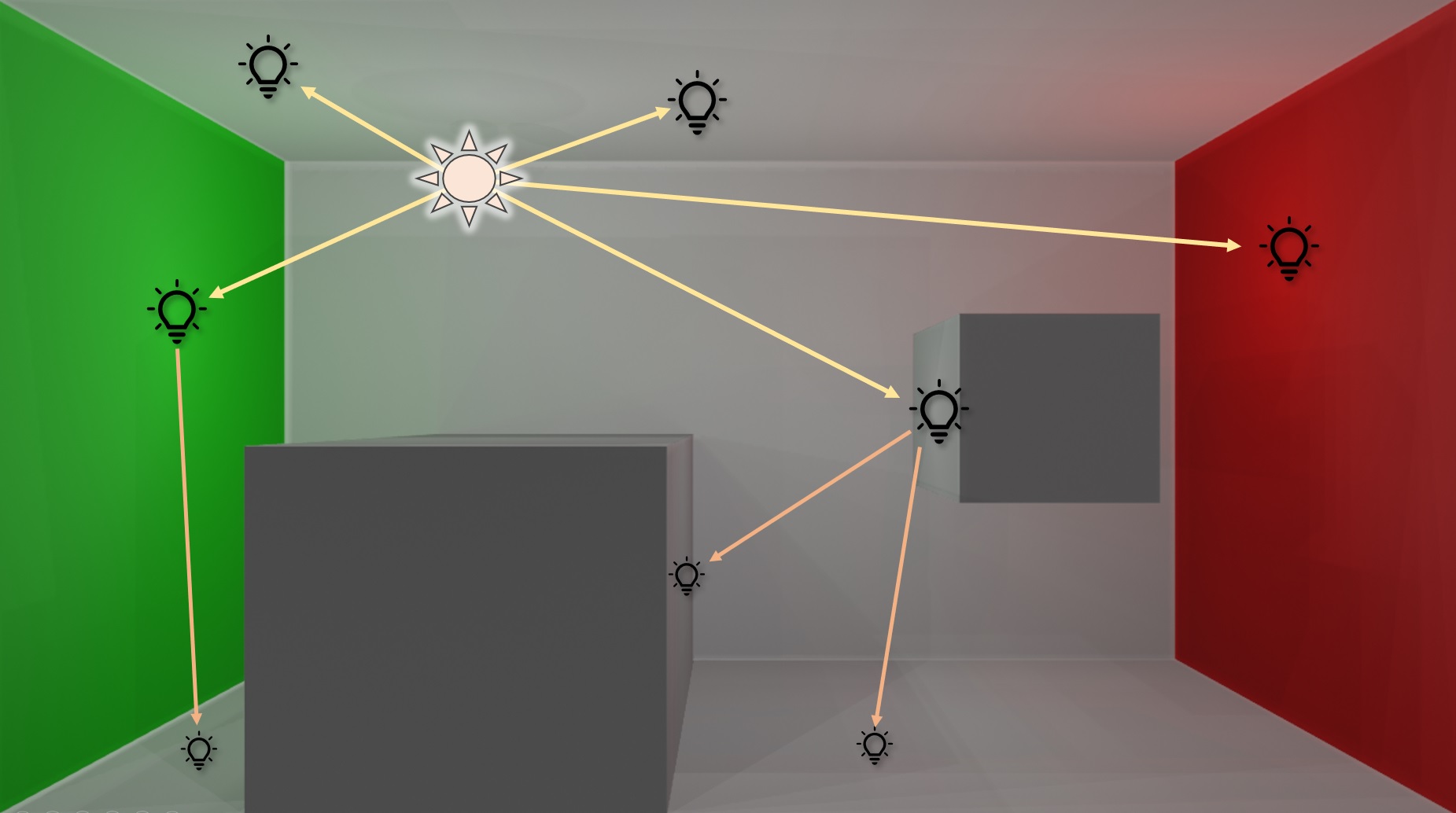}
\decoRule
\caption[]{Concept behind instant radiosity. Rays are scattered from a light-source. On their points of collision we place a new, virtual point-light. The entire scene (including virtual point-lights) can then be rendered using a model like phong.}
\label{instant_radiosity_concept}
\end{figure}

Rays are shot in random directions from light-sources. Then, at their intersection locations with other geometry, \textit{virtual point lights} (VPLs) are created that emit light corresponding with the underlying surface's color and brightness (see fig. \ref{instant_radiosity_concept}). The process can be repeated for each of these light-sources recursively, with the amount of virtual point-lights ultimately determining the quality of global illumination.

Instant radiosity does not require costly pre-computations and can accommodate dynamic scene changes on demand, but generally does not produce the same lighting and shadow quality as regular radiosity, and requires significant amount of GPU power to run in real-time. \textit{Incremental} instant radiosity \cite{instant_radiosity_aalto} allows incrementally adding new VPLs over time, whilst maintaining an even distribution of VPLs across the scene.

Instant radiosity can loosely be classified as a \textit{directional sampling} approach. In regular radiosity, visibility is determined separately for all possible patch-pairs, which results in a complexity of at least $O(n^2)$ in the number of patches. In directional sampling approaches, we isotropically scatter a set of rays 
for each patch, then calculate lighting contribution for each patch hit by a ray. In theory this drops the complexity to $O(n*m)$, where $m$ is the maximum amount of samples taken for each patch.

\section{Visibility Determination}

The surface-based rendering equation in (\ref{RenderingEquation_surface}) defines a function $V(x, x')$ that is equal to 1 if no other geometry lies between the two points $x$ and $x'$, otherwise 0.

Determination of visibility has been a cornerstone problem in computer graphics from its very beginning.
Conundrums such as the \textit{art gallery problem}, \textit{watchman route problem} or graph visibility provide great insight into the theoretical perspective in that the underlying issue is NP-hard \cite{Watchman_Route}.

A common solution in computer graphics is to leverage the celerity of rasterization pipelines to calculate approximate visibility through a Z-Buffer. This method is frequently employed for the calculation of shadows, whilst a binary-space partition (BSP)\footnote{Sometimes referred to as Portal-Engine / Portal rendering} restricts the selection to only relevant geometry \cite{BittnerVisibility}.

\begin{figure}[th]
\centering
\includegraphics[scale=0.75]{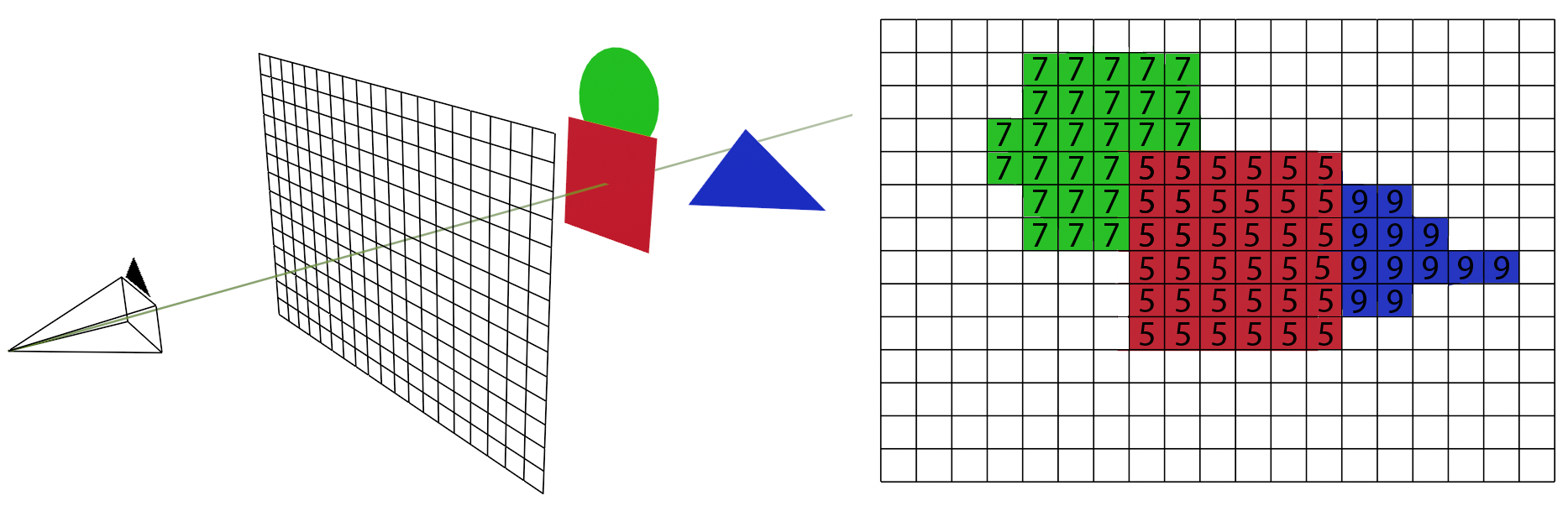}
\decoRule
\caption[]{Rasterization of basic shapes (left) and respective z-buffer (right).}
\label{zbuffer}
\end{figure}

\subsection{Z-Buffering}\label{zbuffering}

A z-buffer (or depth buffer) consists of the clip-space z-coordinates for every pixel of a rendered primitive. These values correspond to the "image depth" or the distance of the painted geometry to the camera \cite{fu_cg} (see \ref{zbuffer} for an example). Z-buffers are regular by-products of the rasterization process.

To determine whether an object is visible to a certain location, a scene can be rasterized from the perspective of the given point, as described by the hemicube approximation (see \ref{ViewFactor_Hemicube}). Instead of the object's colors, it is sufficient to simply store a unique ID for each object into the pixels of the raster image. The result will contain the IDs of all visible objects, barring translucent or small surfaces that occupy less than a pixel \cite{ProducerScroungers, fu_cg}. 

\begin{figure}[th]
\centering
\includegraphics[scale=0.55]{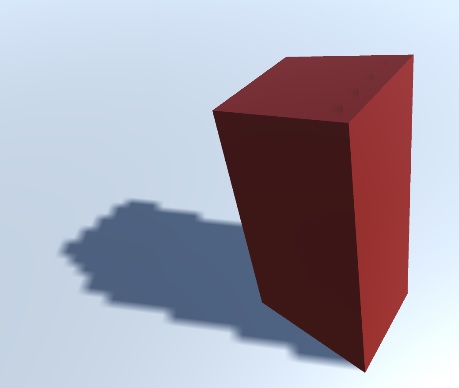}
\decoRule
\caption[]{Red cuboid with a pixelated shadow as a result of z-buffering (Unity Engine 2021.3.21f).}
\label{pixelated_shadows}
\end{figure}

This method is frequently used to generate shadows in real-time, although a low render resolution can impair the quality  and can lead to shadow pixelation (see fig. \ref{pixelated_shadows}).

\subsection{Raytracing for Visibility}\label{RaytracingForVisibility}

Instead of translating a 3D location to a screen position, we can translate a screen position to a 3D location by tracing a ray through the corresponding pixel in the image plane. Likewise, two locations $x_1$ and $x_2$ are mutually visible if a ray launched from $x_1$ can reach $x_2$ unimpeded:

\begin{equation}
    V(x_1, x_2) \leftrightarrow I(x_1, x_2 - x_1) = x_2
\end{equation}

Whilst z-Buffering can be very quick in computing the visibility of several objects from a single location (like with shadows of a point-light), it is not as well-suited for the visibility computation of large sets of arbitrary point-pairs, as is required by radiosity.


\chapter{The Turing Architecture and DXR} 

\label{Chapter3} 



In August 2018 the multinational tech company \textit{Nvidia} introduced the first consumer products capable of genuine real-time raytracing in the form of the \textit{GeForce 20 series} of GPUs.

Built on a newly developed \textit{Turing microarchitecture}, these chips subsume several different types of specialized processor cores to accelerate their respective tasks considerably \cite{turing_whitepaper}. The so-called \textit{RT core} is designed specifically to process the traversal of bounding volume hierarchies and triangle intersection tests, thereby enabling Turing GPUs to execute rudimentary raytracing algorithms at a rate of several billions of rays per second \cite{rtx_gems, turing_whitepaper}.

In this chapter we will review the exact makeup of this architecture and what its key enablers are for the significant boost in graphics performance. Additionally we will outline its embedded solution for real-time raytracing and how it can be used through the DirectX 12 raytracing pipeline.

\section{GPUs and Parallelism}

GPUs (graphics processing units) can be generally defined as specialized processing units designed for the quick computation and management of visual data in a frame buffer.

In essence, their primary intended task is to continuously compute 2D matrices of color values that represent pixels on a screen.
For most intents and purposes (excluding post-processing effects such as anti-aliasing, bloom or blurring), the color of any one individual pixel is independent of its predecessor and neighbours. As such, this task is highly adequate for parallelization.

\begin{figure}[th]
\centering
\includegraphics[scale=0.5]{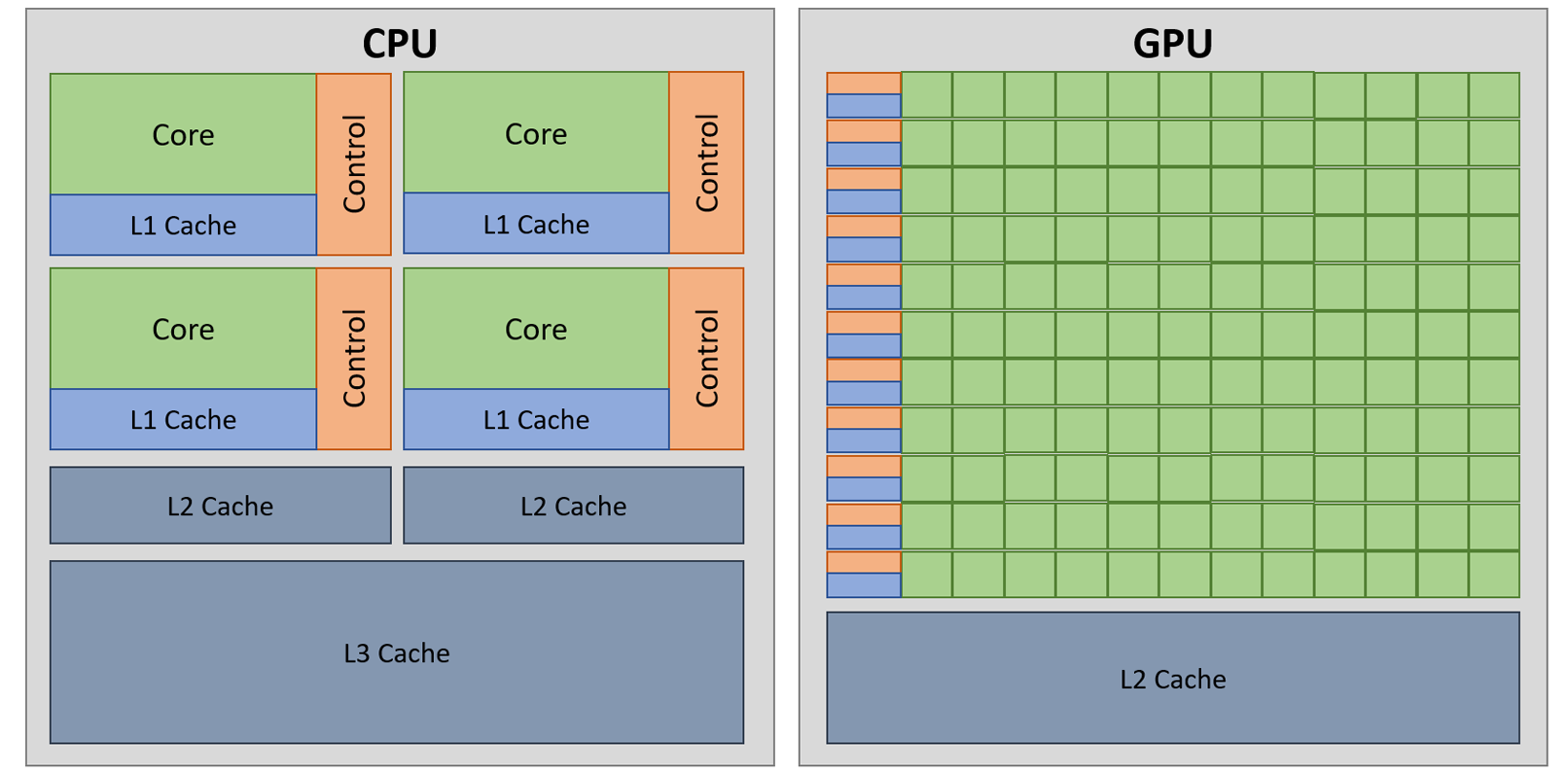}
\decoRule
\caption[]{General architecture of a CPU (left) vs. a GPU (right). Image adapted from the Nvidia CUDA programming guide 2022 \cite{cuda_guide}.}
\label{fig:cpuvgpu}
\end{figure}

Fig. \ref{fig:cpuvgpu} shows the generalized differences in architecture between CPUs and GPUs: The latter contain hundreds to thousands of cores/ALUs (arithmetic logical units) that can run a large set of lightweight threads simultaneously, whilst CPUs are geared towards a small number of highly efficient, general-purpose cores and fast access to system memory.

Due to their immense power in parallel computing, GPUs have found use in many other areas than graphics, such as machine learning and cryptography.

\section{The Turing Architecture}

The Turing GPU microarchitecture (named after \textit{Alan Turing}) was introduced in August 2018 \cite{turing_whitepaper} and is the architecture that Nvidia's newest range of consumer-grade graphics processors are built upon. 

Turing cards inherit large parts of their design from earlier microarchitectures, namely the 2010 \textit{Fermi} \cite{FermiArchitecture} architecture and its successors (\textit{Kepler (2012)}, \textit{Maxwell (2014)} and \textit{Pascal (2016)}).
Turing's key enabler for its new features and improved performance is the overhauled GPU processor which accommodates "improved shader execution efficiency, and a new memory system architecture that includes support for the latest GDDR6 memory technology"\cite{turing_whitepaper}.

\subsection{TU102 GPU Structure}

The TU102 is a a high-end GPU of the GeForce 20 series and is divided into six Graphics Processing Clusters (GPCs) each of which contains a hybrid makeup of computational units \cite{turing_whitepaper}. Below we provide a description of its primary components in line with the \textit{Turing architecture whitepaper} \cite{turing_whitepaper}.

The GPU receives its instructions and data through a PCIe 3.0 interface, which connects it to the rest of the computer and acts as the main access point to the systems main memory \cite{turing_whitepaper}.
Each GPC comes equipped with its own L1 cache, whilst all six GPCs share twelve memory controllers and an additional 512kb of L2 cache \cite{turing_whitepaper}.

An onboard, chip-level \textit{GigaThread Engine} receives command queues from the host via the PCIe bus and executes these by setting up shaders on available hardware \cite{FermiArchitecture}. This is the central command processor that manages the entire chip, including context switches, scheduling and power management \cite{FermiArchitecture}.

\subsection{GPC}

\begin{figure}[H]
\centering
\includegraphics[scale=0.42]{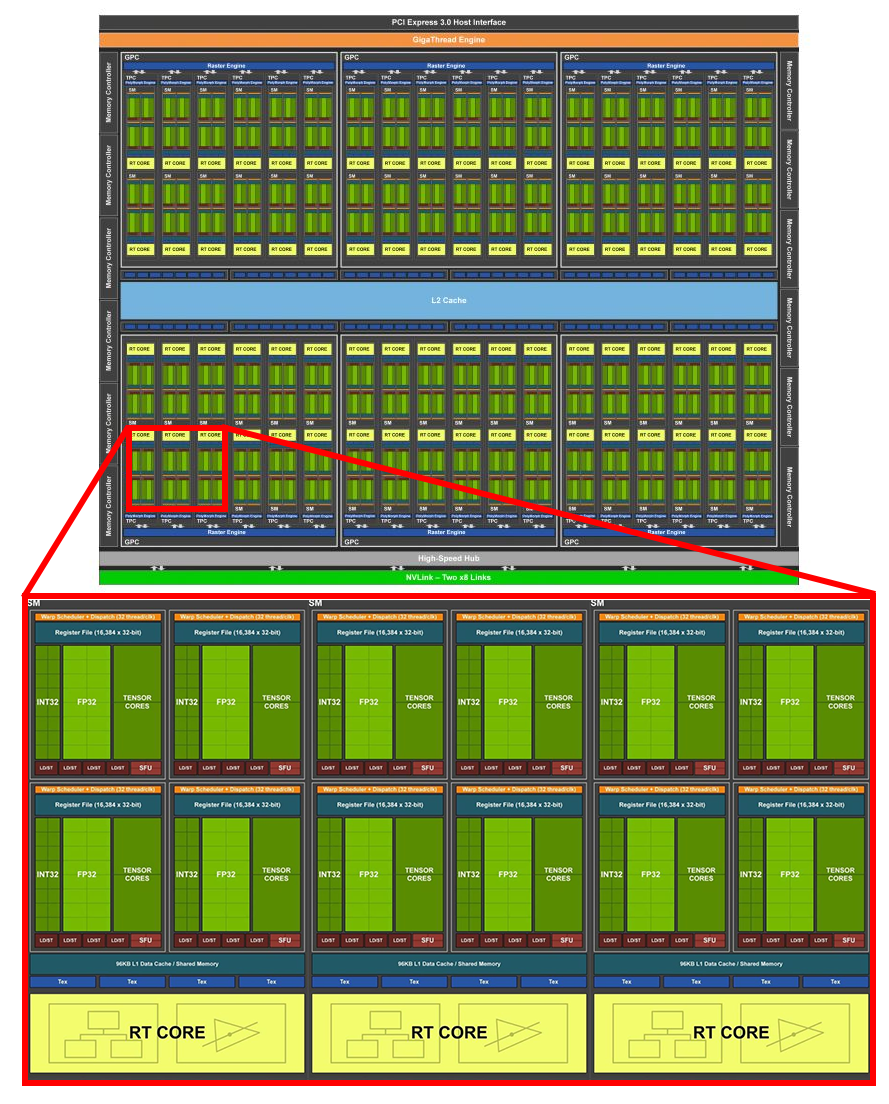}
\decoRule
\caption[]{Architecture of the TU102: 6 GPCs, each consisting of 6 TPCs that contain 2 SMs each. Each SM contains a number of ALUs for integer, floating point and matrix operations as well as a single RT core. Each RT core has a separate core for BVH traversal and triangle intersections respectively. Image adapted from the Turing architecture whitepaper \cite{turing_whitepaper}.}
\label{GPC_overview}
\end{figure}

Each GPC, as depicted in fig. \ref{GPC_overview}, houses six \textit{Texture Processor Clusters} (TPCs), which are nested sub-clusters in of themselves, as well as a dedicated raster engine \cite{turing_whitepaper}. The GPC/TPC subdivision is useful for distributing and allocating of the computational workload evenly, as they can be managed as a single unit \cite{turing_whitepaper}.

Each TPC consists of two Streaming Multiprocessors (SM), which the parent TPC can wake and sleep based on how heavy the incoming workload is \cite{turing_whitepaper}.

\subsection{SM - Streaming Multiprocessor}

The streaming multiprocessor houses all the primary computation units of the GPU itself.
Each one contains its own onboard memory in the form of a 256KB register file, four texture units and 89 KB of L1 shared memory which can be dynamically allocated depending on the workload \cite{turing_whitepaper}.

Each SM contains three different types of processing cores that represent the GPU's entire computational capacity \cite{turing_whitepaper}:

\begin{itemize}
    \item 64 CUDA cores (4608 total)
    \item 2 Tensor cores (576 total)
    \item 1 RT core (72 total)
\end{itemize}

\begin{figure}[th]
\centering
\includegraphics[scale=0.3]{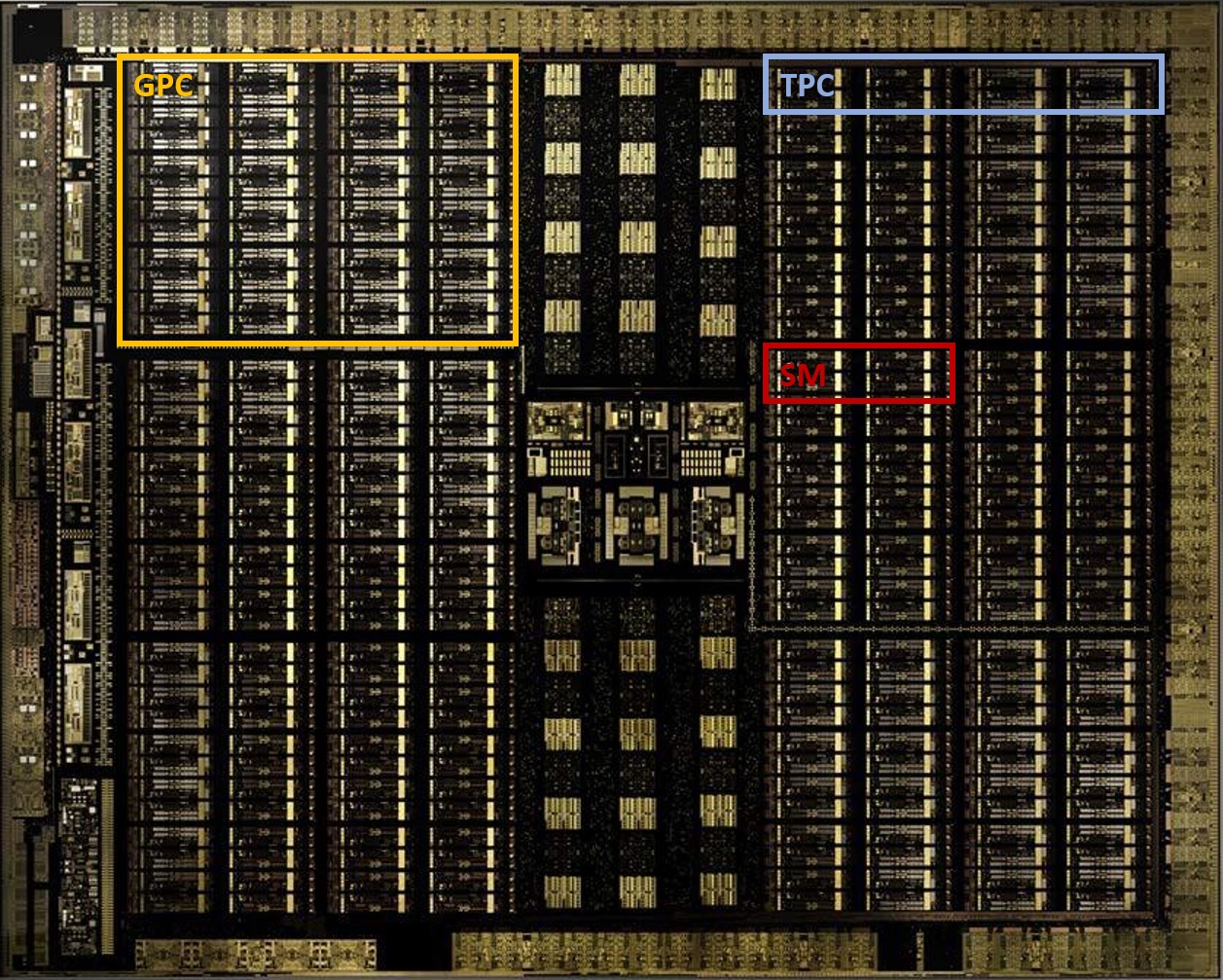}
\decoRule
\caption[]{Turing GPU die. Image adapted from the Turing architecture whitepaper \cite{turing_whitepaper}.}
\label{turing_die}
\end{figure}

All of these components can be identified on the GPU die itself, as shown in fig. \ref{turing_die}.

\subsubsection{CUDA Core}

CUDA cores are basic computational units that perform common integer- or floating point operations \cite{cuda_guide, turing_whitepaper}. These can be used both for rendering but also other parallel computing purposes through the CUDA platform\footnote{Note that we are using the term "CUDA core" as an umbrella term for a GPUs general purpose ALUs. Whilst CUDA is technically an Nvidia-exclusive platform, all GPUs are equipped with equivalent computational units.}.

Unlike its predecessors, the Turing architecture provides separate data-paths for integer and floating-point operations. In previous generations the execution of an integer-based instruction would have blocked floating-point instructions from issuing \cite{turing_whitepaper}.

\subsubsection{Tensor Core}

Tensor cores are specialized execution units that are built specifically and purely to accelerate the process of matrix (or tensor) multiplication \cite{TensorCores}. These find heavy usage in vertex transformation operations of common rendering applications but also for deep learning (neural network) purposes.

Turing's tensor cores have been enhanced for inferencing and equipped with additional integer precision modes \cite{turing_whitepaper}.

\subsubsection{RT Core}

One of the biggest innovations of the Turing architecture is the introduction of \textit{RT cores}, a specialized processing core that exclusively performs BVH traversal and ray-triangle intersections to facilitate real-time raytracing \cite{turing_whitepaper}. The underlying acceleration concepts are not new; parallelization and BVHs have been common practice in raytracing for some time.

\begin{figure}[th]
\centering
\includegraphics[scale=0.31]{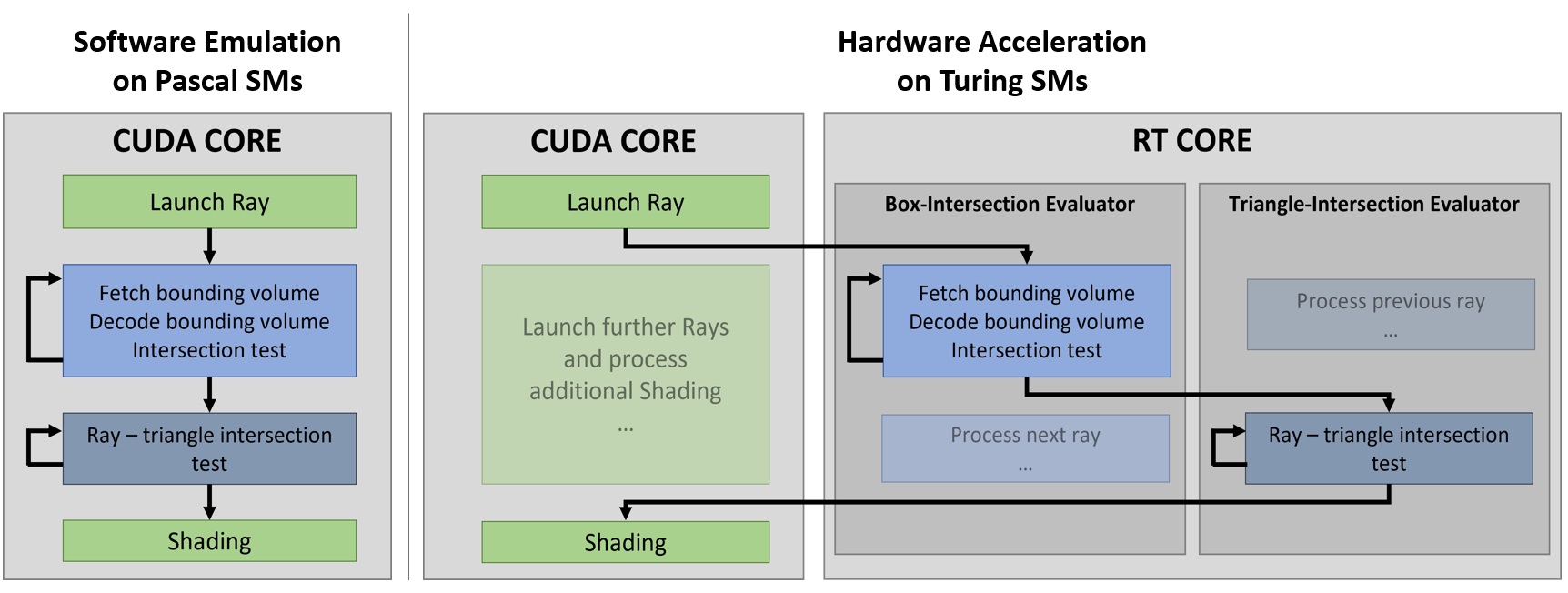}
\decoRule
\caption[]{Software emulated raytracing on Pascal GPUs (left) vs. hardware accelerated raytracing on Turing GPUs (right). Image in line with the depictions in the Turing whitepaper \cite{turing_whitepaper}.}\label{TuringVsPascalRT}
\label{pascal_vs_turing}
\end{figure}

But unlike previous raytracing solutions built on the GPU, RT cores can run autonomously, in parallel to the CUDA cores and thus offload the SM \cite{turing_whitepaper}. This means that when a shader dispatches a ray, it can continue performing other calculations whilst the ray is being traced in parallel. Similarly, once a ray intersects geometry, the RT core can directly move on to tracing the next ray, whilst the CUDA and tensor cores take care of shading and lighting. The horizontal parallelism this concept provides is demonstrated in fig. \ref{pascal_vs_turing}.

RT cores consist of two different units: one for bounding box testing and a second for ray-triangle intersections \cite{turing_whitepaper}. This distinction allows for further horizontal parallelism, as the first unit can move on to the next triangle/input whilst the second unit still finishes the previous input. 

Pascal GPUs have retroactively been made compatible with RTX through the DirectX raytracing API. But since no RT cores are available on these, a software approximation that runs on CUDA cores is employed instead (see fig. \ref{TuringVsPascalRT}). This results in significant performance degradation. Turing GPUs can process 10+ Giga-Rays per second, whilst high end Pascal GPUs only reach approximately 1.1 Giga-Rays per second \cite{turing_whitepaper}.

The general umbrella term for Nvidias real-time raytracing technology is \textit{RTX}, which is an abbreviation for \textit{Ray Tracing Texel eXtreme}.

\section{DirectX}

Rendering applications typically interact with GPU hardware through graphics APIs that span across multiple programming languages and platforms. These describe an abstract programming layer that specifies exactly what result, input and output of each function ought to be.
Minor variations in how computation is performed on a hardware-level, can produce divergent results even when a program is executed on the same API\footnote{For instance, \textit{Qualcomm Adreno 2xx} processors use 24 bit floating precision in fragment shaders, whilst the Nvidia X1 uses the more common 32 bit precision. The OpenGL 4.6 specification states that it "does not guarantee an exact match between images produced"\cite{opengl_46_spec}.}.

The most notable APIs are \textit{OpenGL}, \textit{Vulkan} and \textit{DirectX}, which have all seen industry-wide adoptions across most common graphics cards and applications \cite{ocg_api_comparison}.

Upon their release, Turing's raytracing features were only exposed through DirectX 12, later becoming available on Vulkan and (partially) OpenGL as well \cite{rtx_vulkan_opengl}. Since the compatibility has most matured on DirectX, the remainder of this thesis will focus primarily on DirectX 12.

\subsection{DirectX Rasterization Pipeline}\label{rast_pipeline}

When a scene is rendered into a frame output buffer using rasterization, DirectX employs a programmable graphics pipeline, which sequentially executes a series of highly specialized, fixed-function steps that allow the system to efficiently draw abstract 3D geometry in a given perspective. Each of these steps executes a concrete task with the output of the previous step as its input parameters. Given the consecutive nature of the pipeline, the individual steps can easily be executed in parallel for successive frames to be rendered.

The GPU programs that can be inserted in-between these steps are commonly termed \textit{shaders} and let developers configure custom pipeline behaviour. In a DirectX context, shaders are written in a \textit{high-level shader language} (HLSL), which strongly resembles the syntax used in C-based languages.

\begin{figure}[th]
\centering
\includegraphics[scale=0.285]{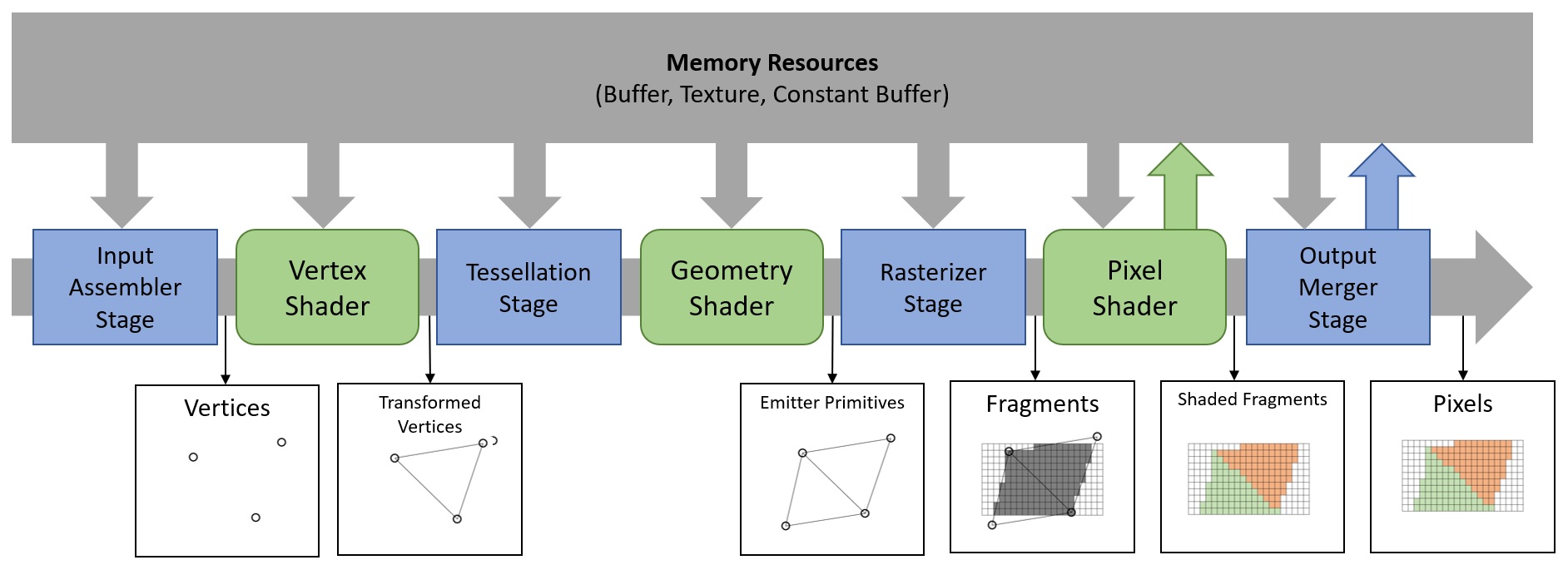}
\decoRule
\caption[]{Stages of the DirectX 12 rasterization pipeline. Programmable shaders are marked in green. Image based on information from the DirectX12 documentation \cite{directx12_docs}.}
\label{directx_rast_pipeline}
\end{figure}

Fig. \ref{directx_rast_pipeline} shows a simplified overview of the individual stages that form the DirectX rasterization pipeline (as described in the DirectX 12 documentation \cite{directx12_docs}).

\subsubsection{Input Assembler Stage}

The Input Assembler Stage reads geometry data from the allocated buffers and assembles it into primitives (usually triangles) that are usable by the other pipeline stages \cite{directx12_docs}. This stage will also attach system-generated values, such as primitive IDs, instance IDs or a vertex IDs, so that subsequent shader stages can reduce processing to only instances, primitives or vertices that have not already undergone processing \cite{directx12_docs}.

The primitives generated by the input assembler stage are subsequently transferred to the vertex shader.

\subsubsection{Vertex Shader}

The vertex shader is the first and arguably most important geometry processing step in any graphics pipeline.
It is an input-output program that is executed on every vertex individually and lets the user transform, modify or otherwise set up vertex-specific data for later pipeline stages \cite{learnopengl}.

The most common operations performed in the vertex shader are the applications of the model-, view- and projection- matrices \cite{fu_cg, learnopengl}. To conserve memory, 3D model data is stored in local coordinates so each instance of any 3D model in a scene comes with a model matrix that describes the object's position, rotation and scale. The model matrix is a 4x4 affine transformation matrix that converts vertex positions from local space into world space.

\begin{figure}[th]
\centering
\includegraphics[scale=0.5]{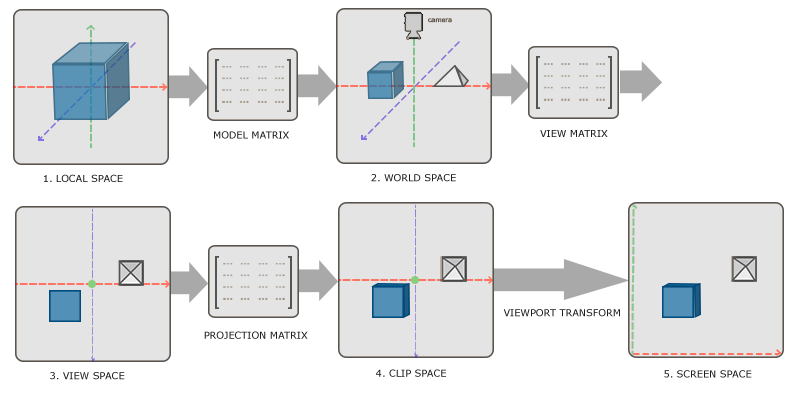}
\decoRule
\caption[]{Coordinate system transformations in the vertex shader, as pictured by Joey de Vries \cite{learnopengl}.}
\label{coord_systems}
\end{figure}

The view matrix will thereafter transform the coordinates into view space and the projection matrix applies the FOV-based distortion of non-orthographic cameras to each vertex. The final vertices find themselves in a $[-1, 1]$-ranged clip space coordinate-system, where vertices outside this range lie outside the camera's view frustum. The entire transformation process is depicted in fig. \ref{coord_systems}.

Furthermore, additional data-points such as a vertex's texture coordinates and normal vector are likewise set up in the vertex shader. By the end of this stage, each vertex will possess its own set of data known as \textit{attributes}, which subsequent pipeline stages linearly interpolate on for respective values anywhere on the spanned primitive.

\subsubsection{Tessellation Stage}

The tessellation stage is an optional stage added in DirectX 12, which allows the user to generate additional vertices directly on the GPU \cite{directx12_docs}. The utility of this stage plays no importance in remainder of this thesis.

\subsubsection{Geometry Shader}\label{GeometryShader}

Another optional step that allows for additional processing is the geometry shader, which is executed on a per-primitive basis.
For triangles, each set of three vertices relayed by the vertex shader are passed into the geometry shader as a triplet, which lets the user remove, subdivide or otherwise transform them in a manifold of ways \cite{directx12_docs, learnopengl}.

\begin{figure}[th]
\centering
\includegraphics[scale=0.4]{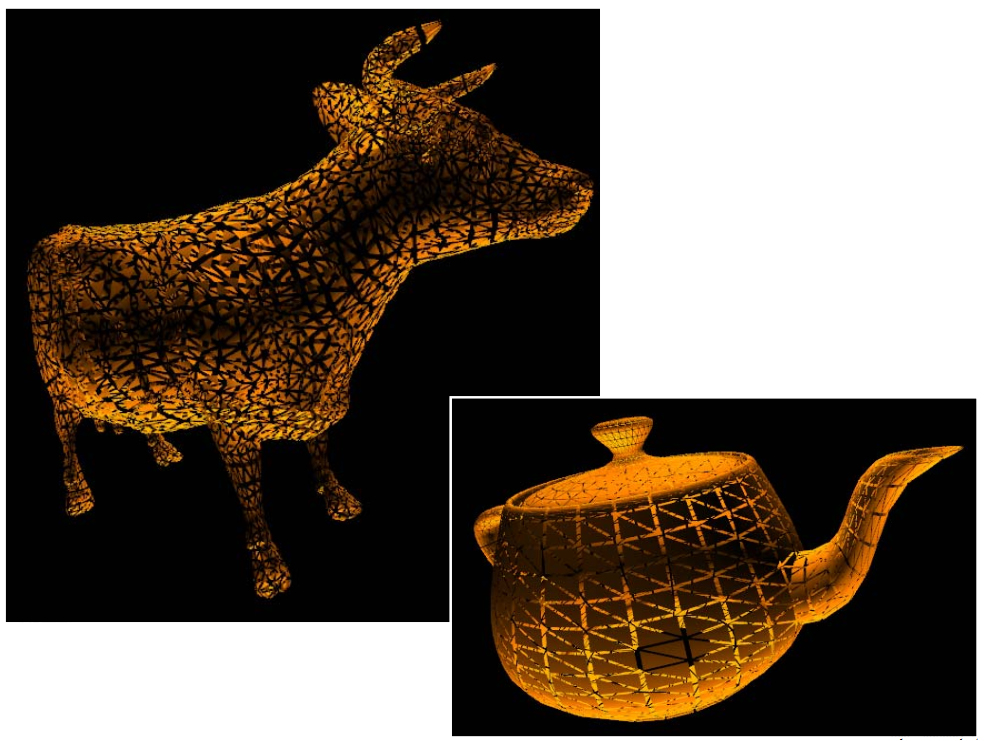}
\decoRule
\caption[]{Example: Shrinking triangles in the geometry shader. Image by Bailey \cite{geom}.}
\label{geom_example}
\end{figure}

This stage is frequently used for triangle-based effects like enlargements or shrinking (see fig. \ref{geom_example}) as well as a vital part for GPU-based voxelization \cite{VXCT, crassin_voxelization1} (see \ref{Voxelization}).

\subsubsection{Rasterizer Stage}

As implied by its name, the rasterizer stage converts each primitive into the set of pixels that it occupies on-screen, whilst interpolating the per-vertex attributes across it, so that each pixel has a corresponding set of values for normal vectors, positions, texture coordinates etc.

If the pipeline is set to utilize multi-sampling (compute several color values per pixel), the individual sub-samples are likewise arranged here \cite{directx12_docs}.
This stage additionally discards any pixels that do not face the camera (face culling), are occluded by other objects (depth clip) or are outside the viewport (scissor clip) \cite{directx12_docs}. This limits the amount of pixels - i.e. fragments - that need to be processed in the next stage.

\subsubsection{Pixel Shader}

The pixel shader (known as \textit{fragment shader} in OpenGL) is executed once for each pixel provided by the rasterization step \cite{directx12_docs, learnopengl}. This is typically the most performance-intensive stage, as it is executed most frequently and includes the vital lighting calculations that produce the final color of each pixel.

Mathematical models like \textit{Phong} \cite{phong} can be used to compute the pixel brightness and color composition equating to the radiance emanated from the fragment position towards the camera. The input data for these will be a position-based linear interpolation between the vertex attributes of each triangle. The factors of this interpolation can be equated with the fragment's barycentric coordinates on the triangle.

\subsubsection{Output Merger Stage}

The final stage of the pipeline utilizes any present depth/stencil buffers to perform depth-testing and blending of transparent objects (alpha testing) with the colors provided by the pixel shader to generate the final image \cite{directx12_docs}.

If the image is being rendered into multiple render targets (DirectX 12 supports up to 8 \cite{directx12_docs}), the writing process is likewise handled by the output merger stage.

\section{DirectX Raytracing}

Since the raytracing procedure fundamentally differs from classic rasterization, DirectX introduces an entirely new graphics pipeline with a whole new set of programmable shaders to accommodate it. 

\textit{DirectX Raytracing}, or simply \textit{DXR}, shares several similarities with its rasterization counterpart, as it aims to strike a balance between fixed-function and programmable stages to maximize both execution efficiency and potential for customization \cite{rtx_spec_ms}.

\subsection{DXR Pipeline}

The system is intended to process rays independently and in parallel. Once a ray hits or misses, it can create further sub-rays, but generated rays that are in-flight can never be dependent on each other \cite{rtx_spec_ms}.

\begin{figure}[th]
\centering
\includegraphics[scale=0.75]{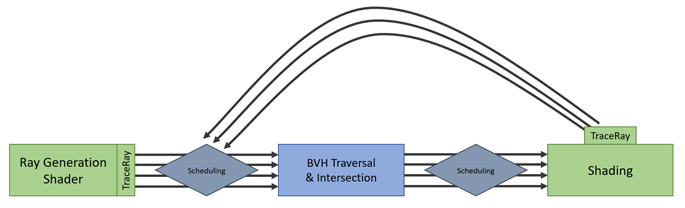}
\decoRule
\caption[]{Simplified overview of RTX pipeline stages. Depiction based on an image from the Microsoft DXR Specification \cite{rtx_spec_ms}.}
\label{rtx_pipeline_simple}
\end{figure}

On a surface level, the raytracing pipeline can be divided into three core components: Ray generation, BVH traversal and shading (see fig. \ref{rtx_pipeline_simple}).
Scheduling functionalities of which ray or shader is processed in which order is an opaque, hardware-bound process and cannot be altered \cite{rtx_spec_ms}. Likewise, the BVH traversal is treated as a single, fixed-function step that is offloaded to the RT cores. In their examination of RTX performance, Sanzharov et al. \cite{RTX_examination} concluded that RTX does some ray grouping and sorting during the GPU work creation process, in order to speed up bundles of rays traversing through the same BVH leaves.

\subsubsection{Rays in DXR}\label{RaysInDXR}

The data structure that represents a ray in DXR closely follows the definition put forth in section \ref{RayDefinition}, consisting of an origin, direction and a min-max distance interval \cite{rtx_gems}. Each of these values needs to be initialized before the ray can be traced:

\begin{lstlisting}
struct RayDesc
{
    float3 Origin;
    float3 Direction;
    float TMin;
    float TMax;
};
\end{lstlisting}

Furthermore, rays in DXR can carry a \textit{payload}, which is a custom data-structure of limited memory that can be accessed on a per-ray basis by any of the programmable shaders \cite{rtx_gems}. An example of how this payload may be utilized is given in section \ref{RTX_code_example}.

\subsubsection{Programmable Shaders}

\begin{figure}[th]
\centering
\includegraphics[scale=0.38]{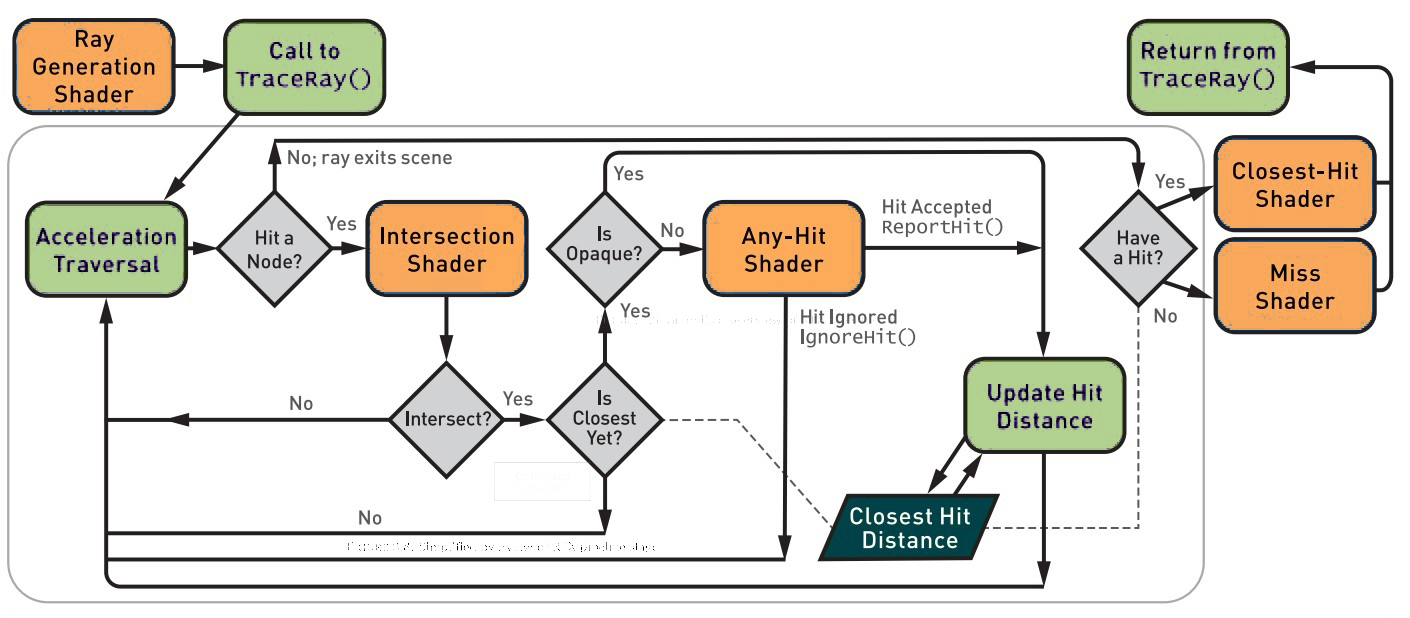}
\decoRule
\caption[]{Overview of the DXR pipeline as depicted in the \textit{Raytracing Gems} book \cite{rtx_gems}. It includes the five programmable shader types marked in orange. BVH traversal is marked by a grey outline.}
\label{rtx_pipeline_detailed}
\end{figure}

The DirectX raytracing pipeline introduces a total of five new programmable shader types that are invoked during the raytracing process based on the flow diagram depicted in fig. \ref{rtx_pipeline_detailed}.

Below we list a brief description for each one in line with how their documentation in \textit{Ray Tracing Gems} \cite{rtx_gems}.

\begin{itemize}
    \item The \textit{ray generation} shader is launched once for every instance of some enumerable index (1D, 2D or 3D grid) and handles the initial ray launches \cite{rtx_gems}. In a traditional raytracing application, this shader would launch the initial rays from the camera through each virtual pixel. An HLSL \verb|TraceRay(...)| function is available for this purpose.
    
    \item \textit{Intersection shaders} define how intersections are calculated with arbitrary primitives \cite{rtx_gems}. By using intrinsic functions like \verb|ReportHit()| or \verb|AcceptHitAndEndSearch()| the user can define which intersections are counted as hits or not. If no intersection shader is provided, the pipeline employs a high-performance default that uses triangles \cite{rtx_gems}. Utilizing intersection shaders instead of the build-in ray-triangle intersection is less efficient but offers far more flexibility \cite{rtx_spec_ms, rtx_gems}.
    
    \item As the intersection shader defines which hits to report and which not to, \textit{any-hit shaders} are executed for all reported hits \cite{rtx_gems}. In addition to running regular HLSL code (like writing data into textures on hit) any hit shaders allow otherwise valid intersections to be discarded, such as transparent surfaces.
    
    \item As the name implies, \textit{closest-hit shaders} are executed at the closest intersection for each ray \cite{rtx_gems}. In a traditional raytracing context, this would recursively launch additional rays to compute the color the hit location has.
    
    \item If no hit is registered for a ray, the \textit{miss shader} is executed \cite{rtx_gems}. This can be used, for instance, to display a background color for pixels not occupied by geometry.
\end{itemize}

This pipeline shares many components and concepts with regular raytracers. Algorithm \ref{DXR_Raytracing} illustrates how a standard DXR pipeline would function if it were executed in sequence.

\begin{algorithm}
    \caption{DXR Raytracing Process (Adapted from Raytracing Gems \cite{rtx_gems})}\label{DXR_Raytracing}
    \begin{algorithmic}[1]
      \For{$x, y \in image.dimensions()$}
        \State $ray \gets createRay(x,y)$\Comment{Ray from $C$ through pixel $(x, y)$}
        \State $closestHit \gets null$
        \While{$leaf \gets findBvhLeadNode(ray, scene)$}\Comment{BVH traversal}
            \State $hit \gets intersectGeometry(ray, leaf)$\Comment{\textit{intersection} shader}
            \If{$isCloser(hit, closestHit) \And isOpaque(hit)$}
                \State $closestHit \gets hit$
            \EndIf
        \EndWhile
        \If{closestHit}
            \State $image(x,y) \gets shade(ray, closestHit)$\Comment{\textit{closest-hit} shader}
        \Else
            \State $image(x,y) \gets miss(ray)$\Comment{\textit{miss} shader}
        \EndIf
      \EndFor
    \end{algorithmic}
\end{algorithm}

\subsection{TraceRay Function}

The DXR-intrinsic \verb|TraceRay(...)| function can be used in programmable shaders to commence a raytracing process on an RT core. It operates under the following parameters \cite{rtx_spec_ms}:

\begin{lstlisting}
TraceRay(
    RaytracingAccelerationStructure AccelerationStructure,
    uint RayFlags,
    uint InstanceInclusionMask,
    uint RayContributionToHitGroupIndex,
    uint MultiplierForGeometryContributionToHitGroupIndex,
    uint MissShaderIndex,
    RayDesc Ray,
    inout payload_t Payload
);
\end{lstlisting}

\begin{itemize}
    \item The first parameter selects the BVH containing the geometry that needs to be traced. In most instances only one BVH of a scene exists, but multiple ones can be defined to trace on different scenes from a single shader.
    
    \item The second parameter is an integer where each bit represents a certain flag that affects ray behaviour. Some notable flags are \begin{spverbatim}RAY_FLAG_CULL_BACK_FACING_TRIANGLES\end{spverbatim} to disregard intersections on triangles not facing the ray, or \begin{spverbatim}RAY_FLAG_ACCEPT_FIRST_HIT_AND_END_SEARCH\end{spverbatim} to terminate the BVH traversal immediately when any geometry is hit.
    
    \item The third parameter is an instance mask that allows skipping geometry on a per-instance basis. Passing a value of \verb|0xFF| (in hexadecimal) would cause all geometry to be tested for intersections, whilst \verb|0x00| would test none.
    
    \item The fourth and fifth parameters define the hit-group for this ray. A hit group consists of an intersection, closest-hit and anyhit shader. Typically each ray-type is separated into a respective hit-group so that the respective shader code is executed. (For instance, shadow-rays, diffuse reflections and specular reflections would each possess their own hit groups)
    
    \item The sixth parameter lets us configure which miss shader to use, irrespective of the hit group.
    
    \item The seventh parameter is the ray description, as defined in \ref{RaysInDXR}, and
    
    \item the eighth parameter is the payload the ray carries over its lifetime.
\end{itemize}

An alternative \verb|TraceRayInline(...)| function exists, that does not use separate shaders, and defers all shading to the caller \cite{directx12_docs}.

\subsection{Code Example}\label{RTX_code_example}

The code below serves as a simple example for the respective shaders a simple DXR program for shadow rays would utilize. The respective anyhit and miss shaders are marked by \verb|[shader("anyhit")]| and \verb|[shader("miss")]| attributes respectively:

\begin{lstlisting}
// Custom ray-payload datatype
struct ShadowPayload {
    bool isVisible;
};

[shader("miss")]
void ShadowMiss(inout ShadowPayload payload) {
    payload.isVisible = false;
}

[shader("anyhit")]
void ShadowAnyHit(intout ShadowPayload payload, BuiltInTriangleIntersectionAttributes attrib)
{
    // Ignore transparent surfaces
    if( isTransparent(attrib, PrimitiveIndex() ) ) {
        IgnoreHit();
    }
}

[shader("raygeneration")]
void rayGen()
{
    float3 origin = CameraPosition();
    
    float3 dir = PixelPosition(DispatchRaysIndex().xy) - origin;

    RayDesc ray = {
        Origin = origin,
        Direction = dir,
        TMin = 0.001f,
        TMax = 1000f
    };
    
    ShadowPayload payload = { true };
    
    TraceRay( scene,
              RAY_FLAG_SKIP_CLOSEST_HIT_SHADER,
              0xFF, 0, 1, 0, ray, payload );
    
    if( payload.isVisible ) {
        OutputTexture[DispatchRaysIndex().xy] = RED;
    }
    else {
        OutputTexture[DispatchRaysIndex().xy] = BLACK;
    }
}
\end{lstlisting}

This example colors every pixel that is occupied by geometry in red. It uses the ray payload to mark whether a pixel is occupied by any non-transparent geometry.
The \verb|isVisible| component is initialized as \verb|true|, but is set to false in the miss shader. The anyhit shader discards intersections with transparent geometry
\footnote{Some functionality given in the example, such as the functions CameraPosition(), isTransparent() and PixelPosition() are not part of DXR and have been abstracted for the sake of simplicity.}.

\subsection{Host Initialization}

The foregoing sections provide a comprehensive overview of the DXR raytracing pipeline and its GPU-side functions and capabilities. However, as with any graphics API, the global pipeline state and execution is managed by the device host (CPU side) through a series of API function calls.

Low level DirectX code is typically highly verbose, with even simple projects requiring hundreds to thousands of lines in C++ code. For the sake of brevity, this thesis will only provide superficial examination of the key functions required for raytracing.

Initialization of a DXR raytracer typically follow these common steps \cite{rtx_gems}:

\begin{itemize}
    \item Initializing the DirectX device (GPU) and verifying that it supports raytracing.
    \item Loading scene geometry and generating a BVH acceleration structure from it.
    \item Loading and compiling the respective HLSL shaders, defining root signatures and shader tables.
    \item Defining a DirectX pipeline state object.
    \item Dispatching a workload to the pipeline.
\end{itemize}

\subsection{Acceleration Structure}

As described in section \ref{BVH_theory}, utilizing a hierarchical acceleration structure can reduce the complexity per ray from linear to logarithmic in the number of triangles.
There are a variety of acceleration structure types available and DirectX does not mandate the use of any particular one, though bounding volume hierarchies (BVHs) are generally best suited \cite{rtx_gems}.

\begin{figure}[H]
\centering
\includegraphics[scale=0.7]{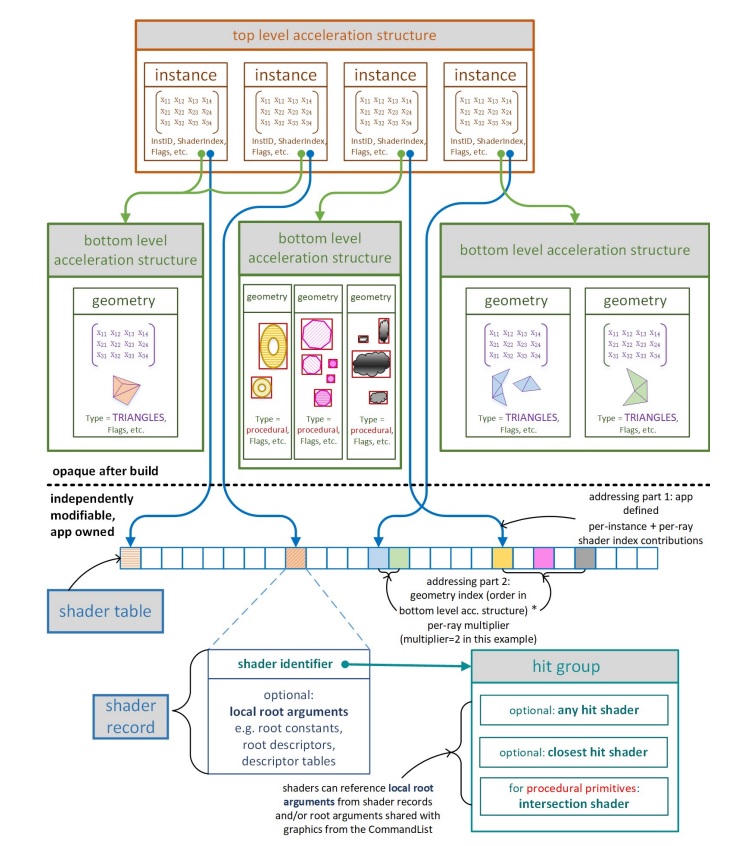}
\decoRule
\caption[]{Key components of an RTX pipeline: TLAS, BLAS and shader binding table. Image from the Microsoft DXR specification \cite{rtx_spec_ms}.}
\label{blas_tlas_table}
\end{figure}

The construction process and data structure of DirectX BVHs is entirely opaque, as they are built and maintained by the device driver on the GPU \cite{rtx_gems}. Different graphics card vendors may choose alternative structures, but DirectX operates on a given set of structural principles \cite{rtx_gems, rtx_spec_ms}:

The acceleration structure consists of two levels: a bottom-level acceleration structure (BLAS), which contains geometric primitives, and a top-level acceleration structure (TLAS), that contains one or more bottom-level structures (see fig. \ref{blas_tlas_table}).

\subsubsection{BLAS}

A bottom-level acceleration structure is usually a BVH in of itself that represents a single geometry type. Ray-triangle intersection tests are performed on BLAS data \cite{rtx_gems}.

If the geometry topology remains fixed, BLAS structures can be refit with scene changes, which is an order of magnitude faster than complete rebuilds. However, repeatedly performing refit operations may degrade the quality and performance of the acceleration structure over time. It is generally recommended to use an appropriate combination of refits and complete rebuilds \cite{rtx_gems}. 

\subsubsection{TLAS}

\begin{figure}[H]
\centering
\includegraphics[scale=0.35]{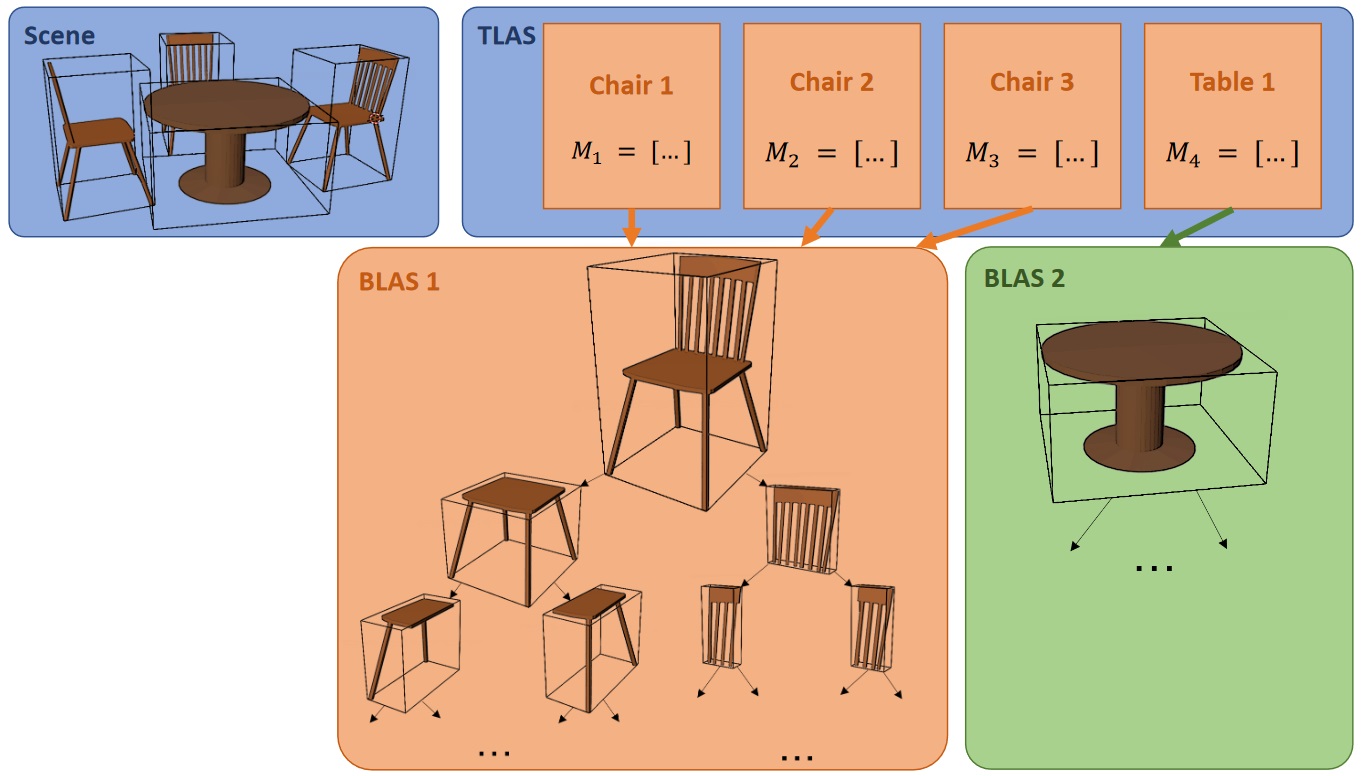}
\decoRule
\caption[]{Simplified illustration of how the individual components of a scene relate to BLAS-TLAS components.}
\label{BlasTlasSimpl}
\end{figure}

Analogously to how single 3D models can be instantiated multiple times with individual model matrices, BLAS instances within a TLAS are referenced with memory pointers alongside a transformation matrix \cite{rtx_spec_ms} (see fig. \ref{BlasTlasSimpl}).
Whilst the reuse of geometry has great benefits to memory requirements, its overuse can impact performance, as the individual BLAS instances ought to overlap as little as possible \cite{rtx_gems}.

A TLAS is, in essence, an acceleration structure of acceleration structures. Pointers to BLAS structures already living in GPU memory are contained alongside an instance matrix as well as other data like shader-index, flags etc.

\subsubsection{Shader Table}

The whens and hows of tracing rays in DXR are not as strictly mandated by a sequential pipeline as in rasterization. As such, all resource bindings and shaders must be simultaneously available during the entire execution time. The selection of which shader to run is treated as any other resource binding and kept as a set of shader records in a contiguous region of memory known as the \textit{shader table}.

The shader binding table is, in essence, a region of 64-bit aligned GPU memory that is owned and managed by the application \cite{rtx_spec_ms}. It links the acceleration structures, hit groups and shader functions together by indicating what programs are executed for which geometry and which resources are associated with it \cite{rtx_spec_ms}.

\section{Status Quo of RTX}

In their examination of RTX technology, Sanzharov et al. conclude that whilst RTX on Turing GPUs performs well, its software-emulated variant that runs without RT-cores is an inefficient and expensive process that "essentially loses to simple and straightforward open source ray tracing" \cite{RTX_examination}, implying that "'the golden age of software' has ended and that 'the golden age of compilers and HW/SW projects' has started" \cite{RTX_examination}.

The visual fidelity that hardware-accelerated RTX can provide has indeed found favour with developers, as implied by its widespread adoption across many products and rendering engines such as software by \textit{Adobe}, \textit{Unity} and a vast catalogue of video games \cite{rtx_catalogue}. Even more predicating is the fact that one of Nvidia's primary competitors on GPU market, \textit{AMD} released their newest line of products, the \textit{Radeon RX 6000 Series}, with an RT-core equivalent component, designed specifically for hardware-accelerated raytracing \cite{RDNA2_explained}.

Despite these successes, lower \textit{frames-per-second}, in addition to price and compatibility constraints, show that RTX remains an \textit{enhancement}, not a replacement, to classical rasterization.
The underlying concept of raytracing remains just as computationally expensive as it has been since its inception in the late 1970s.

In the same time, rasterization-based techniques have come a long way in finding ideal mathematical approximations, shortcuts and simplifications that maximize photorealism. Combining the newest kinds of these techniques can produce images of similar quality to raytracing, albeit at far greater speeds.

\begin{figure}[th]
\centering
\includegraphics[scale=0.2]{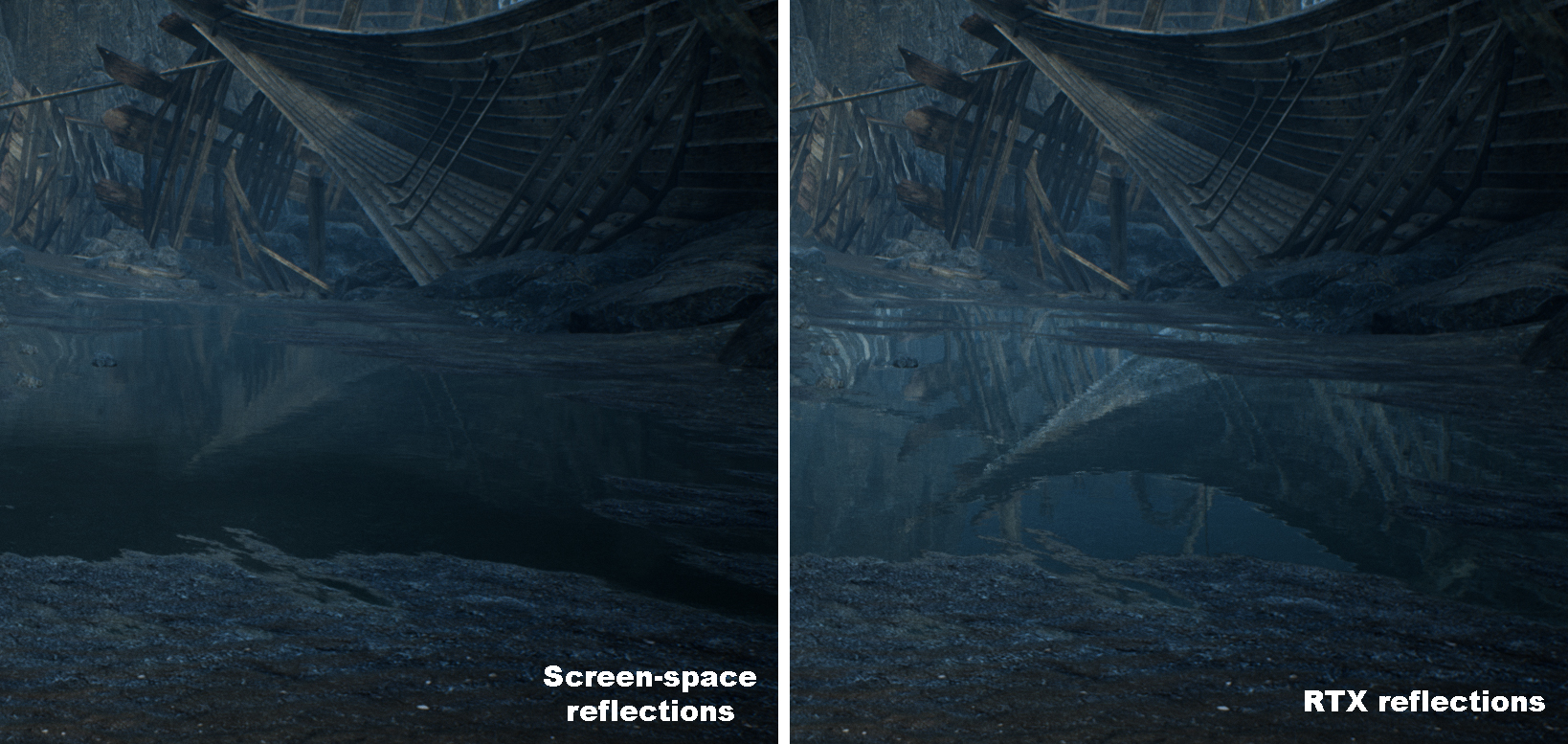}
\decoRule
\caption[]{Screen-space reflections (left) vs ray-traced reflections (right). Screen-space reflections are generally less resource-intensive, but can only reflect geometry that is rendered to the screen itself. Images taken from the PC game \textit{Hellblade: Senua's Sacrifice} by \textit{Ninja Theory}.}
\label{rtx_vs_ssr}
\end{figure}

A great example of how clever and sophisticated these techniques have become can be observed in the form of \textit{screen-space reflections} (fig. \ref{rtx_vs_ssr}). This is a fast method of creating realistic looking reflections by simply taking respective pixel values already rendered into the FBO \cite{SSR}. At minimal performance impact, this requires no triangle-intersections or BVH traversals, but only geometry that is visible on screen can be reflected \cite{SSR}.

Raytracing, on the other hand, provides a more powerful, brute-force approach that costs exponentially more. Consumers may prefer incurring the small cost in photorealism provided by raytracing in return for a more responsive application running at a higher frame-rate.

It is generally recommended by Nvidia themselves, to opt for a hybrid approach by using rasterization as a base and complement it with raytracing where it provides the most benefits (such as specular reflections, refractions and shadows) \cite{turing_whitepaper, Hybrid_RTX_Mader}.

In chapter \ref{Chapter2} we demonstrated that radiosity and raytracing share many fundamental aspects by deriving both from the rendering equation. The underlying implication, which we will examine in the subsequent chapters, is that the performance increase RTX provides for raytracing ought to applicable to radiosity as well.

\chapter{RTX Radiosity} 

\label{Chapter4} 



The preceding chapters delineate how Nvidia's RTX technology functions and how it can be leveraged through the DirectX 12 API.

In this chapter we present a concrete implementation in the form of an RTX-based radiosity application referred to as \textit{RTRad}. We commence by first exhibiting the primary components of regular, progressive radiosity, which we later expand to incorporate \textit{refinement} (and more) in chapter \ref{Chapter5}.

\section{Status Quo of GPU-based Radiosity}

Radiosity variants or derivations are used industry-wide in many real-time rendering engines.
Yet despite their immense potential for paralellization, many radiosity implementations continue to perform better on high core-count CPUs as opposed to GPUs.

In their measurements, Carr et al. found that although CPUs currently perform better in matrix-based radiosity calculations, a GPU's performance scaling is significantly closer to linear, albeit with a fairly constricting upper limit, due to a GPU's limited memory capacity \cite{ray_engine}.

The algorithm finds itself in an unusual predicament, where it is neither particularly well suited for CPU nor GPU execution. 
Each radiosity patch can be processed in parallel on a GPU, but solving the visibility function $V$ requires a data structure to be traversed \textit{sequentially}, a process better suited to the powerful cores of a CPU.

Some solutions, like the one proposed by D’Azevedo et al. \cite{ComplexGPURadiosity}, attempt to tackle this imbalance by dividing the workload onto a hybrid GPU/CPU platform, where only the view-factors are computed GPU-side via a compute-shader, then read back into CPU memory and finally used in a rudimentary CPU-based solution.

The introduction of RT cores strikes a compromising balance between the few, high-performance cores of a CPU and the many, low-performance cores of a GPU. Offloading the visibility component of radiosity onto this new hardware may prove to be an ideal solution to this bottleneck. 

\subsection{Lead-up to RTRad}

Raytracing and radiosity are both based on the rendering equation and, as such, share many fundamental mathematical components.
The primary goal behind \textit{RTRad} is to demonstrate that the performance increase RTX provides to raytracing is applicable to radiosity as well.

In this chapter we set out to alleviate the visibility bottleneck by substituting techniques based on hemicuboid z-buffering with an RTX-based solution, which should prove more adequate for for random rays requiring random memory-access \cite{RTX_examination}.

The introduction of the Turing architecture opens the door to an entirely new set of specialized computation units on GPUs that may well be useful in areas beyond their intended use-case.
Similarly to how regular graphics processors, initially intended purely for 3D rendering, have found themselves beneficial for purposes such as cryptography and machine learning, \textit{RT cores} may also prove themselves useful for applications outside of the real-time raytracing domain.

\begin{figure}[H]
\centering
\includegraphics[scale=0.4]{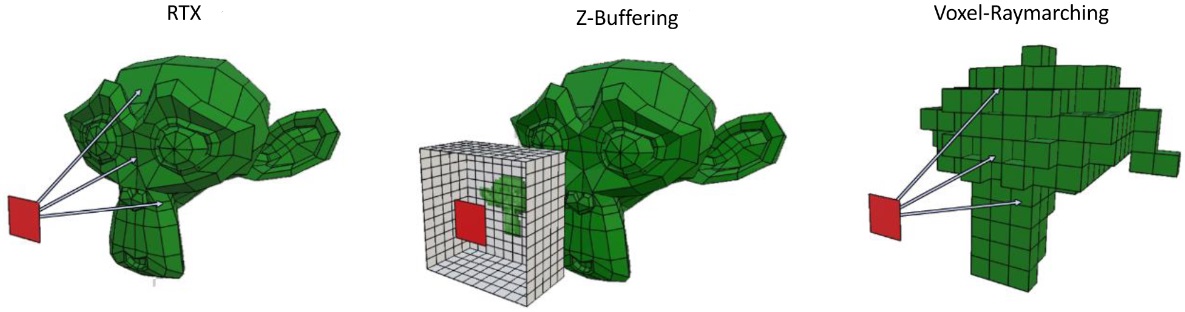}
\decoRule
\caption[]{Different methods for calculating visibility. This thesis examines how RTX and voxel-raymarching hold up against z-Buffering.}
\label{VisMethods}
\end{figure}

With the implementation of an RTX-based radiosity algorithm, we put forth the general argument that RT cores can offer a great computational shortcut for highly accurate visibility simulations in general (see fig. \ref{VisMethods}).

\subsection{Target Use-Case}

The implementation presented in this thesis is fully capable of quickly producing high-quality radiosity lightmaps that can be used as textures for diffuse global illumination.
However, the underlying software is primarily intended as a research project and proof-of-concept, not a consumer-targeted application to be used in production-ready applications. 

The underlying use case an RTX-based lightmapper could cover in practice, is mainly oriented towards developers or designers of 3D environments working on specialized workstations with RTX-compatible graphics cards. The faster computation time would facilitate a more comfortable workflow and, once computed, the lightmaps can be mapped onto geometry and displayed to end-users on any PC at virtually no cost at all.

\section{Previous Work}

The relative novelty of RTX means its application or examination in anything other than real-time raytracing is quite sparse.

According to its documentation, contemporary versions of \textit{Unreal Engine} make use of RTX in their \textit{GPU Lightmass} system \cite{UnrealDocu}, although it is not clear in what manner or capacity.
Shcherbakov et al. hint at the idea of potentially utilizing RTX in the future to accelerate their \textit{Dynamic Radiosity} algorithm \cite{dyn_rad},
and Lin advocates for its usage to accelerate the VPL generation process in instant radiosity \cite{rtx_instant_rad}.

Radiosity implementations running on GPU hardware \textit{without} RTX have existed for some time alongside several noteworthy publications describing them.

\subsection{GPU Radiosity}

As part of the 2005 \textit{GPU Gems 2} book \cite{gpu_gems_2005}, Coombe et al. provide an early implementation of the progressive refinement radiosity algorithm that runs on common GPUs using z-Buffering for visibility.

Alongside a series of other acceleration techniques, this solution can ultimately "compute a radiosity solution of a 10,000-element version of the Cornell Box scene to 90 percent convergence at about 2 frames per second" \cite{gpu_gems_2005}.

This implementation has served as a primary influence on this thesis with several core concepts, such as using GPU-generated mipmaps to decrease shooting resolution, being derived from it.

\subsection{Rapid-Radiosity (RRad)}

Not to be confused with the program presented in this thesis (RTRad), \textit{RRad} \cite{rrad} is a GPU-based implementation that served as a direct lead-up to this thesis, made with the open-source API OpenGL. Completed as a software project part of the computer graphics lecture at the Freie Universität Berlin, it highlights with clarity how visibility is the only major hurdle that prevents the widespread adoption of GPU-based radiosity.

\begin{figure}[th]
\centering
\includegraphics[scale=0.35]{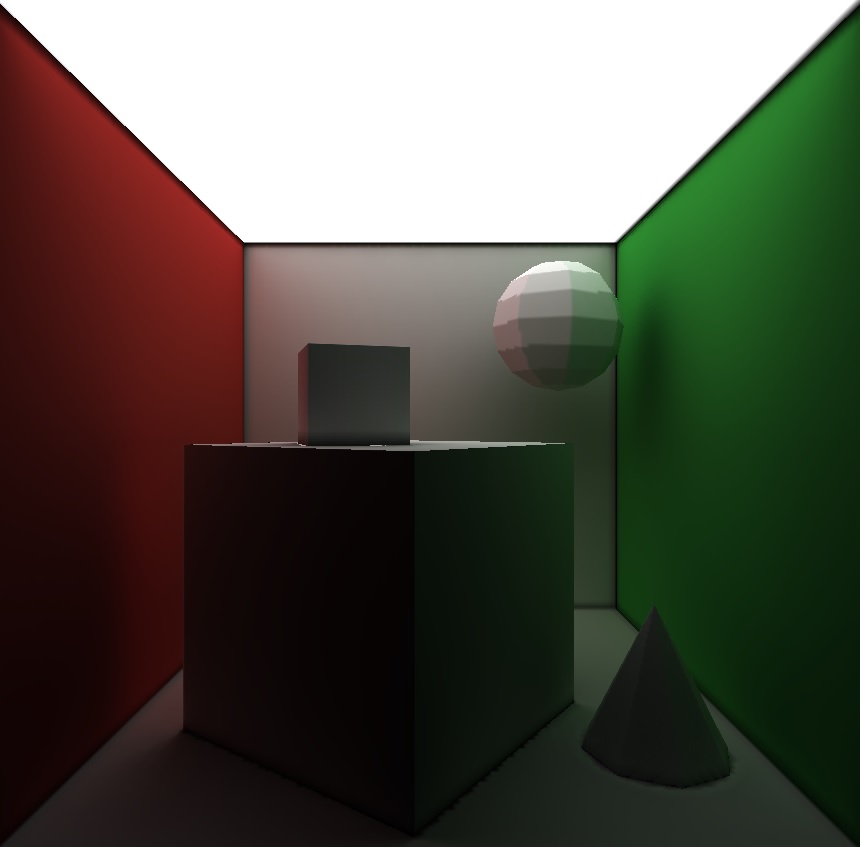}
\decoRule
\caption[]{Simple render done with RRad. The GPU implementation means that performance is decent, but the amount and complexity of the shapes (spheres and triangles) is highly limited, which are all tested for intersections sequentially.}
\label{rrad_example}
\end{figure}

RRad approximates a scene through basic geometric shapes (spheres and triangles) and then loops over each shape and performs a simple, discrete ray-intersection on each. This geometric approximation is hard-coded into the shader's code itself, making the application simple and lightweight, but entirely unsuitable to complex environments. Fig. \ref{rrad_example} shows the default RRad scene with a lightmap of $512\times512$ pixels, for which a single bounce of light requires approx. 2 seconds of computation time on a GeForce RTX 2070S GPU.

Many fundamental design choices presented in this thesis have their roots in RRad, with the most substantial change being the replacement of the discrete, hard-coded raytraces that run on CUDA cores, to a fully flexible RTX solution.

\section{Source Code and Dependencies}

RTRad was developed in C++ 17 with Nvidia's own \textit{Falcor} as the underlying framework. Visual Studio 2022 served as the primary IDE and the program was exclusively tested on a system with an Nvidia RTX 2070S GPU and an AMD Ryzen 3900X CPU.

The complete source code, a demonstration video, as well all executable files for the finished project can be found on the following github repository\footnote{The latest commit ID at the time of writing is \texttt{056e0e1b7cc89190231a6fbb1e81bd04ac6e0701}.}: \url{https://github.com/Helliaca/RTRad}

\subsection{Falcor}

Falcor is an open-source framework intended specifically for rapid prototyping of real-time rendering applications \cite{falcor, RTX_Tuto}.
Maintained as well as internally utilized by Nvidia, it provides a considerable set of advanced graphics features such as stereo rendering for VR, physically based shading and, most importantly for the context of this thesis, built-in RTX support \cite{RTX_Tuto}.

Additionally, there are thin abstraction layers on top of DirectX 12 that reduce the amount of redundant, verbose DirectX code required for a functioning application as well as convenient UI and profiling systems.

RTRad is built on Falcor 4.4, leaving some advanced features available in later versions (Faclor 5.2 being the most recent at the time of writing) such as DLSS aside.

\begin{figure}[th]
\centering
\includegraphics[scale=0.3]{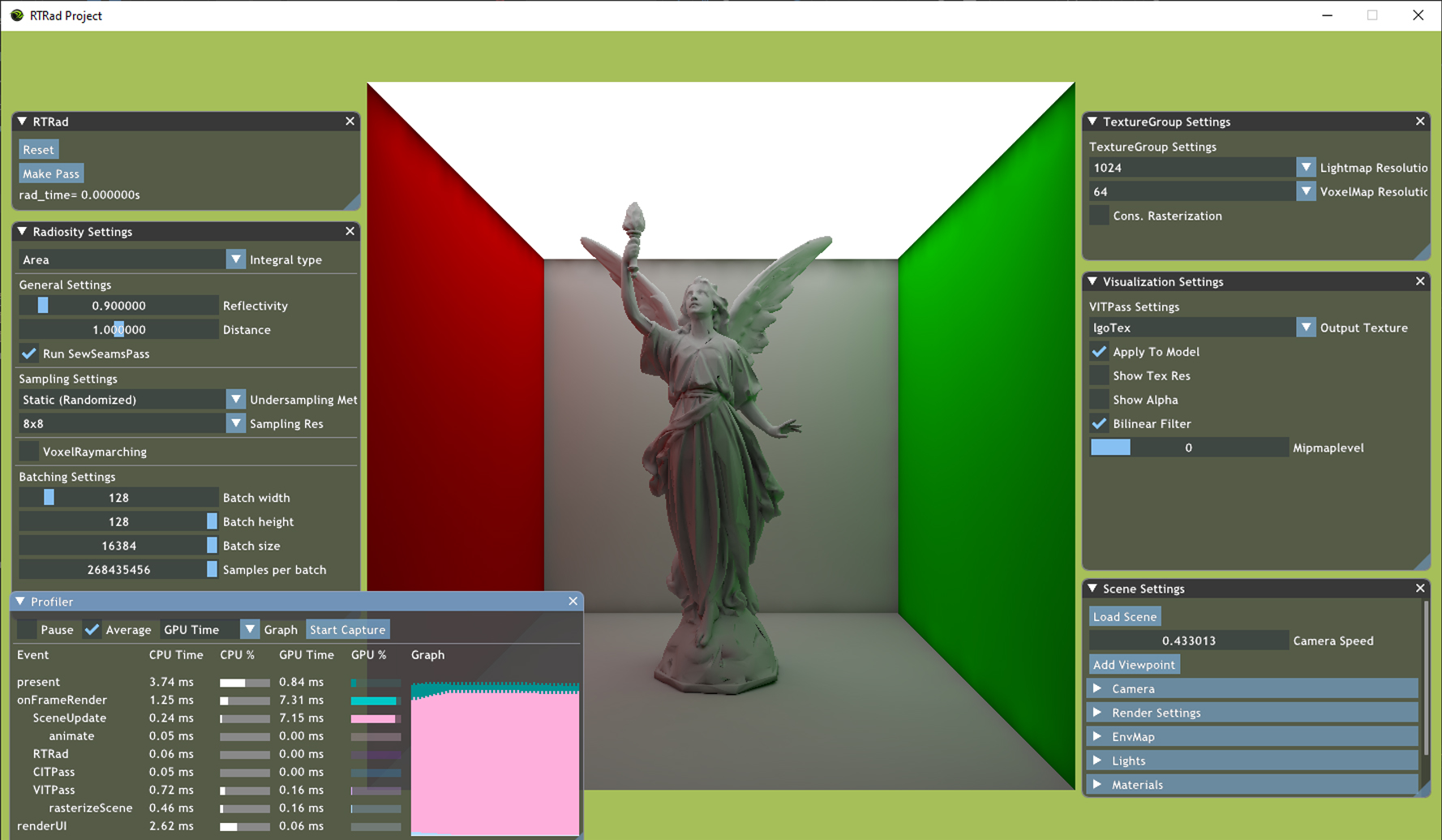}
\decoRule
\caption[asd]{GUI of the RTRad application. Each render pass has its dedicated GUI window to adjust settings, alongside scene controls, pipeline controls and a profiler.}
\label{rtrad_UI}
\end{figure}

\section{Program Structure}

The employed programming patterns were kept consistent with the precedent set in Falcor's source code and respective educational content (see \cite{RTX_Tuto, RTX_Intro, falcor}), including the usage of factory methods to create objects and referencing them with smart pointers to avoid memory leaks.

The core program takes the form of a graphics-pipeline that consists of several graphics passes which are executed and managed by a central manager object based on Falcor's \textit{IRenderer} interface.

\subsection{Class Structure}

\begin{figure}[h]
\centering
\includegraphics[scale=0.6]{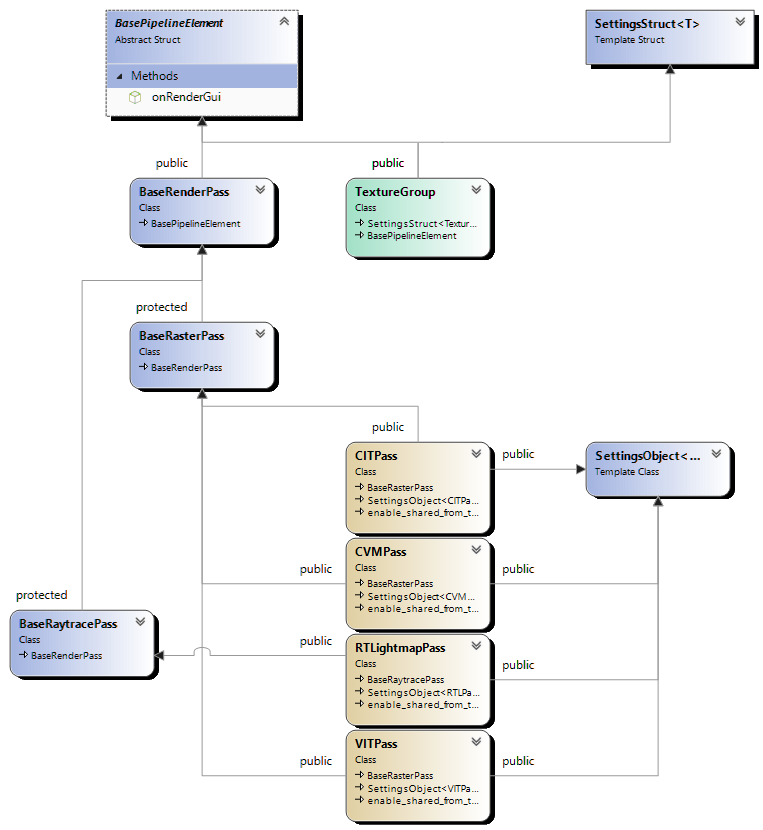}
\decoRule
\caption[]{Simplified class diagram of RTRad. The TextureGroup class (green) represents the input/output data container, which is processed by up to four subsequent passes (orange).}
\label{class_diag}
\end{figure}

RTRad follows the class structure depicted in fig \ref{class_diag}: A \textit{BasePipelineElement} class serves as a base class that provides a custom GUI method, so that each pipeline component can manage their own GUI elements that are used to adjust settings which are bound to their respective parents through the \textit{SettingsObject} template (see fig. \ref{rtrad_UI}).

A \textit{BaseRenderPass} serves as the base class for render passes which form the backbone of the pipeline. A central \textit{RTRad} object manages its execution and data-flow. Each pass self-manages its UI elements, shaders and shader-uniform variables. A \textit{TextureGroup} serves as the input-output data structure which is passed through the pipeline.

\section{Input Data}

The underlying goal is to take a scene, consisting of geometry and materials, and generate a bitmap texture containing diffuse global illumination as the output, where each pixel corresponds to a radiosity patch. 

Mapping each pixel in the output texture to a surface in 3D space is accomplished through the scene's \textit{UV coordinates}.

\subsection{Subdivision through UV Mapping}

The process of ascribing each 3D vertex an additional 2D coordinate on a texture is called \textit{UV mapping}.
Utilizing this process for radiosity patch placement comes with the benefit of streamlining the data-containers for each input. Increasing or decreasing the resolution of the lightmap can occur seamlessly, as every patch points to a texture-coordinate correspondent to it.

There exist a plethora of algorithms to automatically generate UV unwrappings for any 3D model or scene \cite{UV_unwrapping}. In practice, rendering engines tend to come equipped with their own built-in tools specifically meant for automatic lightmap UV generation \cite{unreal_uv_unwrap, unity_docs}.

Making these unwrappings as seamless and efficient as possible is a unique challenge beyond the scope of this thesis. All unwrappings utilized in RTRad were created manually, or with the tools contained in the \textit{Blender} 3D modelling software.

\begin{figure}[H]
\centering
\includegraphics[scale=0.4]{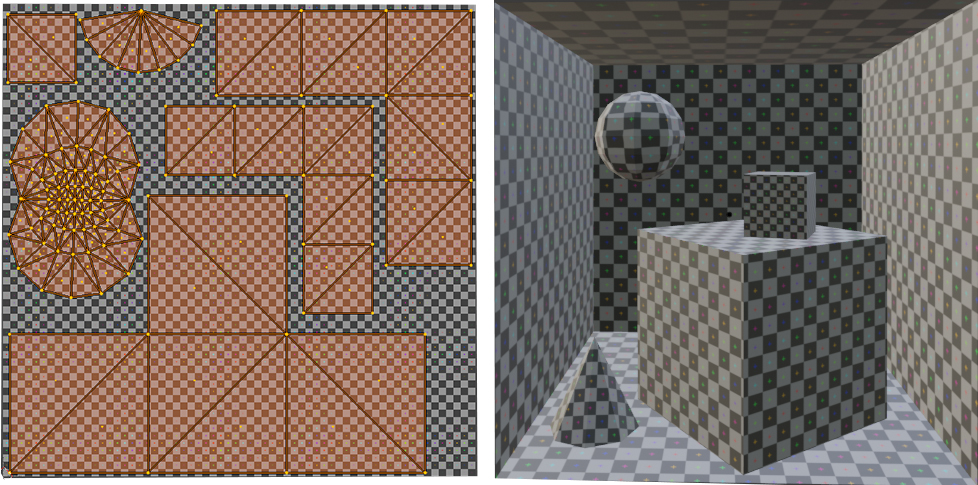}
\decoRule
\caption[]{The default RRad scene \cite{rrad} (right) and its UV coordinates (left) with a checker pattern applied as a texture. Each square of the checker pattern would correspond to a radiosity patch.}
\label{UV_unwrapping}
\end{figure}

\subsection{Input Components}

Recall the Monte-Carlo approximation for view factors established in (\ref{eqn:ViewFactor_MonteCarlo}):

\begin{equation}
    F_{i,j} = \frac{1}{K} \sum_{k=1}^{K} A_j \frac{\cos \theta_{ki} \cos \theta_{kj}}{\lVert x_{ki}-x_{kj} \rVert ^2} V(x_{ki},x_{kj})
\end{equation}

$\cos \theta_{ki}$ can be derived as the dot product between the normalized normal vector of the sample point and the normalized vector pointing from $x_{ki}$ to $x_{kj}$. As such, the exact data required to calculate a form factor $F_{i,j}$ is the following:

\begin{itemize}
    \item Surface area $A_j$
    \item Normal vectors $n_{ki}$ and $n_{kj}$
    \item World-space positions $x_{ki}$ and $x_{ki}$
    \item Visibilities of the the two locations $V(x_{ki},x_{kj})$
\end{itemize}

Since the processing order of rays is not deterministic, it is imperative that all of these data points for all patches are available at all times of a radiosity iteration.
We accomplish this by pre-computing individual textures that contain each data-point for all patches. The conglomeration of these textures is aptly named \textit{TextureGroup}, and serves as the primary input-output data structure for the entire application.

\subsubsection{Texture-Group}\label{TextureGroup}

Each individual texture contains different information, but conforms to the exact same UV mapping.
If, for instance, we require the normal vector of a patch, we simply perform a look-up operation on the texture containing normal vectors.

In total, following textures make up the texture group:

\begin{itemize}
    \item $pos$: World-position
    \item $nrm$: World-space normal vector (normalized)
    \item $mat$: Material properties (color)
    \item $arf$: Surface area of each patch
    \item $lig_{in}$: Input lighting - i.e. emission - values of the current iteration
    \item $lig_{out}$: Output lighting texture to write to
\end{itemize}

Pixels that are not occupied by any geometry at all, are marked as 'non-patches' by setting their alpha value in the $pos$ texture to zero. When sampling for lighting contribution, only pixels with a non-zero alpha value are processed.

\begin{figure}[H]
\centering
\includegraphics[scale=0.42]{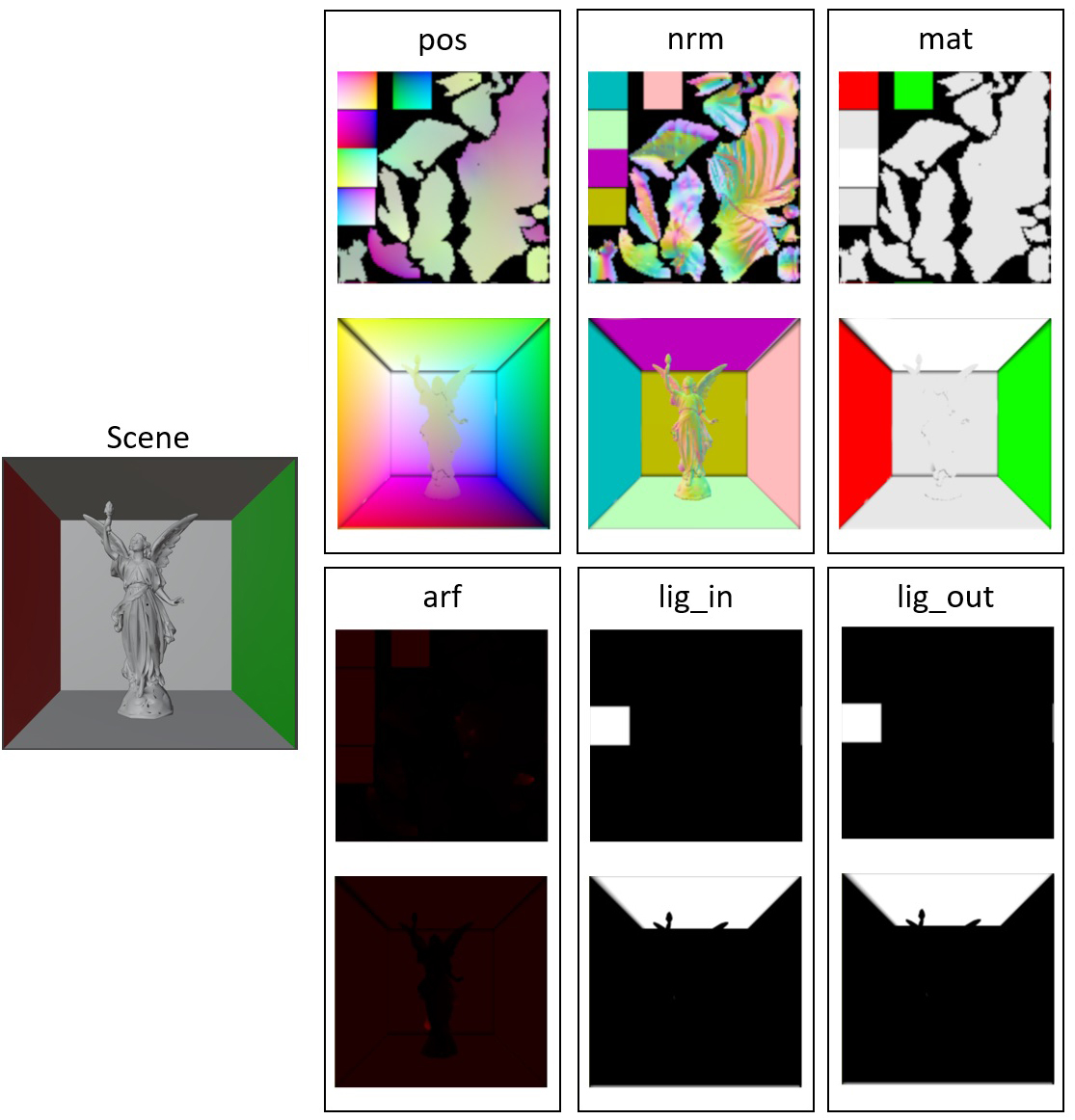}
\decoRule
\caption[]{Original scene (left) and corresponding texture group at a resolution of $128\times128$ (right). The bottom rows corresponds to each texture being mapped back onto the geometry.}
\label{figTexturGroup}
\end{figure}

\section{CITPass}

The \textit{create input textures pass} is a separate, a-priori rasterization pass that generates the above described textures from the scene's 3D geometry.

It involves applying a custom vertex shader that places each vertex into a position in clip-space that corresponds to its UV coordinates. Whilst a typical vertex shader commonly applies an objects model, view and projection matrices like so:

\begin{equation}
    vert(v) = P * V * M * v.pos
\end{equation}

our custom vertex shader for the CITPass simply applies a vertex's UV coordinates:

\begin{equation}
    vert(v) = 
    \begin{bmatrix}
    2 * (v.uv.x - \frac{1}{2}) \\
    2 * (v.uv.y - \frac{1}{2}) \\
    0 \\
    1
\end{bmatrix}
\end{equation}

UV coordinates range from zero to one, whilst clip-space is defined as the $[-1, 1]$ range, hence the subtraction of $\frac{1}{2}$ and multiplication by 2. We assume the UV mapping to contain no overlapping geometry, making the z-coordinate produced by this vertex shader irrelevant.

With the vertex shader in place, setting the output resolution to the desired resolution of the input textures then ensures that the subsequent pixel shader is executed exactly once for each pixel - e.g. each patch -  in these textures.

An example how how this texturegroup looks like after the CITPass can be seen in fig. \ref{figTexturGroup}. The lighting textures $lig_{in}$ and $lig_{out}$ store the emissive values $L_e$ for each patch, which serves as the radiant-exitance for the first radiosity pass.

\subsection{Surface Area}
\label{SurfaceArea}

Whilst normal vectors, positions and material properties are easily passed into the pixel shader through the rasterization pipeline, surface areas of patches require additional information that a simple, barycentric interpolation cannot provide.

\begin{figure}[th]
\centering
\includegraphics[scale=0.5]{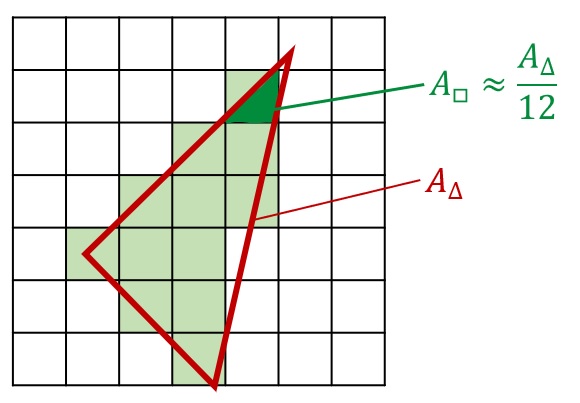}
\decoRule
\caption[]{Approximation of a patches surface area: Since this triangle occupies 12 patches on the lightmap, the world-space surface area of each patch can be approximated by ascribing each patch the surface area of the triangle divided by the amount of patches.}
\label{triangle_surface_area}
\end{figure}

Our textures consist of pixels, effectively squares, which correspond to radiosity patches.
The set of pixels rastered by a triangle, will only \textit{partially} cover or be covered by it. As such it is difficult to calculate an accurate value for the world-space surface area each patch possesses, for which we employ the following approximation:
\textit{The surface area of any one patch is equal to the surface area of the underlying triangle divided by the total amount of patches occupied by said triangle.}

Let $A_{\bigtriangleup}^{w}$ be the world-space surface area of the triangle, with $A_{\bigtriangleup}^{uv}$ being its surface area on the $[0,1]$ bounded UV map.
The amount of patches a triangle occupies can be derived from the product of its UV surface area and the total amount of patches in the texture $n$. For instance, if a triangle occupies half of the UV map, its UV surface area will be equal to $\frac{1}{2}$, which implies that it is represented by $\frac{n}{2}$ patches.

The world-space surface area $A_{\boxdot}^{w}$ of a single patch can then be computed from the relationship between  this product and its world-space triangle $A_{\bigtriangleup}^{w}$ (see fig. \ref{triangle_surface_area}):

\begin{equation}
    A_{\boxdot}^{w} \approx \frac{A_{\bigtriangleup}^{w}}{A_{\bigtriangleup}^{uv} * n}
\end{equation}

\begin{figure}[th]
\centering
\includegraphics[scale=0.32]{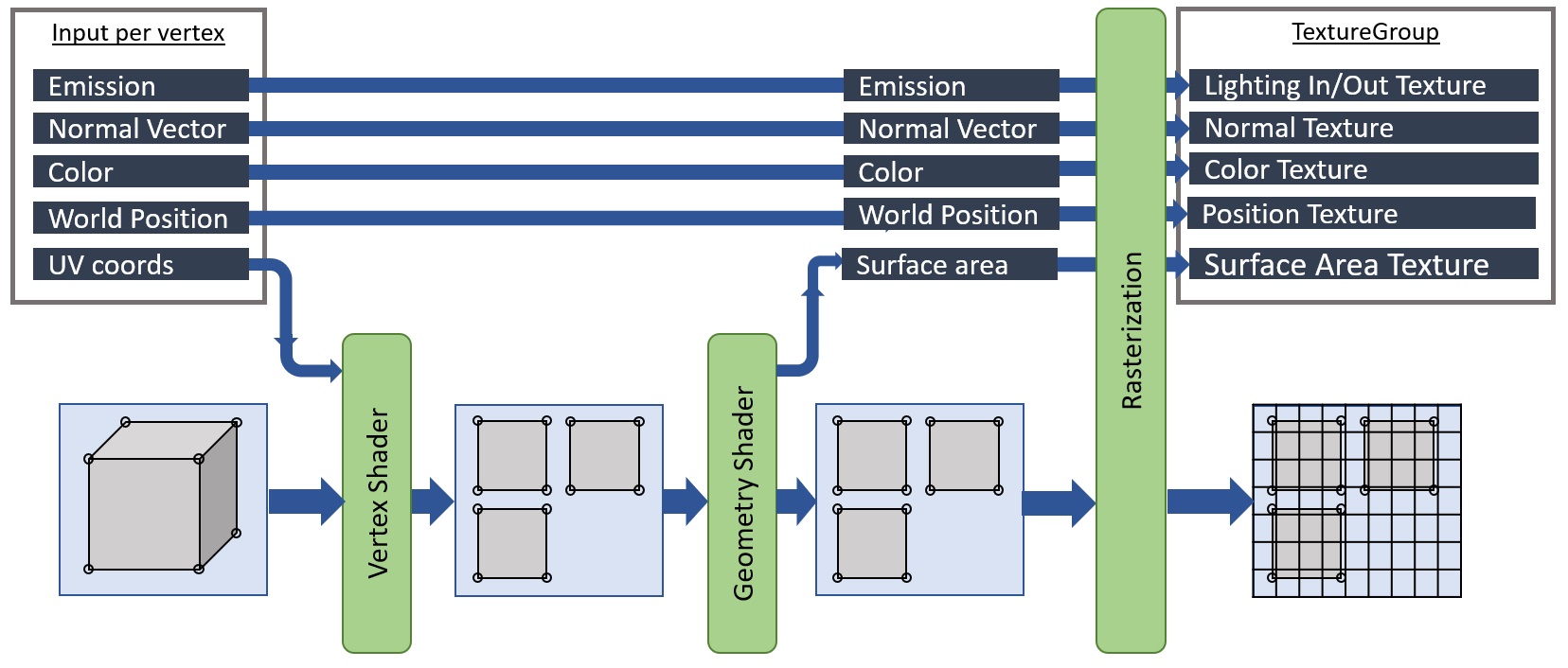}
\decoRule
\caption[]{Dataflow diagram of the CITPass, resulting in the texturegroup that serves as the input-output datastructure for subsequent passes.}
\label{CITPass}
\end{figure}

We perform this operation on a per-triangle basis within the geometry shader of the CITPass. Unlike other common applications of geometry shaders (see \ref{GeometryShader}) we do not modify the output vertices, but merely append the fraction above as an attribute to each vertex, which then gets passed on to the pixel shader that stores it in the surface area texture \textit{arf}.

The overall execution and information flow of the CITPass is visualized in fig. \ref{CITPass}.

\section{RTLPass}

\begin{figure}[th]
\centering
\makebox[\textwidth][c]{\includegraphics[scale=0.3]{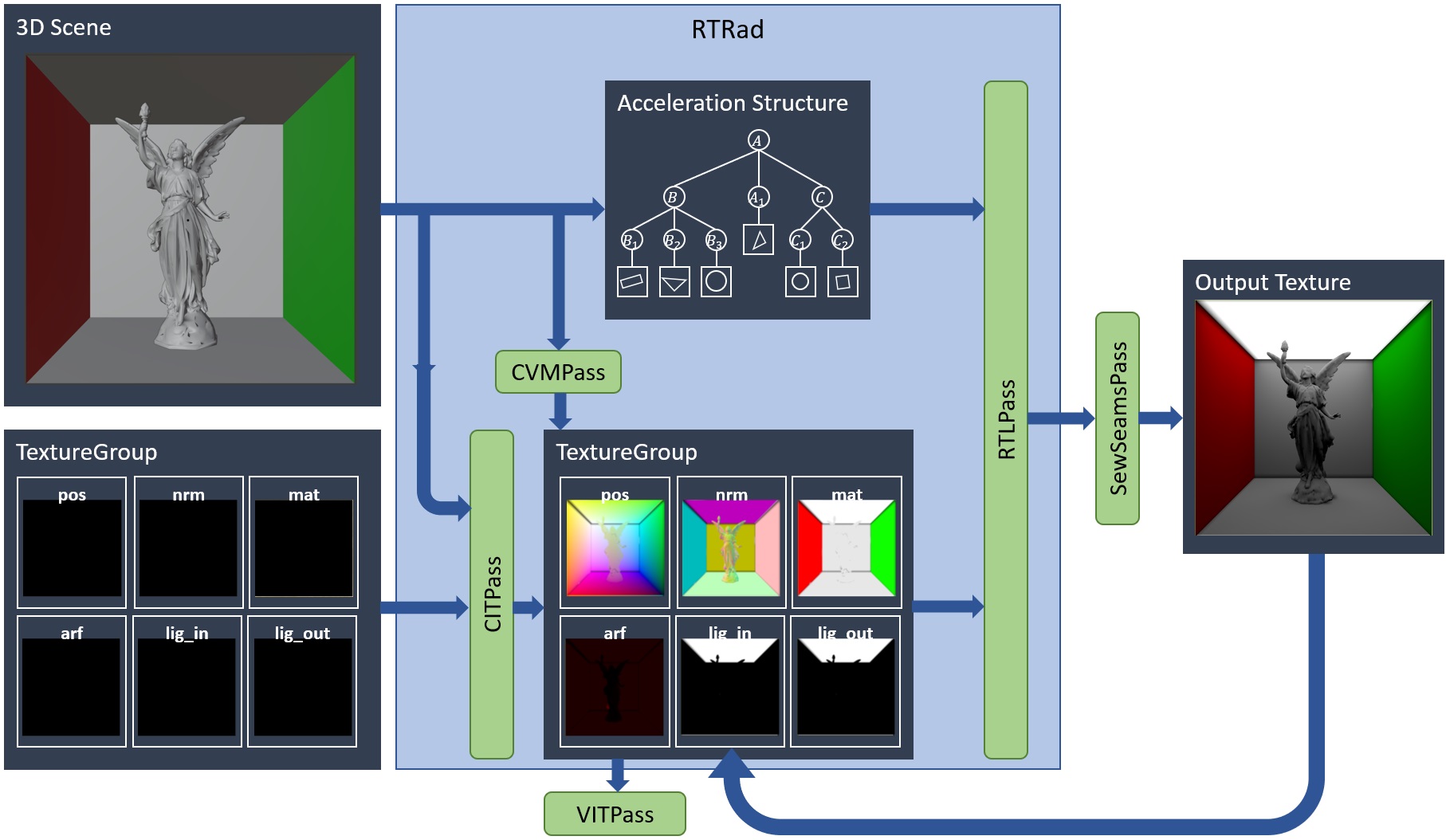}}
\decoRule
\caption[asd]{Generalized overview of information flow in RTRad. The 3D scene is used by the CITPass to populate a blank texturegroup which, alongside the acceleration structure, serves as the input for the RTLPass. The output texture of the RTLPass can then be used as input for the next iteration. Textures can be visualized on screen through a \textit{VITPass} (see \ref{VITPass}). The purpose of \textit{CVMPass} and \textit{SewSeamsPass} are described in sections \ref{VoxelRaymarch} and \ref{SewSeamsPass} respectively.}
\label{RTRad_flow}
\end{figure}

The central component of our application is the \textit{ray-traced lightmap pass} (RTLPass), which takes the generated texturegroup as its input values and produces an output texture correspondent to the lightmap after a single iteration of progressive radiosity. For each subsequent pass the output lighting-texture is swapped and fed back into the algorithm as the input for the next iteration (see fig. \ref{RTRad_flow}).

The RTLPass is the only DXR pass (non-rasterization pass) in the entire application. It launches the ray-generation shader for each pixel in the lighting texture, which commences a separate GPU thread for each radiosity patch which loops over all the other patches to sum up their respective lighting contribution into the output texture\footnote{Note that since visibility is symmetrical under $V(x_1, x_2) = V(x_2, x_1)$, only half the rays given in algorithm \ref{RTLPass_psuedocode} are theoretically required. Unfortunately, in practice this leads to multi-threading memory collisions that are described in section \ref{memory_conflicts}.}.

The overarching procedure of this pass closely resembles the progressive radiosity algorithm given in section \ref{ProgressiveRadiosity} and can be summarized with the following pseudo-code:

\begin{algorithm}[H]
    \caption{RTLPass}\label{RTLPass_psuedocode}
    \begin{algorithmic}[1]
      \For{$i \in [0, n]$}\Comment{For each patch (executed in parallel)}
        \State $L_{out}(i) \gets L_e(i)$\Comment{Set initial lighting value}
        \For{$j \in [0,n]$}
            \If{$j \neq i$} \Comment{For every other patch}
                \State Shoot a ray from $pos(i)$ to $pos(j)$
                \If{no geometry is encountered along the way}
                    \State Calculate view factor $F(i,j)$
                    \State $L_{out}(i) \gets L_{out}(i) + mat(i) * F(i,j) * L_{in}(j)$ \Comment{Add contribution}
                \EndIf
            \EndIf
        \EndFor
    \EndFor
    \end{algorithmic}
\end{algorithm}

\subsection{Visibility Raytracing}

As defined in section \ref{RaytracingForVisibility}, two locations $x_1$ and $x_2$ are mutually visible if a ray launched from one towards the other, arrives at the other unimpeded:

\begin{equation}
    V(x_1, x_2) \leftrightarrow I(x_1, x_2 - x_1) = x_2
\end{equation}

The insight gained from the shader execution order pictured in chapter \ref{Chapter3} (fig. \ref{rtx_pipeline_detailed}) implies that, in order to minimize the required amount of BLAS traversal the ray lengths ought to be kept as short as possible in addition to stopping execution upon any intersections.

As an alternative to the equation above, two points $x_i$ and $x_j$ are also mutually visible if a ray from $x_i$ towards $x_j$ with some offset $\varepsilon$ and of length $|x_i - x_j| - 2\varepsilon$ does not encounter any geometry at all, which in RTX would trigger the execution of the miss shader.

Shortening the rays to a length of $|x_i - x_j| - 2\varepsilon$ and then determining visibility through the \textit{miss shader} requires fewer intersection tests, binding-table lookups and should generally speed up the algorithm, as any non-visible pairs can be quickly discarded through the  \verb|RAY_FLAG_ACCEPT_FIRST_HIT_AND_END_SEARCH| flag.

In our implementation, we use the ray payload to get indices on the origin and destination patches into the miss shader.
Alternatively, one could also store a boolean that represents visibility and then perform lighting calculations in the ray-generation shader itself.

Below we provide a simplified, superficial version of our RTLPass shader code:

\begin{lstlisting}
// Custom ray-payload
struct RayPayload
{
    uint2 origin_coord;     // sampler patch
    uint2 destination_coord;// sampled patch
};

[shader("raygeneration")]
void launchRays()
{
    uint2 origin_coord = DispatchRaysIndex().xy;

    for (uint x = 0; x < lightmap.width(); x++) {
        for (uint y = 0; y < lightmap.height(); y++) {

            uint2 destin_coord = uint2(x, y);

            float3 origin_pos = pos[origin_coord];
            float3 destin_pos = pos[destin_coord];

            RayDesc ray;
            ray.Origin = origin_pos;
            ray.Direction = destin_pos - origin_pos;
            ray.TMin = EPSILON;
            ray.TMax = distance(origin_pos, destin_pos) - 2 * EPSILON;

            RayPayload rpl = { origin_coord, destin_coord };

            TraceRay(scene,
                RAY_FLAG_ACCEPT_FIRST_HIT_AND_END_SEARCH, 
                0xFF, 0, 0, 0
                ray,
                rpl
            );
        }
    }
}

[shader("miss")]
void primaryMiss(inout RayPayload rpl)
{
    // Lighting calculations.
    // Determine what light rpl.destin_coord contributes onto rpl.origin_coord
}
\end{lstlisting}

\subsection{View Factor Calculation}

\begin{figure}[th]
\centering
\includegraphics[scale=0.59]{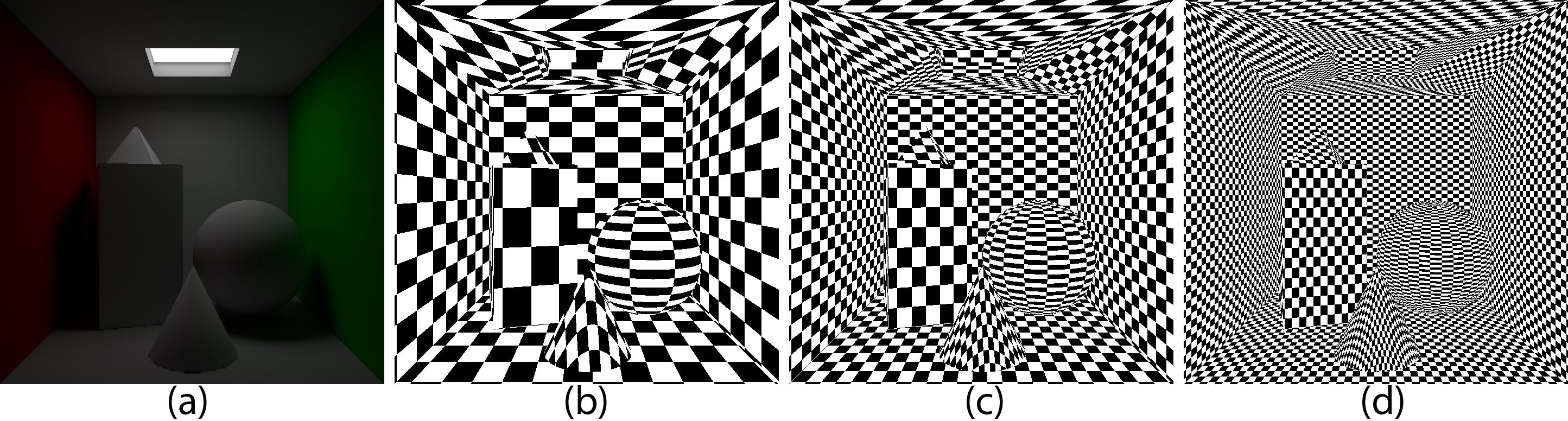}
\decoRule
\caption[]{Original scene (a) and patches for a lightmap of $32\times32$, $64\times64$ and $128\times128$ pixels respectively (b-d). Each black or white rectangle corresponds to a single patch.}
\label{lightmap_sizes}
\end{figure}

From the results of some of our early prototypes we deduced that increasing the amount of samples in a Monte-Carlo based view factor was operationally equivalent with simply increasing the resolution of the underlying lightmap itself. Instead of sampling a single patch multiple times, this would essentially divide the patch up into several smaller ones (see fig. \ref{lightmap_sizes}).

In light of the foregoing, we determined that a Monte-Carlo view factor with $K=1$ was sufficient, where larger values of $K$ can effectively be simulated by larger lightmaps as well as our employed Monte-Carlo \textit{undersampling} (see section \ref{StaticUndersampling}). As such, we are able to compute view factors in a single step through the following formula:

\begin{equation}
    F_{i,j} = \frac{1}{K} \sum_{k=1}^{K} A_j \frac{\cos \theta_{ki} \cos \theta_{kj}}{\lVert x_{ki}-x_{kj} \rVert ^2} V(x_{ki},x_{kj}) \approx  A_j \frac{\cos \theta_{i} \cos \theta_{j}}{\lVert x_{i}-x_{j} \rVert ^2} V(x_{i},x_{j})
\end{equation}

Bearing in mind the individual textures given in section \ref{TextureGroup}, allows us formulate this equation as the exact programmatic steps our shader undertakes to compute the view factors between two patches $i$ and $j$:

\begin{equation}
    F(i, j) = arf(j) \frac
    {(nrm(i) \cdot \frac{pos(j) - pos(i)}{||pos(j) - pos(i)||}) * (nrm(j) \cdot \frac{pos(i) - pos(j)}{||pos(i) - pos(j)||})}
    {||pos(i) - pos(j)||^2}
\end{equation}

where $nrm(i)$ represents a texture-lookup on the normal-vector texture for patch $i$ etc.

\subsection{Lighting Contribution}
\label{RTRad_Lighting}

Combining the factors from the radiosity equation given in \ref{eqn:RadiosityEquation} with the view factor definition given above, yields the total lightflow from a patch $j$ to another patch $i$ as the following:

\begin{equation}\label{eqn:RTRad_Lighting}
    L(j \xrightarrow{} i) = lig_{in}(j) * mat(i) * \frac{\rho}{\pi} * F(i, j)
\end{equation}

where $\rho$ is the reflectivity constant, $mat$ is the color of the material and $lig_{in}$ is the input lighting texture (equivalent to the emission texture on the first pass).

This equation can be seen reflected in the code of our miss shader. A simplified version of said shader is listed below:

\begin{lstlisting}
void AddContribution(uint2 self_c, uint2 other_c) {
    // World positions
    float3 self_wpos = pos[self_c].xyz + minPos;
    float3 other_wpos = pos[other_c].xyz + minPos;

    // Distance
    float3 self_to_other = other_wpos - self_wpos;
    float r = length(self_to_other) * distance_factor;

    // Cosines
    self_to_other = normalize(self_to_other);
    float3 self_nrm = nrm[self_c].xyz;
    float3 other_nrm = nrm[other_c].xyz;
    float self_cos = dot(self_nrm, self_to_other);
    float other_cos = dot(other_nrm, -self_to_other);

    if (self_cos <= 0.0f || other_cos <= 0.0f) return;

    // Form factor
    float F = arf[other_c].r * self_cos * other_cos * (1.0f / (PI * r * r));

    // Apply contribution
    lig_out[self_c] += lig_in[other_c] * mat[self_c] * reflectivity_factor * F;
}
\end{lstlisting}

We found that several custom tweaks not shown in the pseudo-code above, such as clamping lighting contribution and form factors to certain maximum values, significantly improved lighting quality.

\subsection{Indexing and Memory Conflicts}\label{memory_conflicts}

In theory, the reciprocity rule (\ref{eqn:ReciprocityRule}) implies that a view factor between two patches  only needs to be computed once, with its inverse resulting from a simple multiplication in the form of $F_{ji} = \frac{A_i}{A_j}F_{ij}$.
This simple fact has far-reaching implications in that dramatically fewer texture-lookup and ray-trace operations are required, effectively cutting the required computational expense in half. Unfortunately, it does not play out this trivially in practice.

A naive implementation may look as follows:

\begin{algorithm}[H]
    \caption{RTLPass - Reciprocity Rule}\label{RTLPass_pseudocode_naive}
    \begin{algorithmic}[1]
      \For{$i \in [0, n]$}
        \State $L_{out}(i) \gets (0,0,0)$
        \For{$j \in [i+1,n]$}
            \State Shoot a ray from $pos(i)$ to $pos(j)$
            \If{no geometry is encountered along the way}
                \State Calculate view factor $F(i,j)$
                \State $L_{out}(i) \gets L_{out}(i) + F(i,j) * L_{in}(j)$
                \State $L_{out}(j) \gets L_{out}(j) + F(i,j) * \frac{arf(i)}{arf(j)} * L_{in}(i)$
            \EndIf
        \EndFor
    \EndFor
    \end{algorithmic}
\end{algorithm}

Note in particular line 3, where the inner for-loop commences at $i+1$. Under this offset the complexity of the algorithm is lowered to its Gaussian sum $\frac{n^2 + n}{2}$ which, despite ultimately boiling down to $O(n^2)$, still provides a considerable gain in performance.

The crux of this approach emerges from parallelization, as multiple threads modifying a single patch cause \textit{memory write collisions}. Specifically, parallelizing the outer for loop in algorithm \ref{RTLPass_pseudocode_naive} will cause such collisions on line 8, as multiple separate threads adopt the same value for $j$. If the inner loop is parallelized, the same issue will occur on line 7.

\begin{figure}[th]
\centering
\includegraphics[scale=0.39]{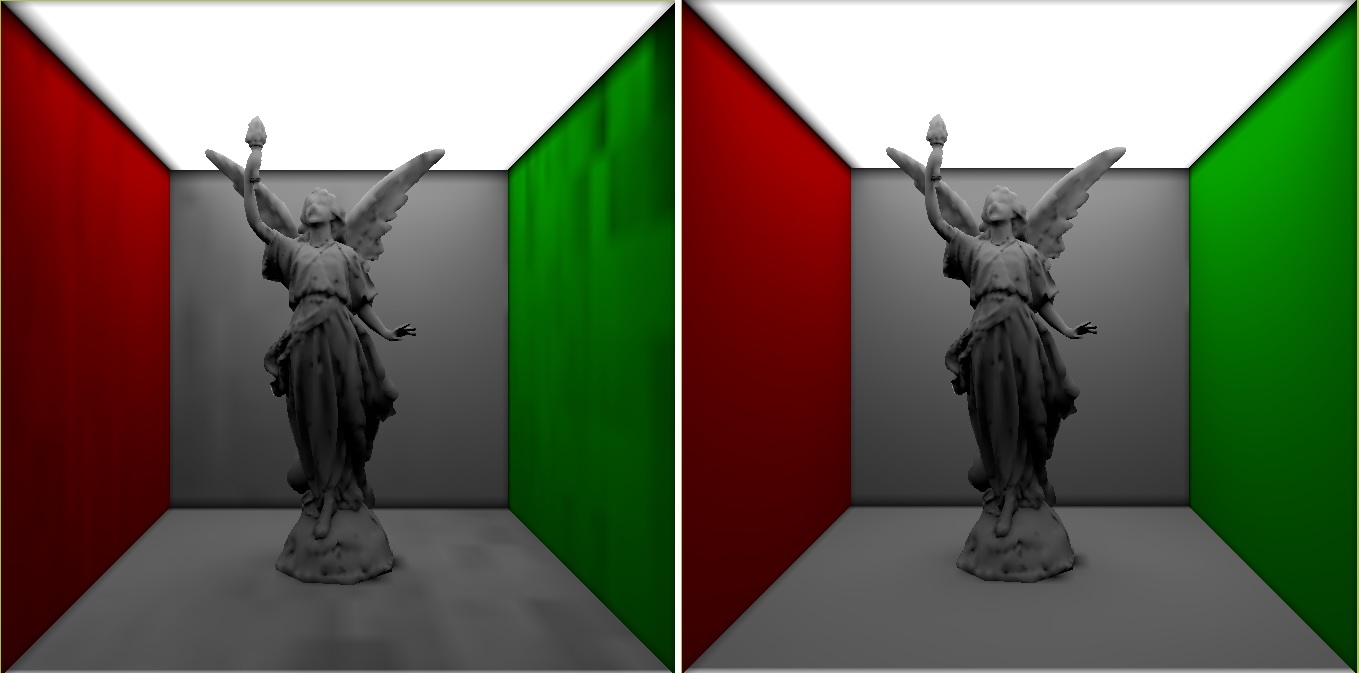}
\decoRule
\caption[asd]{Results when applying a regular algorithm \ref{RTLPass_psuedocode} (right) and when applying the flawed algorithm \ref{RTLPass_pseudocode_naive} (left). Note in particular the grey smudges on the walls that result from collisions of multiple threads writing simultaneously to the same memory address. The location and intensity of the smudges are mostly consistent across multiple runs, but differ depending on input parameters.}
\label{MemoryCollisions}
\end{figure}

Thread safety features, such as mutex locks, come with their own respective performance overheads and are not available in common GPUs.
If ignored, these write conflicts manifest themselves as unusual artifacts depicted in fig. \ref{MemoryCollisions}. The resulting smudges can be allayed under specific conditions using certain batching parameters, though the results are highly unreliable\footnote{The results in fig. \ref{MemoryCollisions} have been produced with a lightmap resolution of $128\times128$ and a batch size of $64\times64$. On our hardware, different batch sizes sway the rate and intensity of these imperfections, but finding the ideal size largely comes down to trial-and-error.}.

For a more reliable solution, following options are available:

\begin{itemize}
    \item Adhering to our original algorithm \ref{RTLPass_psuedocode}, where each thread is assigned a singular patch $i$ that it can write to. No writing conflicts will occur, albeit view factors and rays will have to be computed twice for each patch-pair.
    \item These duplicate calculations can be prevented by temporarily caching their results. For instance, a thread processing patch $i$ would temporarily store the visibility and/or view factor towards another patch $j$ in memory, so that when the thread processing patch $j$ samples patch $i$, the cached value can simply be retrieved. Unfortunately, GPUs have notoriously limited memory capacities and the amount of memory required lies in $O(n^2)$.
    Our testing concluded that caching visibility data \textit{can} be viable for small lightmaps. A detailed account on our findings regarding visibility caching can be found in section \ref{Viscaching}.
    \item Since Turing GPUs contain between 40 to 80 RT cores, a rough limit of 40-80 threads can be assumed to be writing to memory simultaneously. An indexing solution ensuring that any patch-pairs which may potentially write into shared memory are far enough apart in the index-sequence may diminish the effects of memory-conflicts sufficiently to make the final result identical.
\end{itemize}

Most multi-threaded, progressive refinement solutions implement the first solution by calculating all view factors on-the-fly.
Naturally, the performance balance depends on whether the benefits of GPU parallelization can outweigh the cost incurred by doubling the amount of visibility calculations.

Our implementation runs, by default, with the same approach but also allows visibility caching to be enabled for smaller lightmaps (see section \ref{Viscaching}).

\subsection{Batching}

Operating systems typically expect applications to remain responsive in scheduling. Forcing an indefinitely large workload onto the GPU will cause the program to be terminated by the OS after exceeding a certain time threshold, which can be reached rather quickly by a workload as complex as progressive radiosity.

To circumvent this problem, we compute lightmaps in a series of batches, as opposed to all at once.
The batching process is managed by internally the RTLPass object, which allows the user to adjust certain batching parameters through its GUI.
These parameters allow defining the dimensions of the rectangle - or strip - that is computed with each batch. 

By gradually testing various thresholds, we determined an upper limit of $128^4$ rays traced per batch to be adequate for our specific hardware.
Depending on the sampling and resolution settings, the RTLPass object ensures that batching parameters are dynamically adjusted in order to remain at or below this limit.

We recommend maintaining the batch size as close to this limit as possible, because a larger batch count (such as a single patch per batch) inevitably comes with a large overhead cost in setting up each individual render pass.

\section{SewSeams Pass}
\label{SewSeamsPass}

When patches are not aligned with geometry, lights and shadows can leak and produce undesirable protuberances in the form of unrealistic shadows or highlights \cite{FormFactorLecture}. 

Increasing lightmap size or adjusting UV coordinates can eliminate these issues entirely, but is difficult to do automatically. Some re-meshing solutions generate edges along predicted discontinuities of the radiosity function \cite{FormFactorLecture}, though this only functions if each patch is given its own primitive.

\begin{figure}[H]
\centering
\includegraphics[scale=0.4]{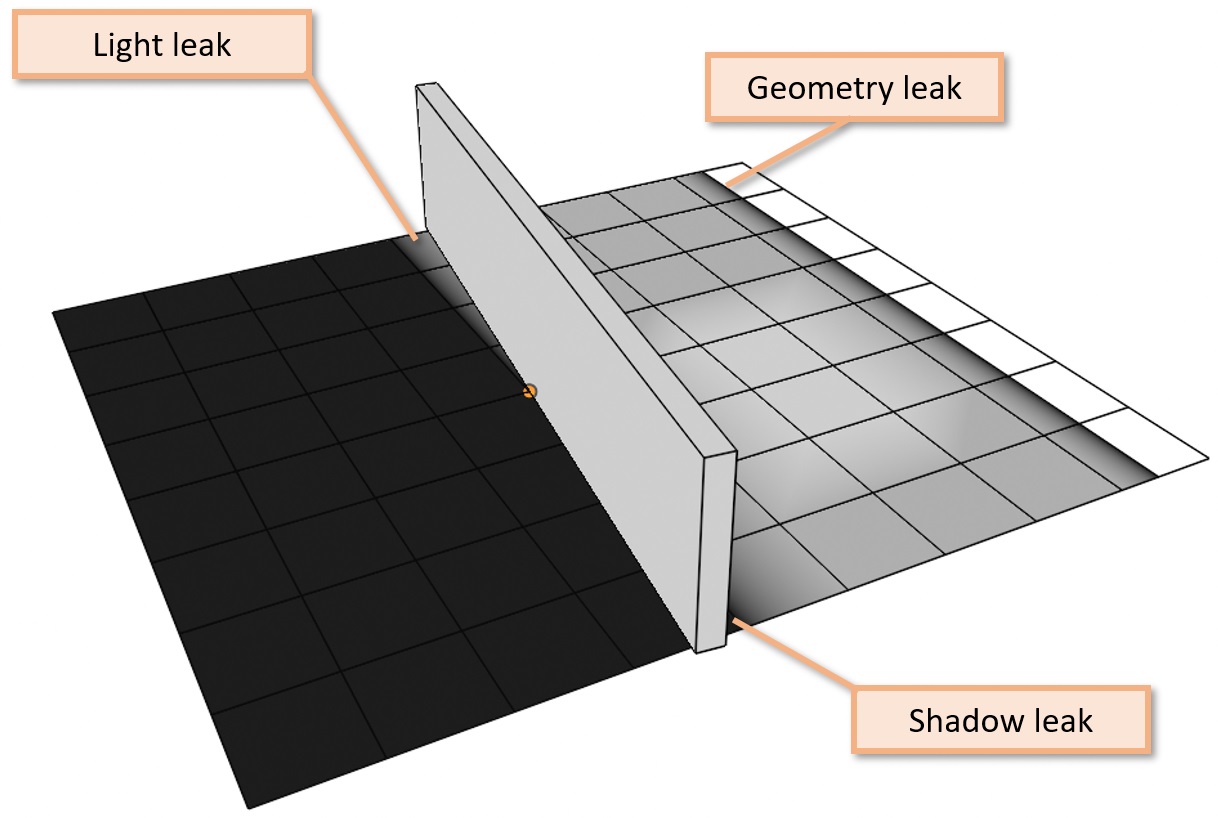}
\decoRule
\caption[asd]{Light and shadow leaks on UV-mapped lighting textures. Geometry leaks occur when a patch/pixel is only marginally covered by a surface. }
\label{leaks_fig}
\end{figure}

Similar inaccuracies (labeled as \textit{geometry leaks} in fig. \ref{leaks_fig}) can result from pixels neighbouring a patch that does not have its center covered by a surface, as these will not be rasterized by the CITPass. These types of leaks can be far more frequent than the former and will not only manifest themselves in corners.

Conservative rasterization prevents this issue, but produces additional problems, as it overwrites patches when occupied by several primitives in addition to annulling our approximation for surface area (see section \ref{SurfaceArea}).

We employed a custom solution in the form of an additional \textit{SewSeams} pass that is executed after each radiosity iteration. 

\begin{figure}[th]
\centering
\includegraphics[scale=0.9]{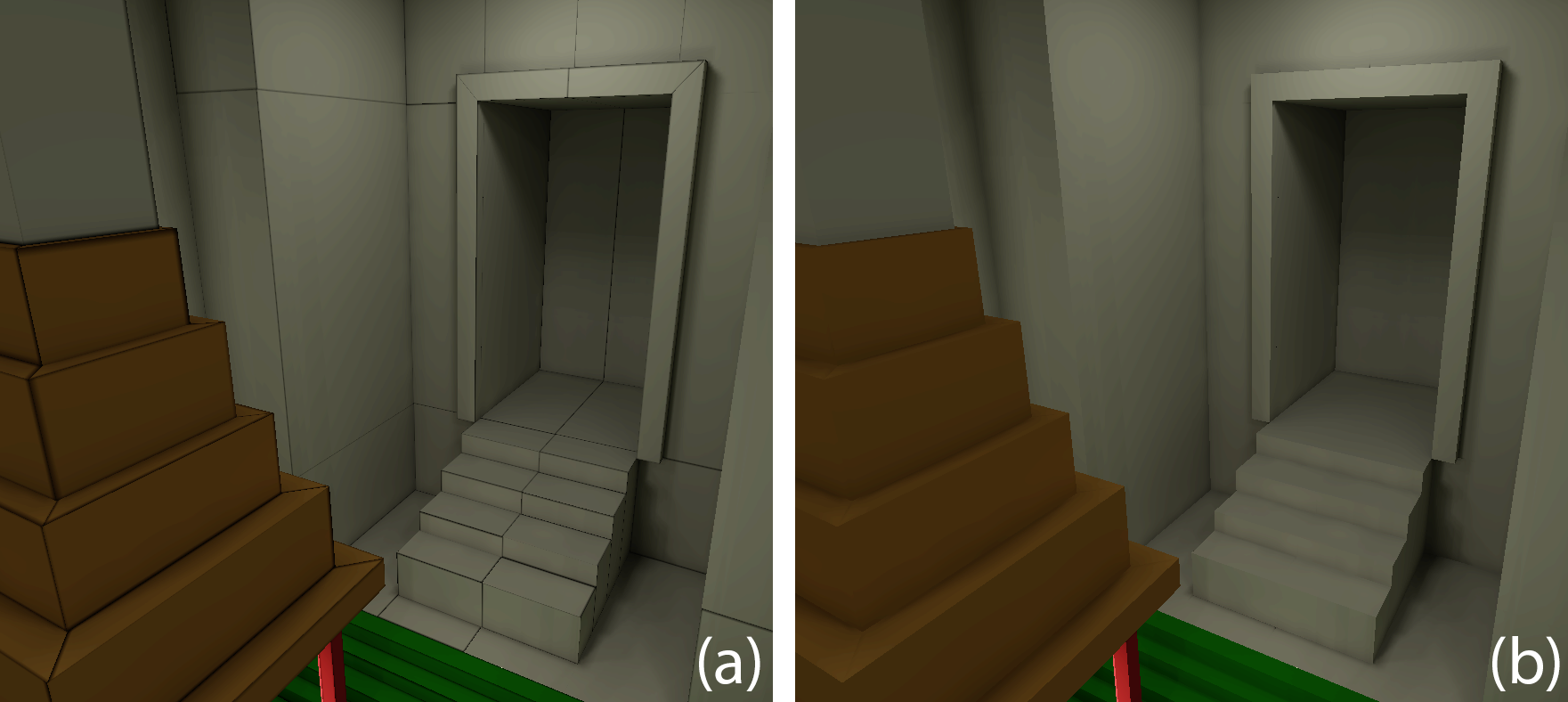}
\decoRule
\caption[asd]{Scene without (a) and with (b) a \textit{SewSeams} pass.}
\label{leaks_fig2}
\end{figure}

This pass runs for every lightmap pixel that was not rastered by the CITPass and thus is not treated as a patch. If any of the pixel's neighbours \textit{is} a patch, then the pixel assumes its color. This process effectively expands the UV mapping of each cohesive shape by a margin of one pixel, which eliminates the vast majority of geometry-based leaks entirely (as shown in fig. \ref{leaks_fig2}).

\section{VITPass}\label{VITPass}

For visualization purposes we employ an additional \textit{visualize input textures pass} (VITPass), which serves as a tool to display and analyze input and output data. Whilst this pass may not be an intrinsic part of our algorithm, it certainly proved invaluable to debug, optimize and analyze our implementation.

We employ a plethora of settings that can be accessed through Falcor's UI to adjust precisely what output should be displayed and in what form. The user can choose which mipmap level of which texture to display and whether to render them as texture, masked texture or applied to the underlying 3D model.

\begin{figure}[th]
\centering
\includegraphics[scale=0.31]{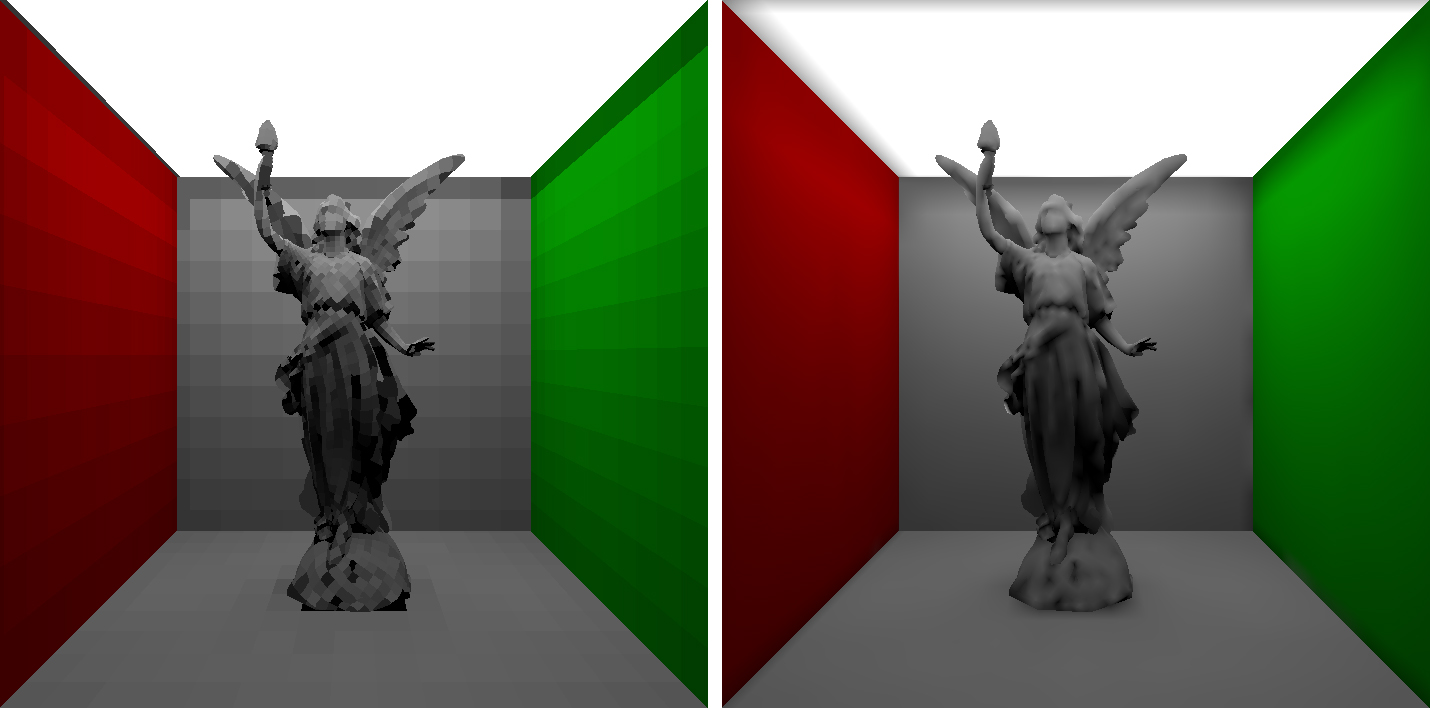}
\decoRule
\caption[asd]{Low resolution lightmap ($64\times64$) without (left) and with (right) a bilinear magnification filter.}
\label{bilinear_filter}
\end{figure}

As is common practice in radiosity, we employ a \textit{bi-linear} magnification filter to smooth out each texture, which can be toggled on or off (see fig. \ref{bilinear_filter}).

\begin{figure}[th]
\centering
\includegraphics[scale=0.83]{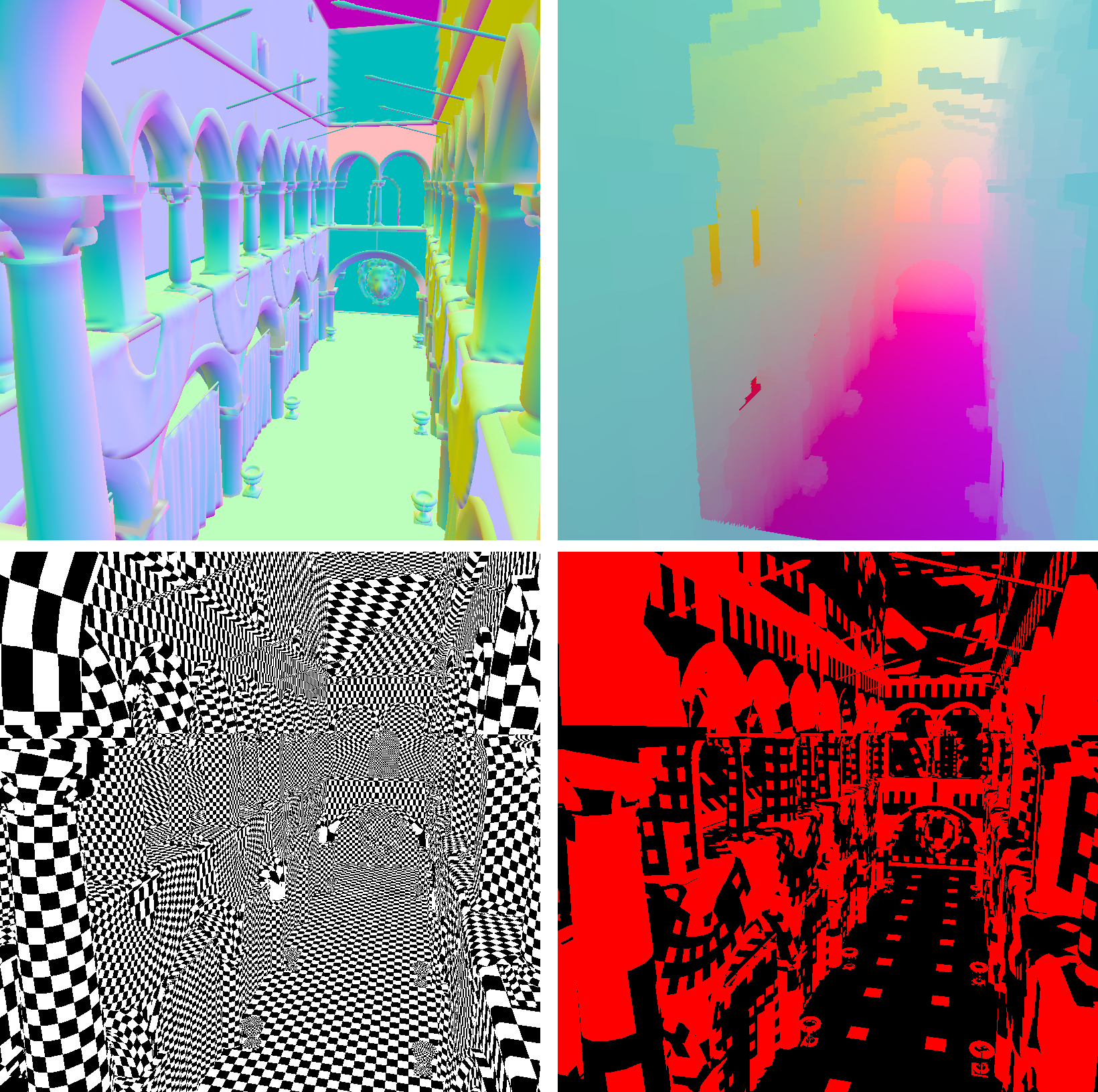}
\decoRule
\caption[asd]{Examples of data visualization through the VITPass. left-to-right, top-to-bottom: Normal vectors, voxel map, texture resolution and quad-tree of a scene.}
\label{visualization_examples}
\end{figure}

The VITPass is built on a custom vertex and pixel shader, which allow for a wide variety of different setups. Additional information such as lightmap resolution, refinement-nodes or voxel-maps can also be adequately visualized (see fig. \ref{visualization_examples}).

All scene renders or raw textures contained in this thesis are produced using the VITPass (unless stated otherwise).

\chapter{Performance Improvements} 

\label{Chapter5} 



Put together, the components outlined in the previous chapter already constitute a fully functional implementation of the progressive radiosity algorithm. In this chapter we expand it to include \textit{refinement} in addition to further performance enhancements and variations commonly seen in other implementations.

\section{Refinement}

In traditional radiosity, patches are assumed to have a uniform radiant exitance across their surface \cite{radiosity_og, progressive_refinement}. 
Since ray-traces are of logarithmic complexity \cite{rtx_gems}, our doubly nested, pair-wise for-loop results in an altogether complexity of $O(n^2\log t)$ where $t$ is the number of primitives and $n$ is the amount of patches.
This complexity suggests that lowering the amount of required samples (e.g. the "contributing" patches $n$) yields a far greater positive impact on performance than diminishing the value of $t$.

\begin{figure}[th]
\centering
\includegraphics[scale=0.57]{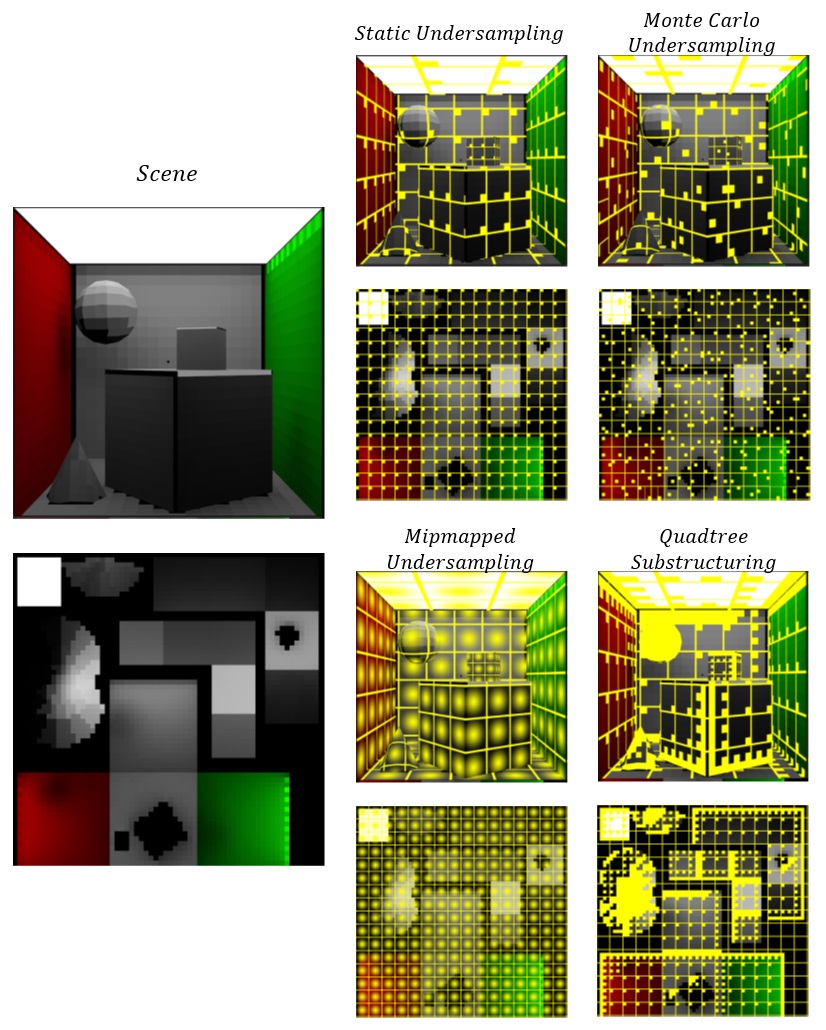}
\caption[]{Different strategies for undersampling on a lightmap of $64\times64$ pixels. One sample is taken for each window of $4\times4$ pixels ($m=16$), with the sampled patches marked in yellow. Mipmapped takes averages and thus has the best coverage, whilst substructuring (e.g. adaptive subdivision) allocates more samples to areas of a high gradient.}
\label{undersampling_strategies}
\end{figure}

There are several different approaches of accomplishing this goal, which were broadly categorized in section \ref{ProgressiveRefinementRadiosity}. In this chapter we primarily focus on methods based on \textit{h-refinement}, namely \textit{static undersampling} and \textit{adaptive subdivision}. Our approaches follow the precedent set by the implementation presented in the GPU Gems 2 \cite{gpu_gems_2005}, by separating the \textit{sampled} patches as a proper subset of the overall lightmap. Our different strategies of generating such a subset are depicted in fig. \ref{undersampling_strategies}.

\subsection{Static Undersampling}
\label{StaticUndersampling}

A crude and simple method of sampling at a lower resolution than that of the lightmap is to simply do so through a static stride. The nature of UV unwrapping implies that neighbouring pixels on the lightmap likely also represent neighbouring patches in world space, thus increasing the viability of discarding samples belonging to these. Whilst no common name has been established for this method, we will refer to it as \textit{static undersampling}.

In its simplest version, we simply discard all but the upper, left-most pixel of each consecutive square consisting of $m$ pixels (the \textit{sampling window}).

This inherently degrades the algorithms' complexity to $O(\frac{n^2}{m}\log t)$, albeit the quality of the resulting lighting may likewise suffer in proportion to $m$. 
The decreased amount of samples needs to be reflected in the lighting contribution values (as given in \ref{RTRad_Lighting}) as a factor of $m$, which is the amount of patches each sample represents:

\begin{equation}
    L(j \xrightarrow{} i) = m * lig_{in}(j) * mat(i) * \frac{\rho}{\pi} * F(i, j)
\end{equation}

\subsubsection{Monte-Carlo Undersampling}

If the edges of a scene's UV map are aligned along one of the texture's axes, then the fixed-step nature of static undersampling may lead to certain colors or surfaces being grossly over-represented in the aggregate lighting contribution. This effect can be mitigated by selecting a \textit{randomized} pixel from each $\sqrt{m} \times \sqrt{m}$ square.

Naturally, the employed randomization ought to be seeded on the patch-index of the contributor, not the shooter.

\subsection{Mipmapped Undersampling}

The information loss incurred by static undersampling can be placated by computing averages for each set of pixels. Some GPU implementations utilize \textit{mipmapping} as a fast and convenient tool to accomplish this \cite{gpu_gems_2005}.

\textit{Mipmaps} are a sequence of pre-calculated, down-scaled versions of a given texture. With each \textit{level}, the image resolution is a factor of four smaller than the previous level. Their applications are primarily centered on \textit{texture filtering} with the intent to reduce aliasing artefacts at long render distances.

\begin{figure}[H]
\centering
\includegraphics[scale=0.27]{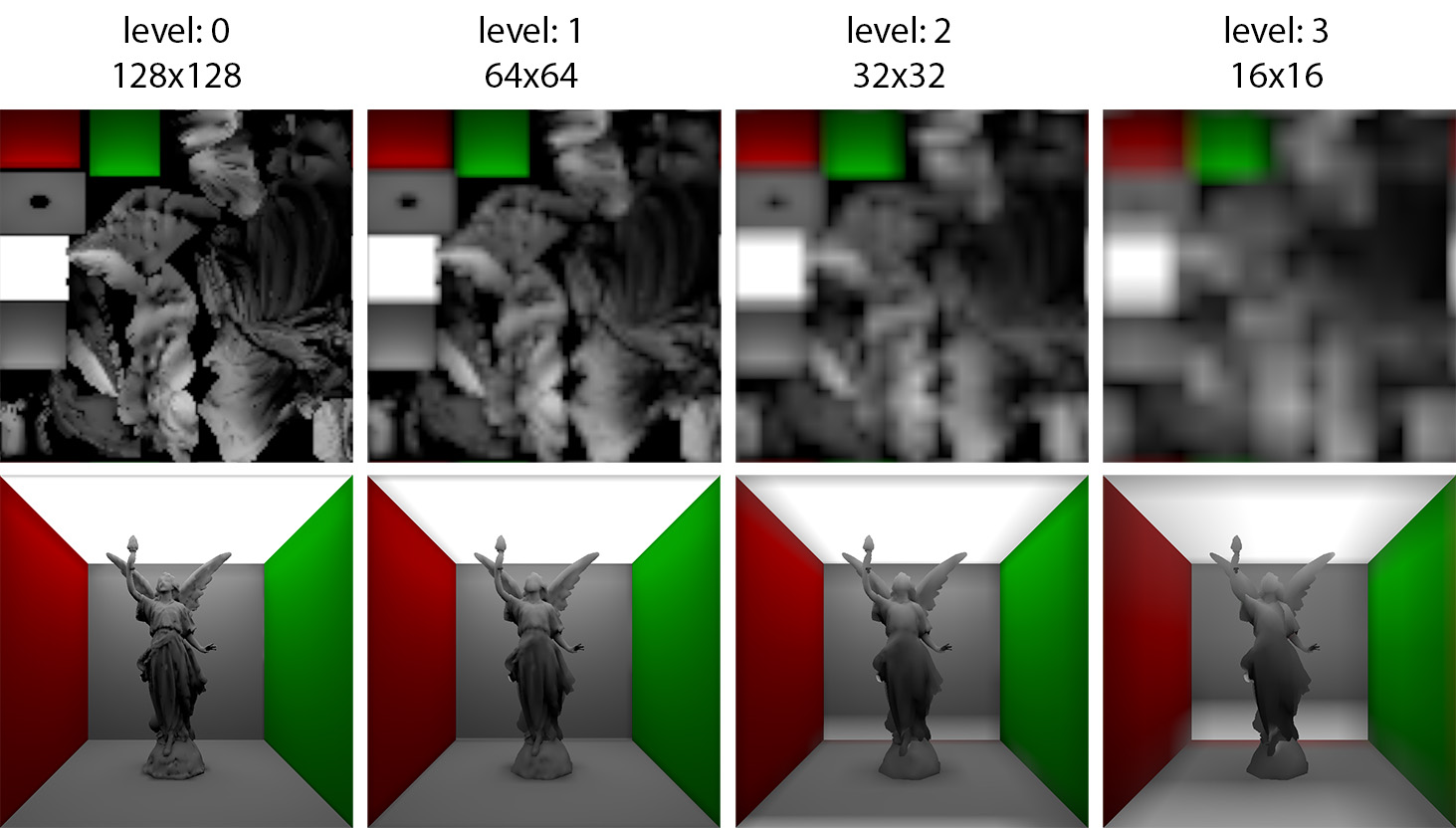}
\decoRule
\caption[]{Bilinear mipmaps of a lightmap (under a bilinear magnification filter). These are the lightmaps sampled by mipmapped undersampling with a value of $m=4$, $m=16$ and $m=64$ respectively.}
\label{mipmaps}
\end{figure}

The GPU employs a \textit{minification filter} to dictate how mipmaps are derived from the original texture. The most commonly used types are the \textit{nearest neighbour} and \textit{bilinear} filters \cite{VXCT, learnopengl}. Bilinear mipmaps are the functionally equivalent inverse of the bilinear \textit{magnification} filter which we touched on in section \ref{VITPass}: Each pixel in the minified image corresponds to the average of a $2\times2$ square in the level below (see fig. \ref{mipmaps}).

Since generating mipmaps on a GPU is a near-instantaneous process, they can be used as a tool to sample texture areas for their average color values conveniently \cite{VXCT, Crassin, gpu_gems_2005}. Computing the average color of a $\sqrt{m} \times \sqrt{m}$ square simply corresponds to a texture lookup on the mipmap of level $\log_2(\sqrt{m})$. 

Mipmapped undersampling ensures no vital patches (such as light-sources) are discarded, but can lead to colors leaking from one surface to another when sampled.

\subsection{Adaptive Subdivision}

\begin{figure}[th]
\centering
\includegraphics[scale=0.3]{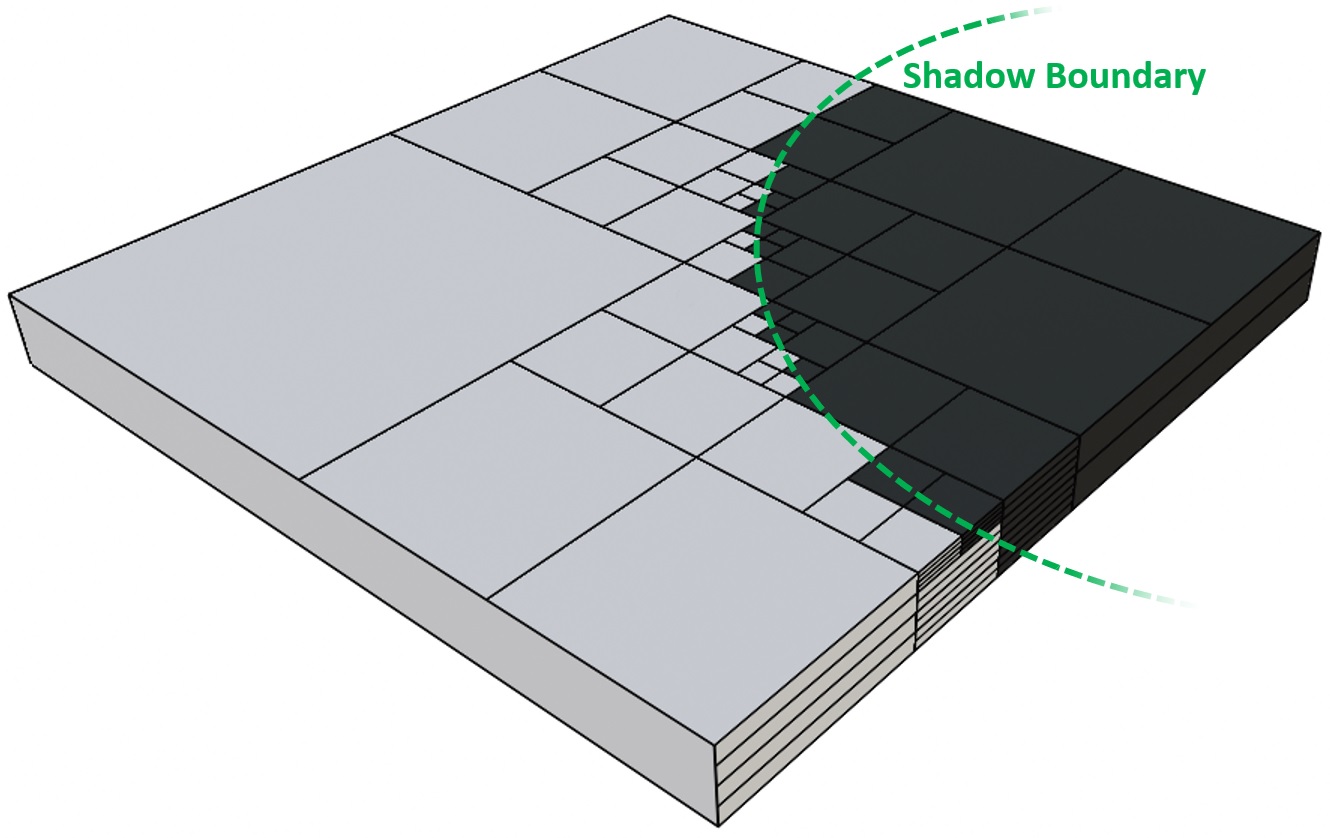}
\decoRule
\caption[]{Quad-tree substructuring of a surface. More subdivisions are made in areas of high detail.}
\label{quad_tree_shadowboundary}
\end{figure}

\textit{Adaptive subdivision} is a more sophisticated counterpart to static undersampling and is generally regarded as a core part of progressive refinement radiosity \cite{progressive_refinement}. Patches that have a high gradient across them, such as shadow boundaries and penumbras, are subdivided into finer grids, whilst low detail areas are represented by a single patch \cite{progressive_refinement, Weimar_radiosity} (see fig. \ref{quad_tree_shadowboundary}). 

Our implementation of this concept in RTRad is loosely based on the hierarchical, quad-tree approach used by Willmott et al. \cite{empirical_comparison} as well as the "progressive accuracy" solution employed by Elias \cite{HugoElias_Radiosity}, whilst we somewhat diverge from other GPU implementations that map each patch to a specific scene polygon and maintain a quad-tree datastructure inside a static buffer \cite{gpu_gems_2005}.

Given our emphasis on handling data directly within UV-wrapped textures, we opted to store meta-information on our quad-tree structure within the alpha-channel of the lightmap texture itself.
This approach allows us to keep the algorithm shown in the previous chapter mostly intact, whilst letting us sample at a significantly lower rate.

\subsubsection{Alpha-Embedded Substructuring}

Section \ref{TextureGroup} briefly mentions that RTRad uses the alpha channel of the position-texture to mark patches that are not occupied by any geometry and thus have no respective surface in the scene.
Our quad-tree employs a similar scheme in that the alpha value of each pixel equates to the amount of patches its color value represents in sampling.

Before commencing a radiosity iteration, we run a fullscreen pixel shader on the input lighting texture that constructs the appropriate quad-tree inside the alpha channel.
The tree is constructed bottom-up, with each pass of the shader processing a single level. For a maximum node size of $16\times16$ pixels, the required passes would thus be $\log_2(16) = 4$, equivalent to the height of a quad-tree that can represent $16\times16$ values under a single root.

Each pass examines the gradient of every four neighbours and determines whether these are to be merged into a single node or not. If the colors hardly deviate, the upper left-most node is designated to represent the entire group, with its siblings being dropped from the sampling pool.

\begin{figure}[th]
\centering
\includegraphics[scale = 0.46]{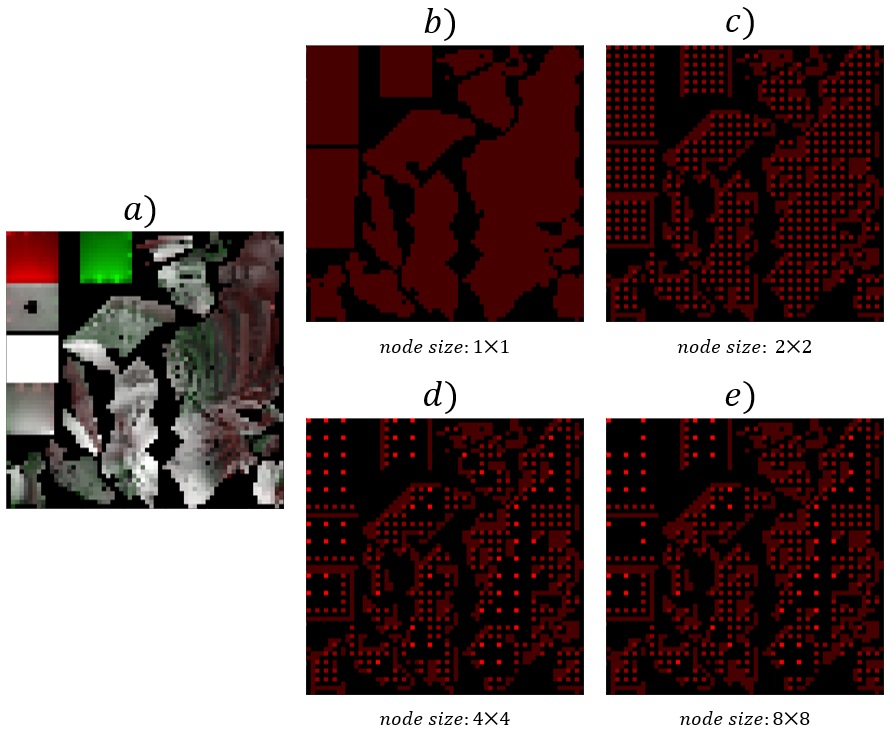}
\caption[]{A $64\times64$ RGB lightmap (a) and its alpha channel (b-e) after 0, 1, 2 and 3 substructuring passes respectively with a gradient threshold of $0.2$. }\label{fig:quad-tree-construction}
\end{figure}

\newpage
The algorithm functions in accordance to the following pseudo-code:

\begin{algorithm}[H]
    \caption{Alpha - Substructuring}\label{Alpha_Substructure_pseudocode}
    \begin{algorithmic}[1]
        \For{each pixel $(x,y) \in lightmap$}\Comment{Initialize all alpha values as 1}
            \State $alpha(x,y) \gets 1$
        \EndFor
        
        \For{$step \in \{ 2, 4, 8 ... \}$}
            \For{each pixel $(x,y) \in lightmap$}
                \If{$x$ and $y$ are divisible by $step$}\Comment{For every $step\times step$ pixels}
                    \State $child_1 \gets (x,y)$\Comment{Get 4 neighbouring child-nodes}
                    \State $child_2 \gets (x+\frac{step}{2},y)$
                    \State $child_3 \gets (x,y+\frac{step}{2})$
                    \State $child_4 \gets (x+\frac{step}{2},y+\frac{step}{2})$
                    
                    \If{all child-nodes have $alpha(child_x) \geq (\frac{step}{2})^2$}\Comment{Only merge nodes that have no children of their own}
                        \State $g \gets gradient(child_1, child_2, child_3, child_4)$\Comment{Calculate gradient}
                        \If{$g < threshold$}\Comment{Merge children}
                            \State $alpha(child_1) \gets \sum_{i=1}^{4} alpha(child_i)$
                            \State $alpha(child_2) \gets 0$
                            \State $alpha(child_4) \gets 0$
                            \State $alpha(child_5) \gets 0$
                        \EndIf
                    \EndIf
                    
                \EndIf
            \EndFor
        \EndFor
    \end{algorithmic}
\end{algorithm}

As shown in fig \ref{fig:quad-tree-construction}, all occupied geometry will commence with an alpha value of one, then each $2\times2$ square with a low gradient is merged by transferring the alpha value of all pixels inside the square into the its upper, left-most member. The same process is repeated for $4\times4$ and $8\times8$ squares respectively.

Once the alpha-embedded quad-tree is constructed for the $lig_{in}$ texture, the RTLPass will loop over all other patches, but disregard those with alpha values of zero and multiply the lighting contribution of the rest by their alpha values. The following algorithm amends our original algorithm \ref{RTLPass_psuedocode} from the previous chapter:

\begin{algorithm}[H]
    \caption{RTLPass with Adaptive Subdivision}\label{RTLPass_substructuring}
    \begin{algorithmic}[1]
      \For{$i \in [0, n]$}
        \State $L_{out}(i) \gets (0,0,0)$
        \For{$j \in [0,n]$}
            \If{$j \neq i$ and $alpha(j) > 0$}
                \State Shoot a ray from $pos(i)$ to $pos(j)$
                \If{no geometry is encountered along the way}
                    \State Calculate view factor $F(i,j)$
                    \State $L_{out}(i) \gets L_{out}(i) + alpha(j) * mat(i) * F(i,j) * L_{in}(j)$
                \EndIf
            \EndIf
        \EndFor
    \EndFor
    \end{algorithmic}
\end{algorithm}

The alpha channel of a standard GPU texture can typically only hold 8 bits of information which, under our algorithm, limits the maximum size of a node to $\sqrt{2^8}^2 = 16\times16$ pixels, amounting to a maximum quad-tree height of four. Although this limitation is negligible, as one could simply store the logarithmic or fractional alpha value instead, the remainder of this thesis will work with an upper node limit of $16\times16$.

\subsubsection{Gradient Calculation}

Determining an approximate gradient for a lighting texture can be accomplished in a number of ways and can incorporate data such as color values to variations in normal vectors. We based our function largely on Willmott et al. \cite{empirical_comparison}, which uses the standard deviation of each patch's radiosity value. We additionally include deviation in normal vectors, to ensure the edges of cohesive surfaces are weighed more heavily in the gradient:

\begin{equation}
\begin{aligned}
grad(a, b, c, d) = & \frac{1}{2}\lVert lig(a) - mean(lig(a), lig(b), lig(c), lig(d))\rVert \\
+ & \frac{1}{2}\lVert nrm(a) - mean(nrm(a), nrm(b), nrm(c), nrm(d))\rVert
\end{aligned}
\end{equation}

where, in our case, $a$ is the upper left-most child or pixel.

\begin{figure}[h]
\centering
\includegraphics[scale=0.5]{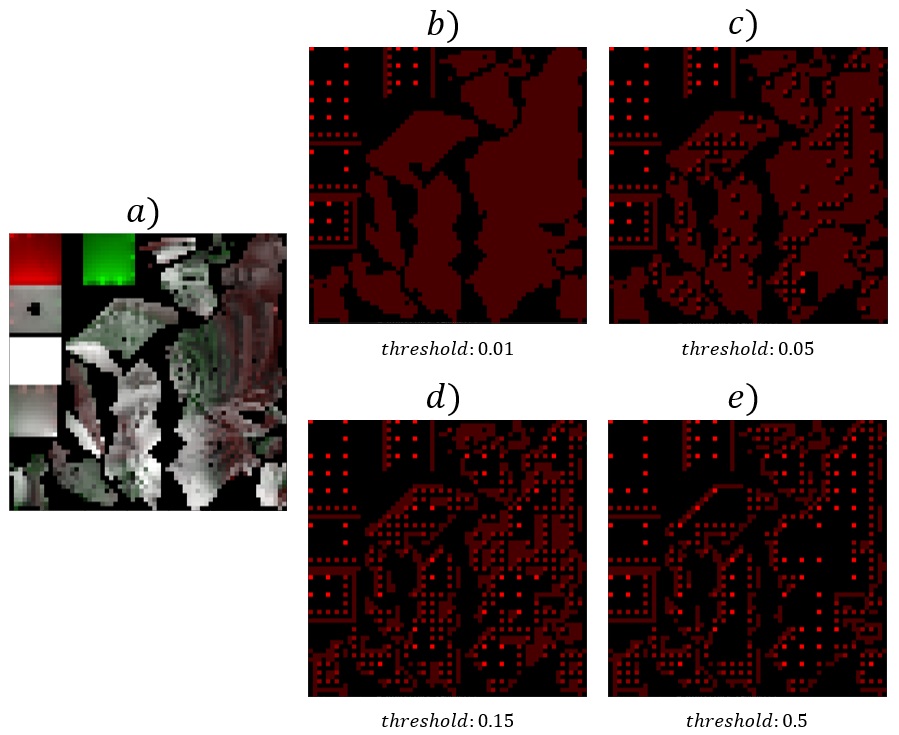}
\caption[]{A $64\times64$ RGB lightmap (a) and its alpha-embedded quad-tree for different gradient thresholds (b-e) (black pixels are not sampled). A larger threshold will lead to more aggressive merging.}
\label{quadtree_threshold}
\end{figure}

Adjusting the gradient threshold respectively leads to more or fewer samples taken (see fig. \ref{quadtree_threshold}). For small lightmaps, we found a value of $0.05$ to provide a decent balance between performance and accuracy. Given that each consecutive RTLPass iteration has a lesser effect on lighting than the previous, a reasonable approach would be to ramp this value up successively with each pass.

\newpage
\section{Visibility Caching}\label{Viscaching}

In this section we describe a visibility caching method used to complement the RTRad algorithm. The resulting performance shows promise for smaller lightmaps and is analyzed in greater detail further ahead (see section \ref{viscaching_analysis}).

\subsection{Memory Complexity}

The required memory for all pairs of $n$ radiosity patches lies in $O(n^2)$. In a naive implementation, a modest lightmap of 256x256 pixels would thus require $256^4$ bits (approx. 0.5GB) of storage, which already borders a critical threshold of what lower grade GPUs can accommodate.

Fortunately, not every pair of patches needs to be considered. Self-referencing pairs of the form $(0,0), (1,1) ... (n,n)$ as well as mirrored pairs ($(x,y)$ or $(y,x)$) can be discarded, bringing the required raw memory down to $\frac{1}{2}*n^2-n$ bits.
Built-in compression algorithms are unlikely to alleviate this problem, as their intended use is centered on compression of geometry, textures or z-Buffers \cite{GPU_Compression}.

Our implementation stores visibility information as an uncompressed, contiguous buffer of bits which we append to our existing texturegroup. We chose an upper limit of $2^{32}$ bytes, as common GPUs tend not to have more than 8GB of onboard memory, and Falcor disallows the allocation of larger buffers. Theoretically, a buffer of this size can provide full coverage of lightmaps with sizes up to $512\times512$.

Figuring out which memory address (index) each patch pair is assigned in the visibility buffer, requires a bijective mapping function between unique pairs and respective indeces.

\subsection{Cantor Pairing Function}

\begin{figure}[th]
\centering
\includegraphics[scale=0.5]{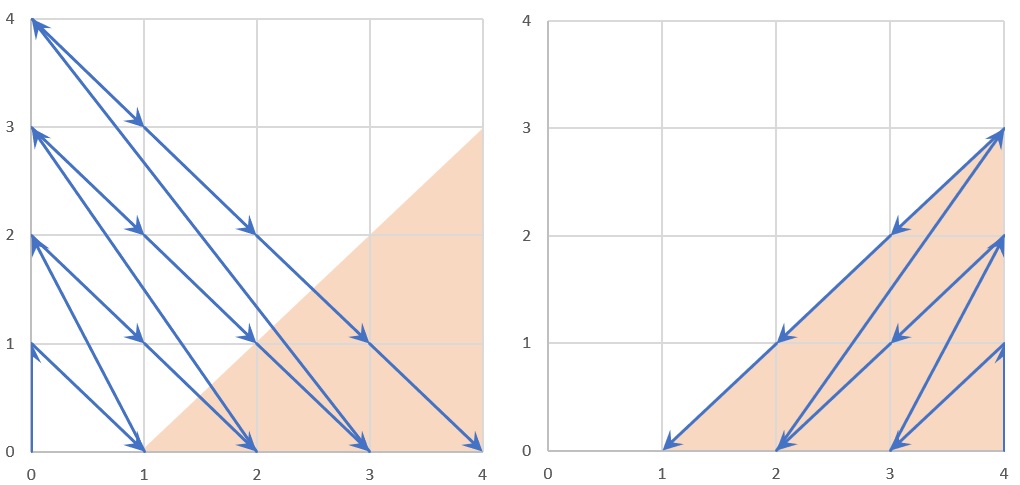}
\decoRule
\caption[asd]{Vanilla Cantor pairing function (left) and our custom pairing function (right) on a $5\times5$ grid. The area marked in orange contains all unique, non self-referencing pairs. Up to the index of 10, our pairing function excludes both mirrored and self-referencing pairs, thus encompassing exactly this area.}
\label{pairing_functions}
\end{figure}

\textit{Pairing functions} uniquely encode two natural numbers into a single one. There are a wide variety of them each with their own use cases and respective advantages \cite{elegant_pairing_function, Other_Pairing_Function}. For visibility caching we encode each unique pair of patches to an index in a cohesive memory sequence, which makes the \textit{Cantor pairing function} a good choice, as it traverses a 2D grid in a triangular shape which can be modified to exactly cover each unique pair and no more (see fig. \ref{pairing_functions}).

A vanilla cantor pairing function for non-negative integers follows this formula \cite{Other_Pairing_Function}:

\begin{equation}
    Cantor(x,y) = \frac{x^2 + 3x + 2xy + y + y^2}{2} = x + \frac{(x+y) (x+y+1)}{2}
\end{equation}

We ensure pair uniqueness by sorting $x$ and $y$ in ascending order. We also mirror the x-axis to ensure that at a given cutoff point (namely $\frac{1}{2}*n^2-n$) only unique pairs have been covered:

\begin{equation}
    address(x,y) = Cantor(n_x - min(x,y), max(x,y))
\end{equation}

where $n_x$ is the lightmap resolution along the $x$ axis.

In this memory-sequence function, the parameters (patches) are encoded as a single number, whilst on the lightmap they are given by two coordinates. Converting between each format is a trivial process:

\begin{equation}
    patch_{1D} = patch_{2D}.x + patch_{2D}.y * n_x
\end{equation}

\subsection{Visibility Buffer}

Visibility data is stored inside a standard DirectX \textit{RWBuffer} of unsigned integers with a maximum size of $2^{32}$ bytes. The modified cantor pairing function defined above ascribes each pair of patches a corresponding index inside the buffer. If the result exceeds the maximum value of $8*2^{32}$, no cached data is available and the visibility will have to be computed on-the-fly through raytracing.

Accessing the individual bits of a buffer directly is not possible in HLSL, so we employ equivalent bit-shifting functions that operate on 32-bit unsigned integers instead\footnote{Since 32bit integers are read and written to simultaneously, this leads to similar memory collision inaccuracies as in section \ref{memory_conflicts}. In our testing their effects were mostly negligible for the smaller lightmaps this algorithm functions on, though became noticeable on a $512\times512$ resolution.}.

In theory this approach may also be used to \textit{partially} store visibility data on larger lightmaps, though based on our measurements, this only lead to performance degradation. The visibility data required by a lightmap of $1024^2$ pixels would require upwards of 500 billion bits, of which the entirety of our 4GB buffer would only cover less than 7\%. The cost of having to calculate the cantor-index for \textit{all} patch pairs heavily outweighs the advantage of being able to skip the raytraces for just 7\% of them.

\section{Voxel Raymarching}\label{VoxelRaymarch}

The primary mechanism behind Nvidia's RTX technology is the complete parallelization of BVH traversal, triangle intersecting and shading. Given the immense number of rays required for a full radiosity iteration, we can safely assume that the limited amount of RT cores still pose a meaningful bottleneck within this system.

We devised a simple, low memory alternative for visibility calculations that can be executed on regular CUDA cores. The underlying idea consists of idle CUDA cores tapping into the unprocessed workload when all RT cores are otherwise busy.

The aforementioned visibility alternative has its roots in voxel cone tracing (\cite{Crassin, VXCT}).

\begin{figure}[h]
\centering
\includegraphics[scale = 0.3]{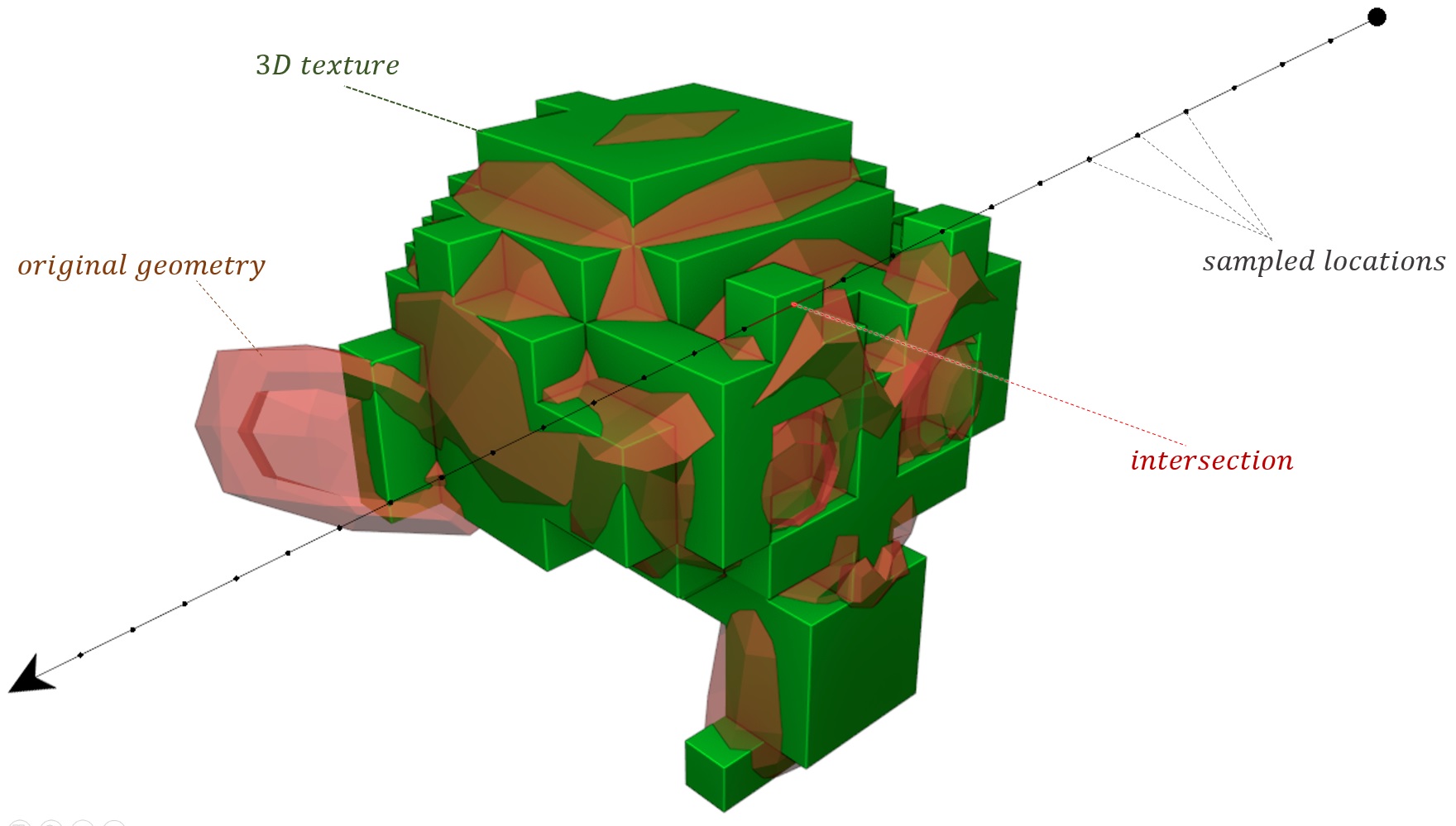}
\caption[]{Raymarching through a voxelized 3D scene. On each of the sampled locations the respective value in the 3D texture is read. In this example the ray will be marked as obstructed (i.e. not visible), because one of the sampled locations contains a value larger than zero.}
\label{voxel_raymarching}
\end{figure}

\subsection{Raymarching and 3D Textures}

Intuitively, \textit{3D textures} function just like regular textures only with an additional depth dimension \cite{VXCT}. The pixel equivalents are aptly named \textit{voxels}, resulting from a combination of the words volume and pixel.

\textit{Raymarching} is a technique usually employed when surface functions are not easily solvable. Unlike raytracing, this process "marches" forward along the ray direction in a series of steps, sampling each point along the way \cite{VXCT}. Raymarching has a wide range of variations such as sphere tracing or SDF ray marching, which require signed distance functions \cite{SDF_raymarching}.

Storing a voxelized representation of a scene into a 3D texture and sampling that texture whilst marching along a ray serves as a crude approximation to a regular ray-trace (see fig. \ref{voxel_raymarching}).

This approach is subject to a number of limitations mostly related to its geometric inaccuracies, yet the performance remains independent of a scene's triangle count and is instead tied to the overall size of the employed 3D texture \cite{VXCT}.

\subsection{Scene Voxelization}\label{Voxelization}

To voxelize a scene into a 3D texture we employ an algorithm virtually identical to the one described by Crassin et al. \cite{Crassin, crassin_voxelization2, VXCT}, which allows an entire scene to be voxelized in a single, lightweight rasterization pass.

We run this pass as an extension to the CITPass, the \textit{CVMPass} (\textit{create-voxel-map-pass}), which consists of a unique vertex, geometry and pixel shader:

\subsubsection{Vertex Shader}

To ensure all surfaces are voxelized, the first step scales the scene to fit entirely into the rendered clipspace.
The formula applied to each vertex simply follows from:

\begin{equation}
vert(v) = 2 \frac{v - P_{min}}{P_{max} - P_{min} - (1, 1, 1)}
\end{equation}

where $P_{min}$ and $P_{max}$ are the minimum and maximum world positions the scene encompasses. In essence, all scene vertices are linearly interpolated from a $[P_{min}, P_{max}]$ interval into the clipspace interval $[-1, 1]$.

The viewport resolution is set to be equal to the width and height of the voxelmap which, when rendered, corresponds to each primitive being projected along a texture axis \cite{VXCT}. The process of choosing the ideal axis of projection is performed by the geometry shader.

\subsubsection{Geometry Shader}

\begin{figure}[th]
\centering
\includegraphics[scale=0.5]{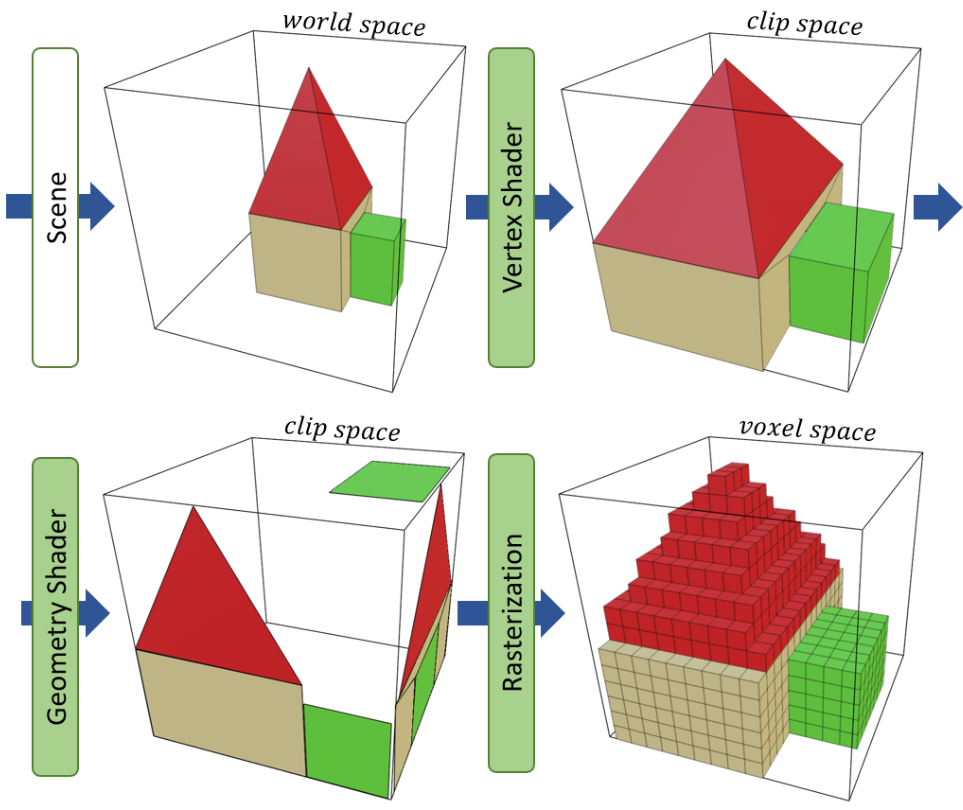}
\decoRule
\caption[]{The individual steps of the GPU voxelization algorithm proposed by Crassin et al. Note that the geometry shader technically does not \textit{project} the geometry, but merely rotates it so that the dominant axis is aligned with the axis of projection.}\label{DominantAxisSelection}
\end{figure}

Rasterizing all surfaces along a single axis may leave gaps in the resulting projection if a surface's normal vector subtends a steep angle with the axis of projection \cite{crassin_voxelization1, crassin_voxelization2, VXCT}. For instance, if the green cuboid in fig. \ref{DominantAxisSelection} were to be rasterized along a single axis, only one of the three surfaces would contain pixels.

In order to ensure a voxelization that covers all surfaces fully, one can either repeat the process for each axis individually or, preferably, rotate each triangle so that the dominant component of its normal vector is aligned with the axis of projection \cite{Crassin, VXCT, takeshige}.

This process is aptly named \textit{dominant axis selection}, and is easily performed in a geometry shader. The dominant axis of a triangle corresponds to the axis that its normal vector shares its largest component with. Commonly, rasterization applications project triangles along the z-axis, so a respective geometry shader would function like so:

\begin{algorithm}[H]
    \caption{Dominant Axis Selection (Adapted from VXCT \cite{VXCT})}
    \begin{algorithmic}[1]
    \Procedure{ProcessTriangle}{$v_1, v_2, v_3$}
        \State $\harpoon n \gets \left| (v_2 - v_1) \times (v_3 - v_1) \right|$ \Comment{Calculate normal vector}
        \For{$v \in \{ v_1, v_2, v_3 \}$}\Comment{Rotate each vertex to maximize z}
            \If{$n.x = \max(n.x, n.y, n.z)$}
                \State $v.xyz \gets v.zyx$
            \ElsIf{$n.y = \max(n.x, n.y, n.z)$}
                \State $v.xyz \gets v.xzy$
            \Else{ $n.z = \max(n.x, n.y, n.z)$}
                \State $v.xyz \gets v.xyz$
            \EndIf
        \EndFor
    \EndProcedure
    \end{algorithmic}
\end{algorithm}

\subsubsection{Pixel Shader}

\begin{figure}[th]
\centering
\includegraphics[scale=0.3]{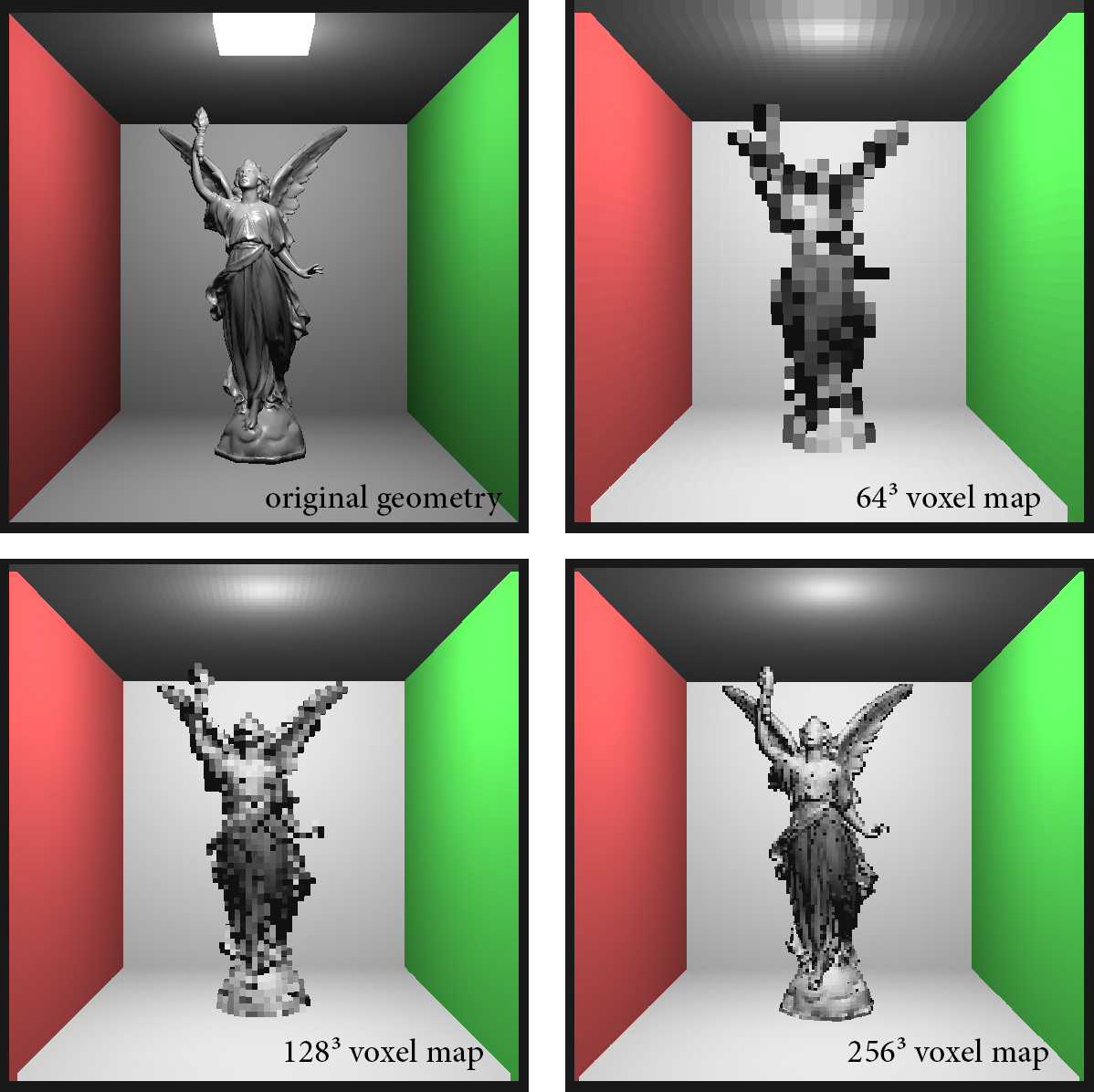}
\decoRule
\caption[]{Phong-model direct light values stored in voxelmaps of different resolutions, as depicted by Benjamin Kahl \cite{VXCT}.}
\label{voxelmaps}
\end{figure}

The rasterization pipeline delivers the world position that each fragment had before the geometry shader's rotations took place. For each rendered pixel that position now corresponds to a coordinate within the 3D texture that is set to be a "solid" voxel (see fig. \ref{voxelmaps}).

In voxel cone tracing, each voxel stores information on direct lighting calculated by the phong model \cite{VXCT}. For our purposes it is sufficient to simply store a 0 for empty space and 1 for geometry.

We store the voxelmap alongside our other textures in the texturegroup. Marching through it along a ray until a voxel is solid becomes a trivial procedure that can be directly called instead of the RTX \verb|TraceRay| function.

In RTRad, we allow the user to set a custom ratio of RTX ray-traces that are replaced by voxel-raymarches. We found that, for small voxelmaps, this could indeed lead to an improvement in pass-time but would significantly deprecate lighting quality. Nevertheless, raymarching does serve as a simple and lightweight fallback for graphics cards that do not support RTX. A more detailed account on our findings is listed in section \ref{VoxelRaymarch_Analysis}.

\section{Directional Sampling}

Some radiosity implementations rely on the hemicube approximation for visibility and do not calculate view factors explicitly \cite{gpu_gems_2005}. Instead, gathered intensity is estimated by generating samples on a hemicube and determining which patch a ray incoming from that direction would have originated on. The total gathered intensity can thus be estimated as the average for each of these directions. This process corresponds, in essence, to the same distribution that a classical raytracing program would sample for diffuse reflections. We broadly categorized this approach as "directional sampling" in section \ref{instant_radiosity_section}.

To analyze the viability and performance of RTX in these types of implementations, we included a modified version of our algorithm that relies on direction-based sampling.

For each patch, a number of rays are shot through a surrounding hemisphere. Whichever surfaces these encounter are sampled for their color and added into the final lighting sum. Instead of weighing lighting contribution by patch surface area, we compute a rudimentary average over the number of samples $|\Omega|$.

Mathematically, this approach is described by the equation for raytracing, as given in (\ref{RenderingEquation_Raytracing}):

\begin{equation}
\begin{aligned}
L_o(x, \harpoon\omega)
= L_e(x, \harpoon\omega) + \frac{1}{|\Omega|}\sum_{\harpoon\omega_i \in \Omega} f_r(\harpoon\omega_i, \harpoon\omega, x)  L_o(I(x, \harpoon \omega_i), -\harpoon \omega) \frac{\harpoon \omega_i \cdot \harpoon {n_x}}{\lVert x - I(x, \harpoon \omega_i) \rVert^2}
\end{aligned}
\end{equation}

Since radiosity is only concerned with diffuse reflections, the BRDF $f_r$ collapses into a generic diffuse BRDF of $\frac{\rho}{\pi}$:

\begin{equation}
\begin{aligned}
L_o(x)
= L_e(x) + \frac{\rho}{\pi|\Omega|}\sum_{\harpoon\omega_i \in \Omega} L_o(I(x, \harpoon \omega_i)) \frac{\harpoon \omega_i \cdot \harpoon {n_x}}{\lVert x - I(x, \harpoon \omega_i) \rVert^2}
\end{aligned}
\end{equation}

The equation above can be adapted fairly easily into our existing architecture by using DXR's \textit{closest-hit} shader which, in essence, corresponds to our intersection function $I(x, \harpoon\omega)$.

The closest-hit shader allows one to retrieve information on the closest intersection, including UV coordinates which lets us access all required data through our existing texturegroup.
The lighting value sampled by each ray can be blurred by sampling a higher mipmap level in order to avoid sporadic illumination effects due to small texture details.

Algorithm \ref{RTLPass_psuedocode_hemispheric} outlines the respectively modified RTLPass for this approach. 
For the sake of brevity, we kept the exact operations (such as mipmap-sampling, accounting for patch-size, tangent-space transformation etc.) obfuscated, but the entire closest-hit shader can be viewed on the RTRad open source repository (see \cite{RTRad_Repository}).

\begin{algorithm}[H]
    \caption{RTLPass}\label{RTLPass_psuedocode_hemispheric}
    \begin{algorithmic}[1]
      \For{$i \in [0, n]$}\Comment{For each patch (executed in parallel)}
        \State $L_{out}(i) \gets L_e(i)$\Comment{Set initial lighting value}
        \For{$\harpoon \omega \in \Omega$}
            \State $\harpoon \omega_t \gets$ $\harpoon \omega$ in tangent-space of patch $i$
            \State Shoot a ray from $pos(i)$ in direction $\harpoon \omega_t$
            \If{the ray hits another patch $j$}\Comment{Closest-hit shader}
                \State $g \gets \frac{\harpoon \omega_i \cdot nrm(i)}{\lVert pos(i) - pos(j) \rVert^2}$\Comment{Geometric factor}
                \State $L_{out}(i) \gets L_{out}(i) + g * mat(i) *  \frac{\rho}{\pi|\Omega|} * L_o(j)$\Comment{Add contribution}
            \EndIf
        \EndFor
    \EndFor
    \end{algorithmic}
\end{algorithm}

\subsection{Hemispheric Direction Generation}

The algorithm above assumes that a set of directions $\Omega$ is given.
To sample directions uniformly across a hemisphere, we utilize the same method that Laine et al. use in \textit{incremental instant radiosity} \cite{instant_radiosity_aalto} to distribute additional VPLs from a light-source.

Laine et al. represent samples as 2D points inside a unit circle (rather than points on a hemisphere) \cite{instant_radiosity_aalto, instant_radiosity_2}, which can subsequently be projected onto a hemisphere using the following operation:

\begin{equation}
\begin{pmatrix}x\\y\end{pmatrix} \longrightarrow \begin{pmatrix}x\\y\\\sqrt{1 - (x^2 + y^2)}\end{pmatrix}
\end{equation}

\begin{figure}[h]
\centering
\includegraphics[scale=0.6]{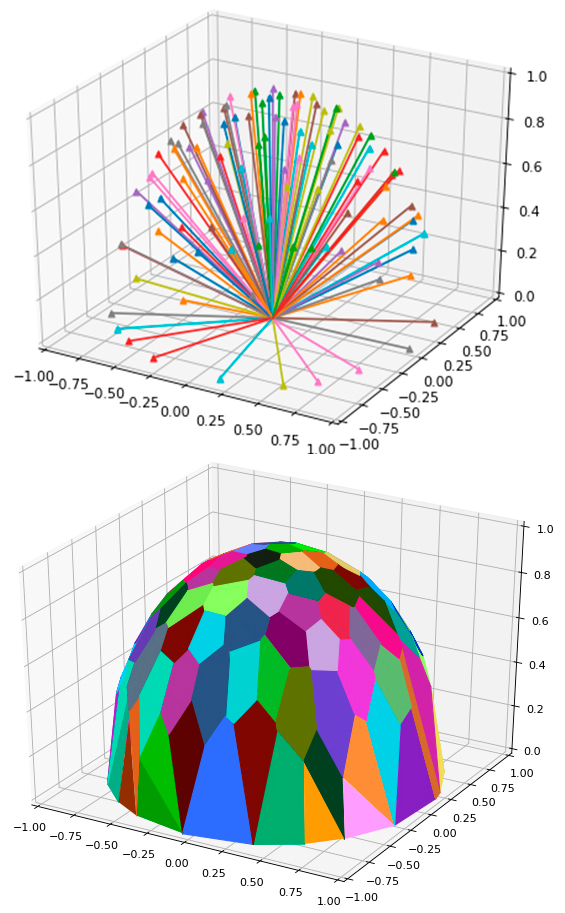}
\decoRule
\caption[]{One hundred generated directions (top) and their corresponding Voronoi diagram projected onto a hemisphere (bottom).}
\label{vpl_adding2}
\end{figure}

As can be observed in fig. \ref{vpl_adding2}, evenly scattered points on a circle result in a lower density of vectors that strongly deviate from the surface's normal vector. This, in essence, induces the geometric cosine-term applied in sampling \footnote{Our implementation follows the pseudo-code listed in algorithm \ref{RTLPass_psuedocode_hemispheric}, meaning we apply the geometric cosine-term (dot product), \textit{despite} the projection onto a hemisphere already inducing the same effect. Leaving this factor out may produce different, potentially better, results. }.

In incremental instant radiosity, the direction a new VPL is shot towards gets determined by finding the largest empty circle within the unit circle and placing a new sample at its center \cite{instant_radiosity_aalto}. This helps maintain an even distribution of VPLs, without prior knowledge on how many samples will be generated.

Geometrically, the largest empty circle inside a sampled area must be centered at either

\begin{itemize}
    \item a vertex in the Voronoi diagram that touches connects three Voronoi regions, 
    \item the intersection between an infinite Voronoi endge and the bounding polygon or
    \item a vertex of the bounding polygon \cite{instant_radiosity_aalto}.
\end{itemize}

With last option being irrelevant when the boundary is a perfect unit circle \cite{instant_radiosity_aalto}.

Our algorithm, which pre-computes a set of evenly distributed samples $\Omega$, is largely based on this principle by adhering to the steps listed in algorithm \ref{sample_generation_pseudocode}.

\begin{algorithm}[H]
    \caption{Directional Sample Generation}
    \begin{algorithmic}[1]
        \State $\Omega \gets $ 4 random points in a unit circle
        \For{$n$ iterations}
            \State Calculate the Voronoi diagram of $\Omega$
            \State $V \gets $ Set of all Voronoi vertices inside the unit sphere as well as all intersections between Voronoi edges and the unit circle boundary.
            \State Find the point in $V$ that has the largest distance to its nearest neighbour and add it as a new sample to $\Omega$.
        \EndFor
    \end{algorithmic}
    \label{sample_generation_pseudocode}
\end{algorithm}

\begin{figure}[h]
\centering
\includegraphics[scale=0.5]{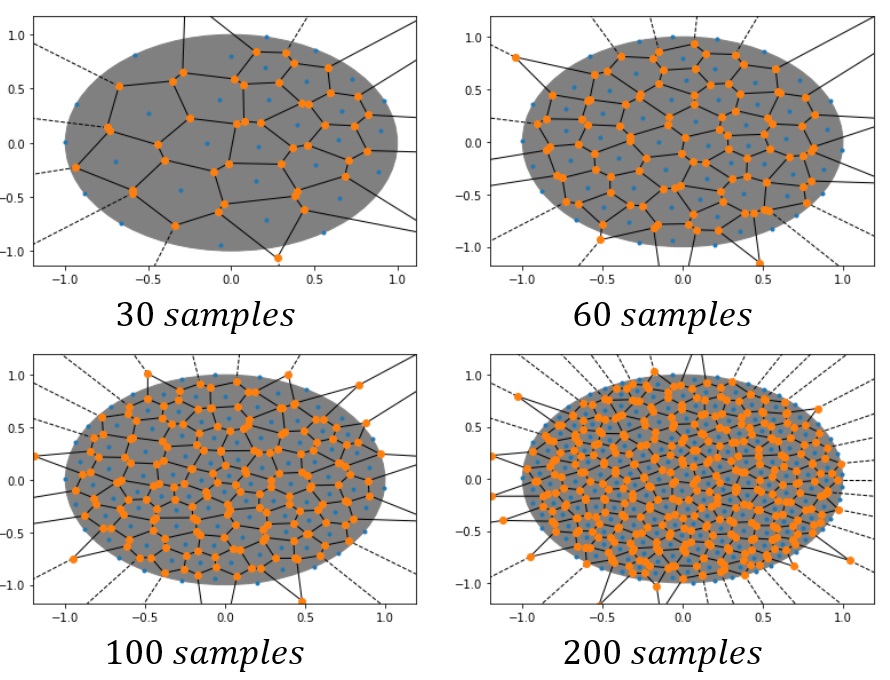}
\decoRule
\caption[]{Unit circles with various amount of samples (blue) alongside their respective Voronoi diagrams (black/orange).}
\label{vpl_adding}
\end{figure}

The results of this algorithm can be observed in fig. \ref{vpl_adding}, whilst fig. \ref{vpl_adding2} shows their respective 3D directions. Incremental instant radiosity is intended to maintain relatively even distributions under a growing sample count. We chose this method to ensure users can select the exact amount of samples to be used for radiosity, whilst not having to worry about those samples being unevenly distributed.

In order to cut down on unnecessary computations during runtime, we use a separate program written in \textit{Python} (included in the RTRad repository \cite{RTRad_Repository}) to pre-compute a list of directions, which are then simply read by the RTLPass shader in the exact same order they were added to the set.

Coding directions in this static manner allows us to omit the complexities of generating Voronoi diagrams during runtime, thus greatly increasing performance. The resulting downside is that the maximum amount of samples becomes restricted to the amount that was pre-generated, as well as the maximum DirectX array size. Our chosen upper limit was 1024, although much larger sets could theoretically be accommodated using GPU data-buffers.

\chapter{Evaluation} 

\label{Chapter6} 




The advent of the Nvidia RTX platform allows us to offload the costly visibility computations of radiosity onto RT cores. By combining RTX and progressive refinement radiosity we seek to provide a competitively fast algorithm that does not compromise on visual fidelity.

This chapter aims to determine if and to what degree RTX can accelerate existing GPU radiosity implementations, as well as which quirks and enhancements are most beneficial.
We will discuss our findings and assessments of the presented implementation and compare its performance with other industry-standard lightmap generators.

\section{RTRad Overview}

In chapter \ref{Chapter4} we presented our implementation of an RTX-based progressive radiosity lightmap generator, which was subsequently expanded upon in chapter \ref{Chapter5}.

\begin{figure}[H]
\centering
\includegraphics[scale=0.38]{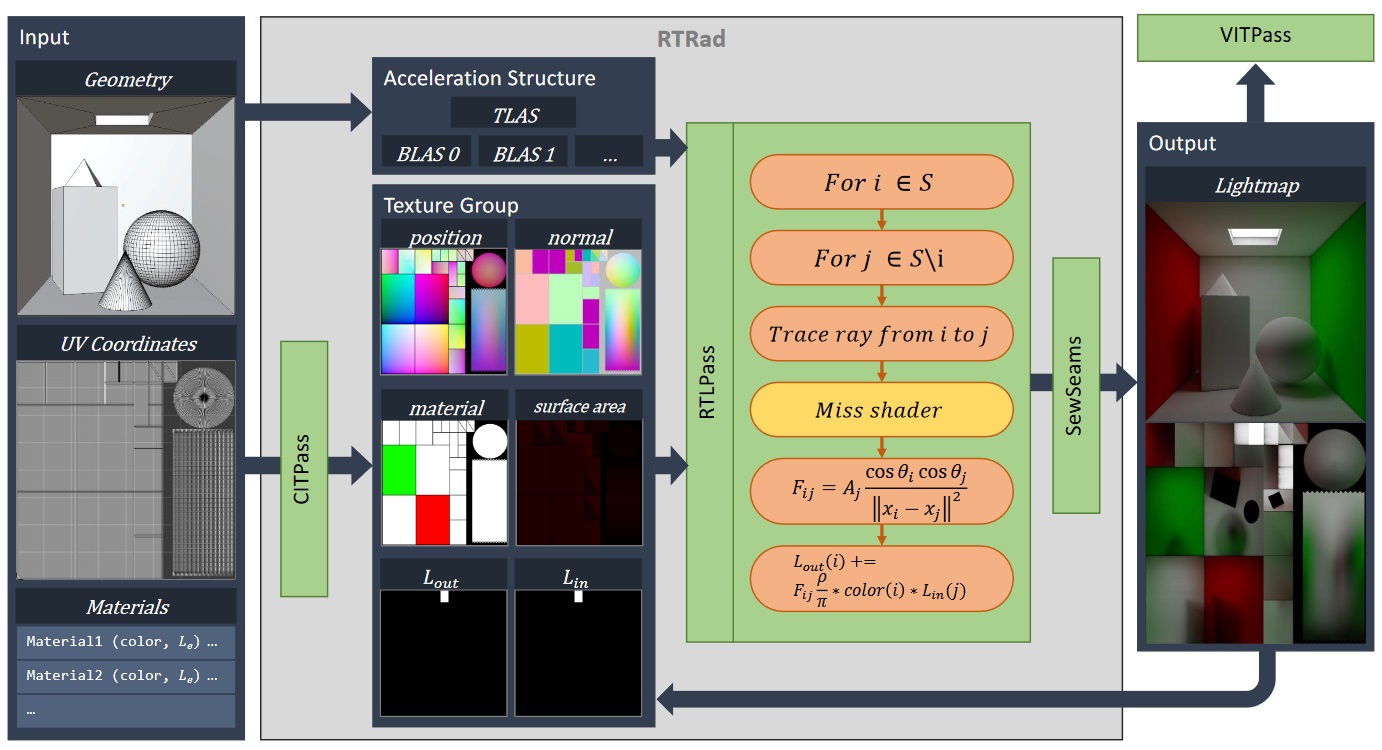}
\decoRule
\caption[]{Simplified overview of the RTRad rendering passes. For a more detailed overview refer to fig. \ref{RTRad_flow}.}
\label{RTLPass_flow_simple}
\end{figure}

The proposed algorithm operates entirely on textures with a series of swift pre- and post-processing rasterization passes that translate the scene into a usable format, generate meta-information on refinement quad-trees and ameliorate leaks on UV seams, as shown in fig. \ref{RTLPass_flow_simple}.

The most important component of our pipeline is the \textit{RTLPass}, which is executed for each patch and uses the \verb|TraceRay| function to test for visibility of other patches. A single-sample Monte-Carlo approximation of the view factors provides the lighting contribution for any patch that passes the test. Once completed, the input and output lighting textures are swapped for the next pass.

\begin{figure}[H]
\centering
\includegraphics[scale=0.5]{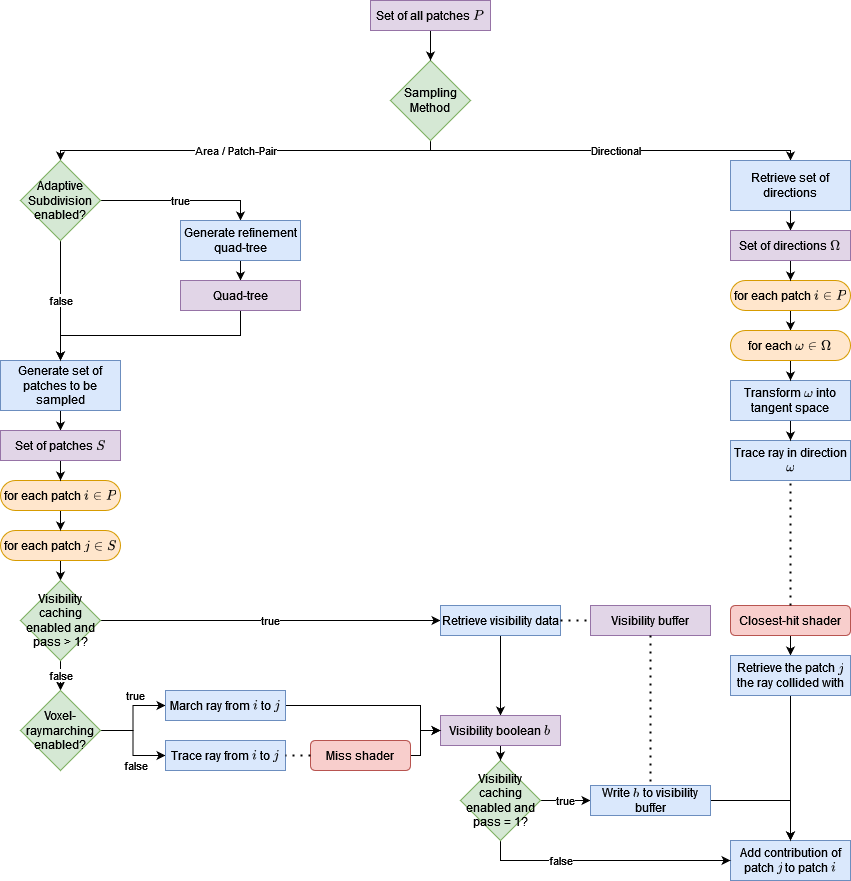}
\decoRule
\caption[]{Flow-diagram of the RTLPass. Most condition branches are implemented using preprocessor directives and do not cost computation time.}
\label{RTLPass_overview}
\end{figure}

The RTLPass can be configured in manifold ways, including visibility caching, directional sampling and various methods for reducing sample counts. Fig. \ref{RTLPass_overview} shows an overview of these various configurations. 

\section{Evaluation Method}

The measurements we exhibit in this chapter were exclusively performed with the built-in Falcor profiler, which uses the most accurate available CPU and GPU timers to measure given functions as an event hierarchy \cite{falcor}.

Our benchmarking system consists of an AMD Ryzen 3900X CPU and an RTX 2070 Super graphics card\footnote{The 2070S contains a total of 40 RT Cores. Higher end RTX GPUs contain up to 82 and would perform respectively faster. We expect that lower end cards, which start at 30 RT cores, would still provide comparable results.} running on Windows 10.0.19044 with Nvidias GeForce Driver version 516.40.

\subsection{Pass-Time}

Our primary reference for performance is the \textit{pass-time}, which constitutes the total of GPU time utilized by the RTLPass for all batches of a pass. 

Our application automatically extracts this information from the profiler and outputs it at the end of a pass. We generally deemed pass-times of under a second to be suitable for real-time applications.

\subsection{DFPR}

To assess general lighting quality, this chapter provides images at the primary points of contention. In addition, we employ a self-conceived unit for \textit{deviation from pure radiosity}, henceforth shortened as \textit{DFPR}. This value measures how close the results of an optimized radiosity variant are to its un-optimized counterpart:

\begin{equation}
    dfpr(L) = \frac{1}{n} \sum_{i=0}^{n} \lVert L(i) - P(i) \rVert
\end{equation}

where $L$ is the lightmap in question and $P$ is a lightmap of the same size, generated with pure, progressive radiosity for the same scene.

The DFPR of an image is computed as its euclidean distance from a correspondingly large lightmap computed with pure radiosity, averaged across all pixels.
A "perfect" DFPR of zero would equate to the image being an exact copy, whilst a theoretical worst DFPR would be $\sqrt{3}$ (the average RGB distance between a completely white and a completely black image).

We found that a DFPR value of 0.025 served as a very conservative threshold at which differences became noticeable to a human observer.

\begin{figure}[H]
\centering
\includegraphics[scale=0.75]{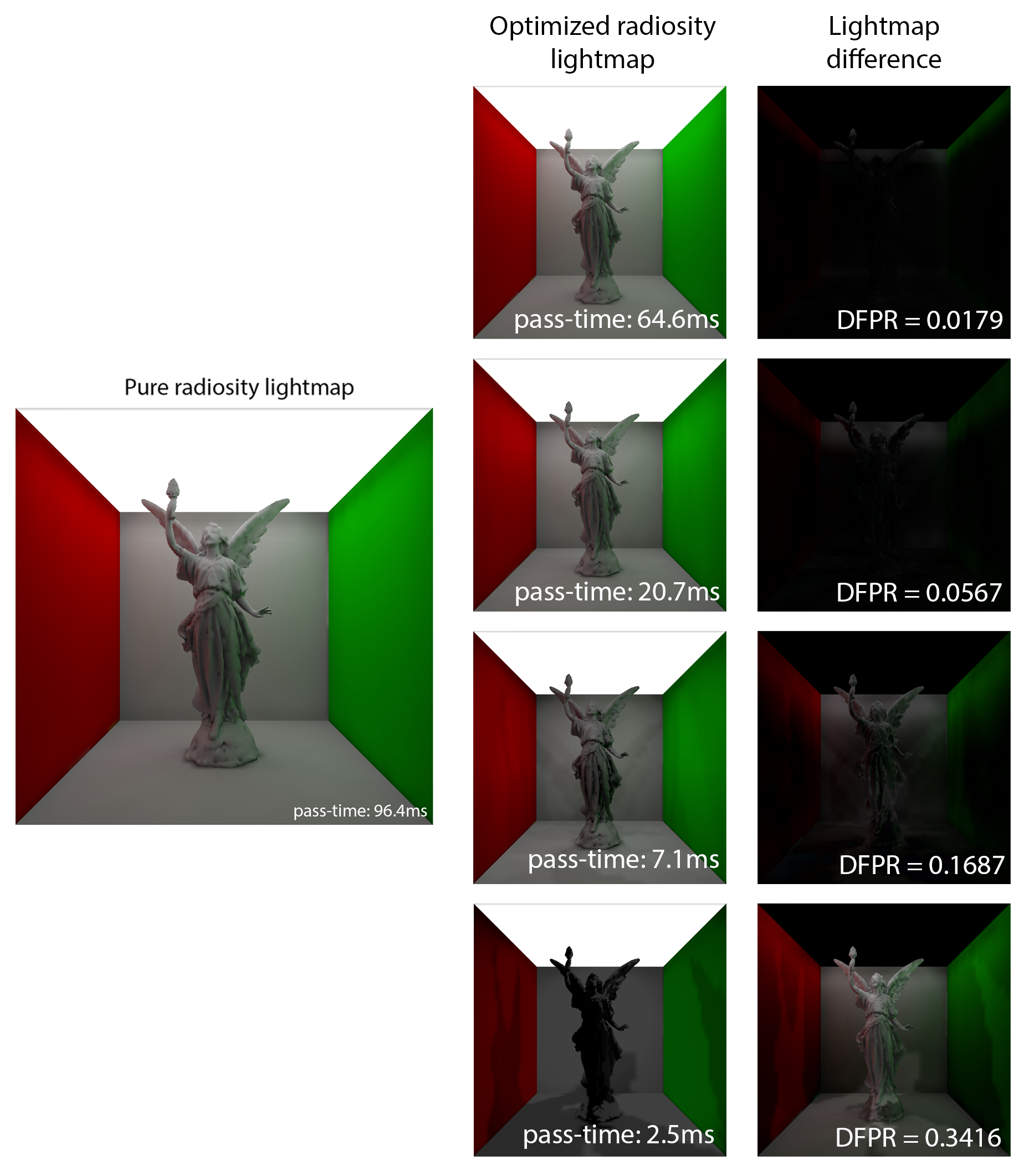}
\decoRule
\caption[]{Examples of how the DFPR is calculated on lightmaps of varying quality.\protect\footnotemark The DFPR of a lightmap results from the average magnitude of all pixels in the subtraction of the lightmap itself and a corresponding "pure" lightmap.} 
\label{DFPR_concept}
\end{figure}

Fig. \ref{DFPR_concept} illustrates this concept with a number of examples: Optimized radiosity algorithms run significantly faster, but also deviate from the results of pure radiosity. This deviation is given by the DFPR and serves as a measurement for visual fidelity under a given lightmap resolution.

Theoretically, dividing the DFPR by the corresponding pass-time could serve as a general measure for quality per cost, but we found that this did not accurately reflect empirical results.

\footnotetext{The lightmaps in this image are all of the size $128\times128$. The optimized variants are generated using Monte-Carlo undersampling with sampling windows of $2\times2$, $4\times4$, $8\times8$ and $16\times16$ respectively.}

\subsection{Scenes}

\begin{figure}[H]
\centering
\includegraphics[scale=0.55]{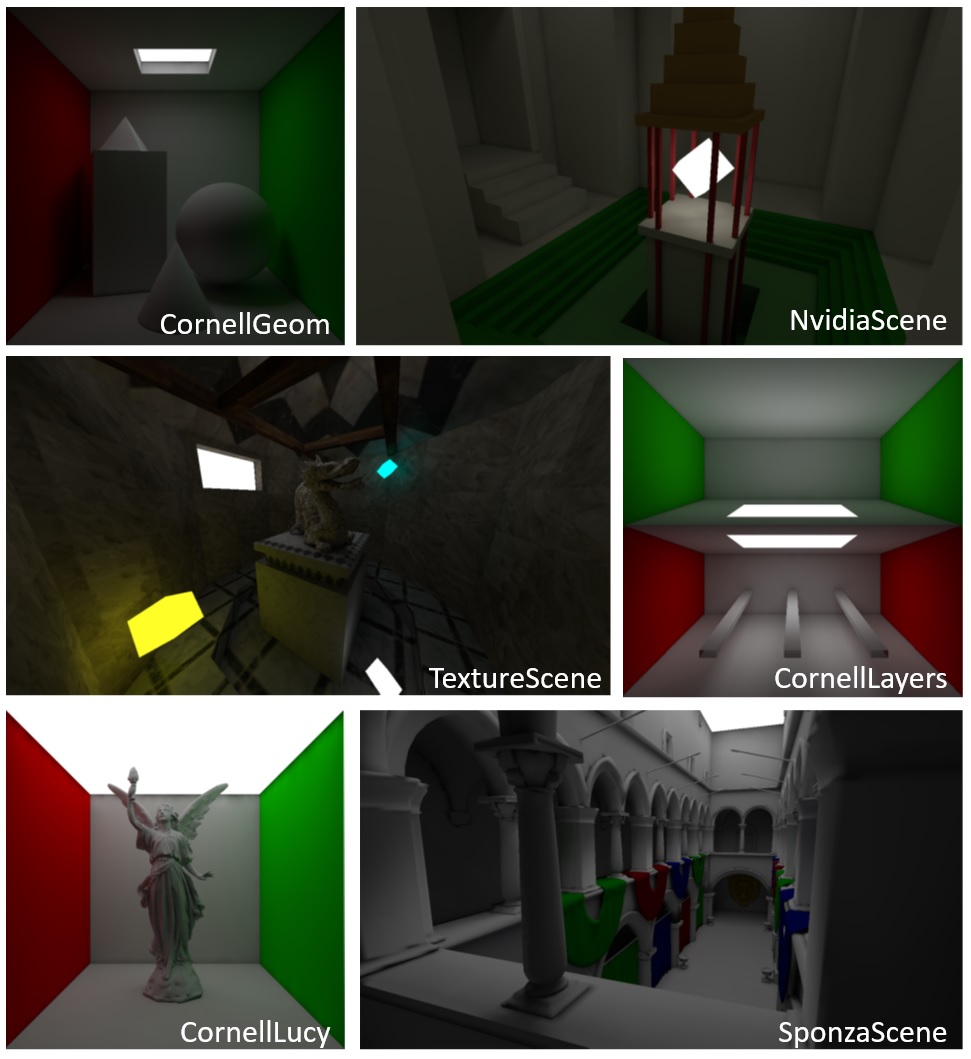}
\decoRule
\caption[]{Our employed testing scenes rendered through RTRad with a lightmap of $1024^2$ pixels.}
\label{Testing_scenes_fig}
\end{figure}

We employed a total of six different scenes for testing our algorithm and analyzing its performance, all of which are rendered through RTRad in fig. \ref{Testing_scenes_fig}.
Three of these are based on the \textit{Cornell box} and are meant to directly demonstrate the system's capabilities in realistic global illumination. A further three scenes are significantly more complex and are meant to simulate a realistic use-case workload:

\begin{itemize}
    \item \textbf{CornellLucy}: Consists of a simplified version\footnote{The same version that was used by Benjamin Kahl in VXCT \cite{VXCT}.} of the "Lucy" statue from the \textit{Stanford 3D scanning repository} \cite{stanford} inside a Cornell box with a large light-source above. Used to test indirect light on a high-detail 3D model with abundant light.
\end{itemize}
    
\begin{itemize}
    \item \textbf{CornellGeom}: Consists of a series of simple geometric shapes inside a Cornell box with a narrow light-source. Used to test inter-reflection in low-light environments and hard shadows.
    \item \textbf{CornellLayers}: A modified Cornell box that is split down the middle with a wedge. Used to ensure no color leaks from one section to the other as well as to test soft shadows.
    \item \textbf{SponzaScene}: A commonly employed graphics benchmarking scene created by \textit{CryTek} \cite{SponzaScene}. To ensure compatibility with lower resolution lightmaps our version is somewhat simplified and texture-less, but still consists of over 150 thousand triangles.
    \item \textbf{NvidiaScene}: A reconstruction of the scene employed by Pharr et al. in the 2005 \textit{GPU Gems 2} book \cite{gpu_gems_2005}. It works well for testing small-scale light-sources as well as a general point of comparison to the GPU-based progressive refinement solution presented by Nvidia in the aforementioned book.
    \item \textbf{TextureScene}: A custom-built scene that consists of several colored light-sources inside an octagonal room with a (simplified) dragon statue from the Stanford repository \cite{stanford}. This scene is, uniquely, covered with detailed color-textures.
\end{itemize}

The overall triangle counts of these scenes range from from 60 (CornellLayers) to over 150 thousand (Sponza). Further scenes (with an adequate UV map) can be brought into RTRad quite easily in the form of FBX-formatted 3D models.

Based on Falcor's scene information, all acceleration structures consisted of a single TLAS, but were subdivided into multiple opaque BLAS geometries.

\section{Results}

\subsection{Pure Progressive Radiosity}

The core algorithm provided in chapter \ref{Chapter4} constitutes an instance of progressive radiosity in of itself. This is a very crude, brute-force approach that does not compromise on realism but comes at a respectively high computational cost. A humbly sized lightmap of $256\times256$ pixels, for instance, would involve up to $256^4$ visibility tests, resulting in well over 4 billion rays being traced. 

\begin{figure}[th]
\centering
\includegraphics[scale=0.63]{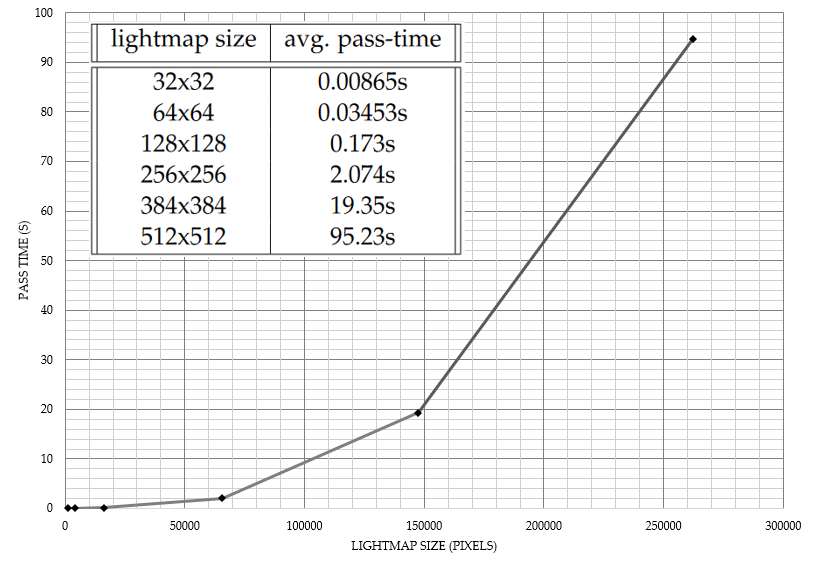}
\decoRule
\caption[]{Average pass-time across three measurements of the same progressive radiosity workload for different lightmap sizes. The underlying scene (CornellLucy) has a UV mapping with a coverage of approx. 86\%. The chosen batchsize is $64\times64$.} 
\label{data1}
\end{figure}

This quadratic growth is clearly reflected in our measurements depicted in fig. \ref{data1}. The approximate 4 billion rays are processed in approx. 2 seconds.

The pass-times required for low resolution lightmaps lie well within the leniencies of real-time applications, despite the costly nature of pure progressive radiosity.

Inevitably, the exponential cost overwhelms the leeway enabled by parallelization as resolutions reach the $512\times512$ mark\footnote{The largest measurement we took for pure progressive radiosity was a lightmap of $768\times768$ pixels, which took 867 seconds to complete.}, which is equivalent to an $n$ of 260.000 (70+ billion visibility tests).

\subsection{Undersampling}\label{Undersampling_Analysis}

Although the scalability of quadradically complex algorithms can never be fully nullified, the cost of higher resolution lightmaps can be ameliorated by lowering the samples performed for each patch.

\begin{figure}[H]
\centering
\includegraphics[scale=0.7]{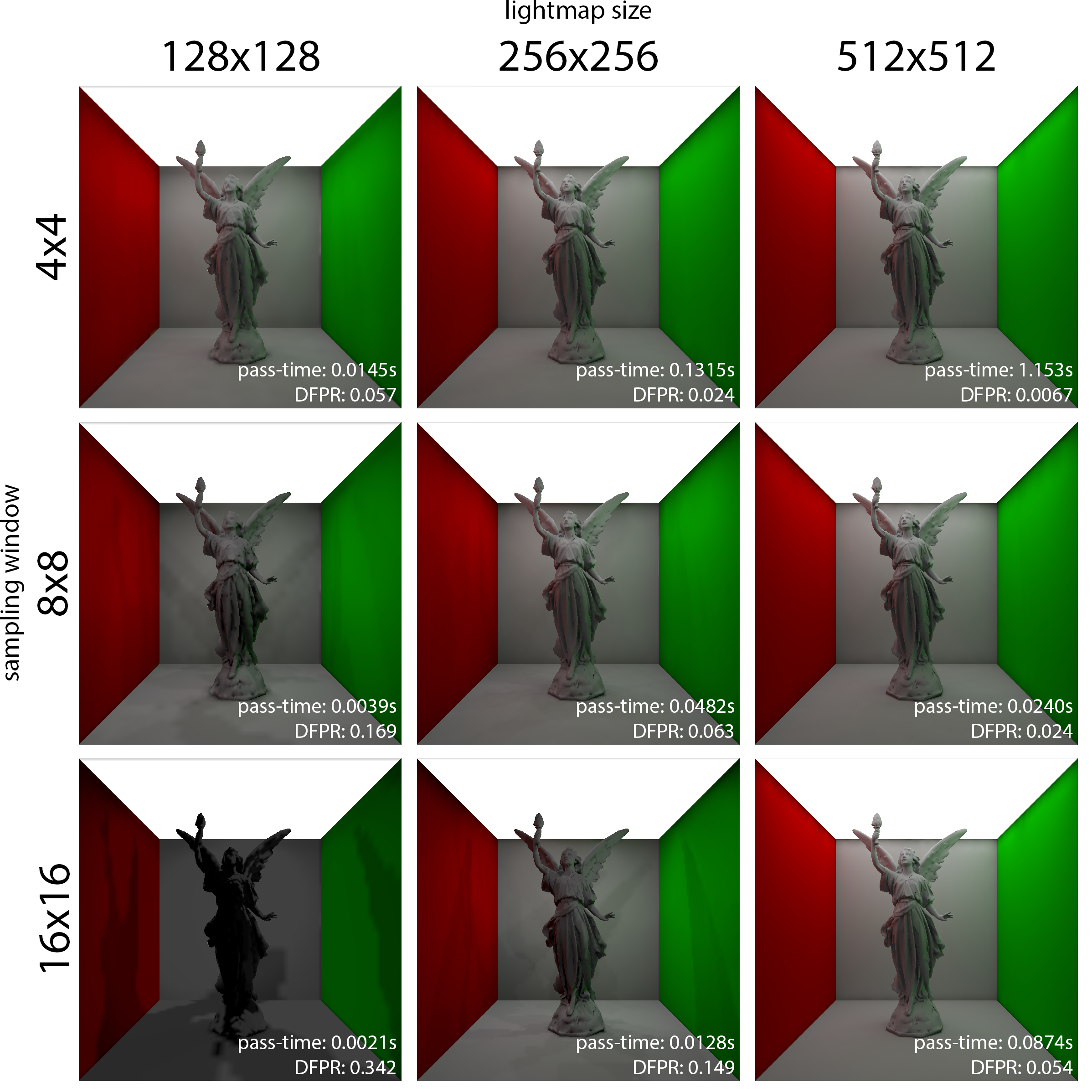}
\decoRule
\caption[]{Results of Monte-Carlo undersampling after two passes for different combinations of texture resolution and sampling window. Anomalies only become noticeable when the sampling window is very large in relation to the lightmap size (towards the lower left corner).} 
\label{undersampling_results}
\end{figure}

Undersampling, as layed out in chapter \ref{Chapter5}, has an enormous positive impact on pass-times.
Under the correct parameters, even larger lightmaps, such as $1024\times1024$, can be fully computed in under five seconds without any noticeable differences to their pure radiosity counterpart (DFPR < 0.025).

All three undersampling methods have near identical pass-times, except for Monte-Carlo undersampling, which requires 5 to 10 percent longer, presumably due to the large amount of pseudo-random number generation.

\begin{figure}[th]
\centering
\includegraphics[scale=0.6]{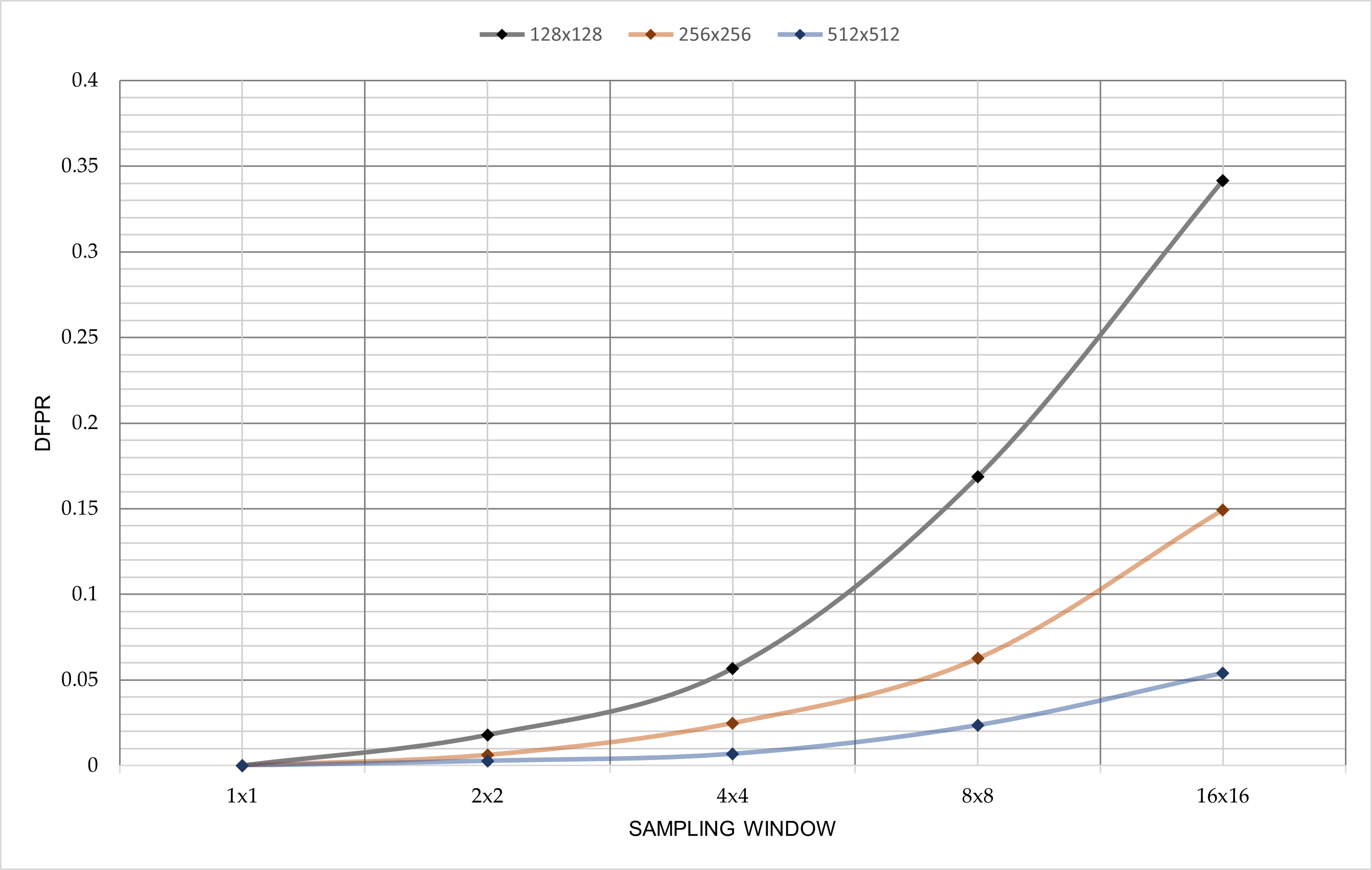}
\decoRule
\caption[]{DFPR of Monte-Carlo undersampling for different resolutions and sampling windows (lower is better). Differences become noticeable at a threshold between 0.025 and 0.1, depending on the scene.}
\label{DFPR_Undersampling}
\end{figure}

Differences in visual fidelity only become noticeable under extreme parameters where the sampling window is large in relation to the lightmap resolution (see fig. \ref{undersampling_results} and fig. \ref{DFPR_Undersampling}). For smaller sampling windows the pass-time benefits massively whilst only producing minimal, indistinguishable differences in the final yield.

We found particular success with a ratio of $\frac{1}{128}$ between lightmap texture and sampling window, which produces adequate pass-times and DFPRs for most lightmap resolutions.
In table \ref{recommended_stettings} we list our measurements under these recommended settings.

\begin{table}[H]
    \centering
    \centerline{
    \begin{tcolorbox}[tablestyle_main,tabularx={X||Y|Y|Y}]
         \hline
         \rowcolor{gray!50!black}
         \textcolor{white}{Lightmap Size} & \textcolor{white}{Sampling Window} & \textcolor{white}{DFPR} & \textcolor{white}{Pass-Time} \\ [0.5ex] 
         \hline\hline
         $128\times128$ & $1\times1$ & 0.0000 & 0.151s \\ 
         \hline
         $256\times256$ & $2\times2$ & 0.0063 & 0.163s \\
         \hline
         $512\times512$ & $4\times4$ & 0.0069 & 0.977s \\
         \hline
         $1024\times1024$ & $8\times8$ & <0.0196\footnoteref{note1} & 4.267s \\
         \hline
         $2048\times2048$ & $16\times16$ & <0.0345\footnotemark & 16.480s \\ [1ex] 
         \hline
    \end{tcolorbox}
    }
    \caption{Pass-time and DFPR with Monte-Carlo undersampling using our recommended ratio of $\frac{1}{128}$.}
    \label{recommended_stettings}
\end{table}

\footnotetext{We used a $512\times512$ lightmap as the pure reference to calculate this DFPR, which \textit{overestimates} the value.\label{note1}}

\subsubsection{Undersampling Method Differences}

Despite being somewhat more costly, Monte-Carlo produces the highest quality lighting of the three undersampling methods. Whilst the differences are not substantial in most scenes, lightsources with small UV geometry can produce uneven shadows when lit with one of the alternatives.

\begin{figure}[th]
\centering
\includegraphics[scale=0.9]{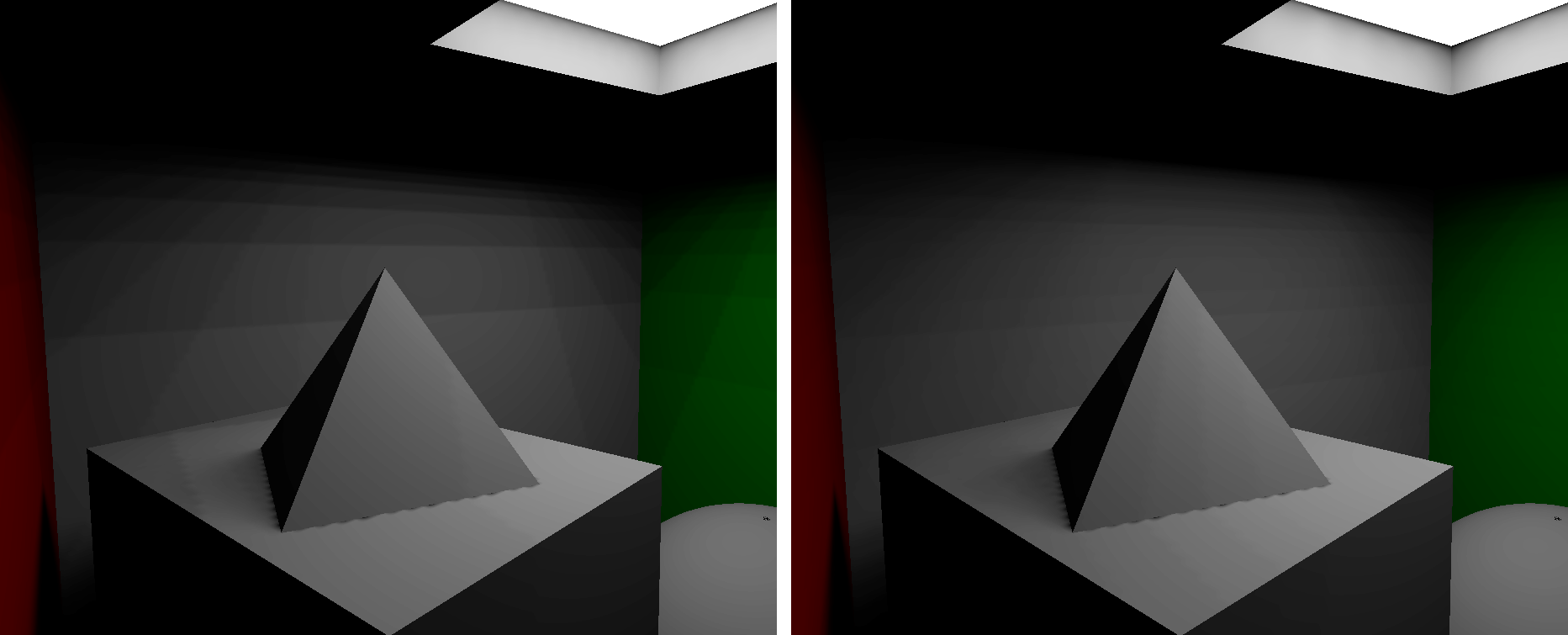}
\decoRule
\caption[]{Uneven shadows produced by mipmapped or static-stride undersampling (left) as opposed to Monte-Carlo undersampling (right) in low-light environments.} 
\label{cascadingshadows}
\end{figure}

\begin{figure}[th]
\centering
\includegraphics[scale=0.43]{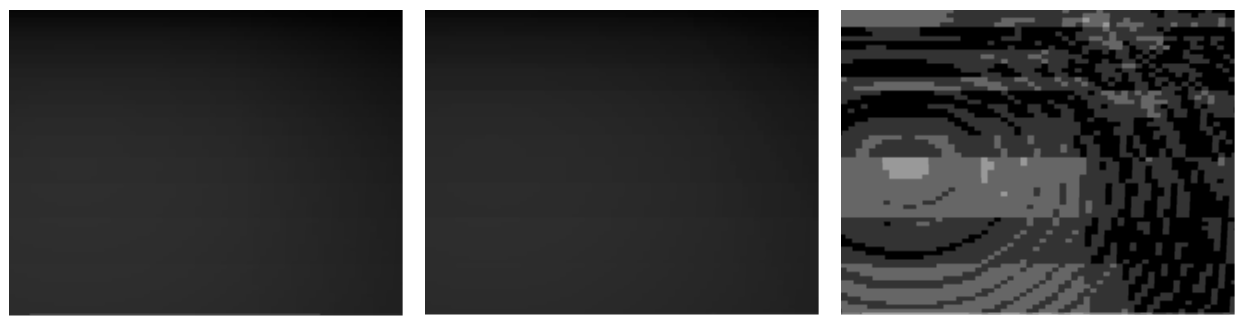}
\decoRule
\caption[]{A single wall in CornellGeom lit with pure radiosity (left), Monte-Carlo undersampling (middle). To the right is the difference between the two textures, with its brightness multiplied by 50 to make the noise visible.} 
\label{Monte-Carlo-U-Noise}
\end{figure}

The uneven shadows seen in fig. \ref{cascadingshadows} are spread out by the randomization of Monte-Carlo undersampling which instead produces a minimal amount of noise (see fig. \ref{Monte-Carlo-U-Noise}), unnoticeable under most circumstances.

We found that most other types of artifacts (such as unnatural highlights along corners) prominent in the latter two can almost entirely be eliminated by clamping the lighting contribution of any one patch to a maximum magnitude of 0.05.

\newpage

\subsection{Adaptive Subdivision}

Unlike undersampling, adaptive subdivision does not produce noise and is less prone to the cascading shadow effect highlighted in fig. \ref{cascadingshadows}.

\begin{figure}[th]
\centering
\includegraphics[scale=0.7]{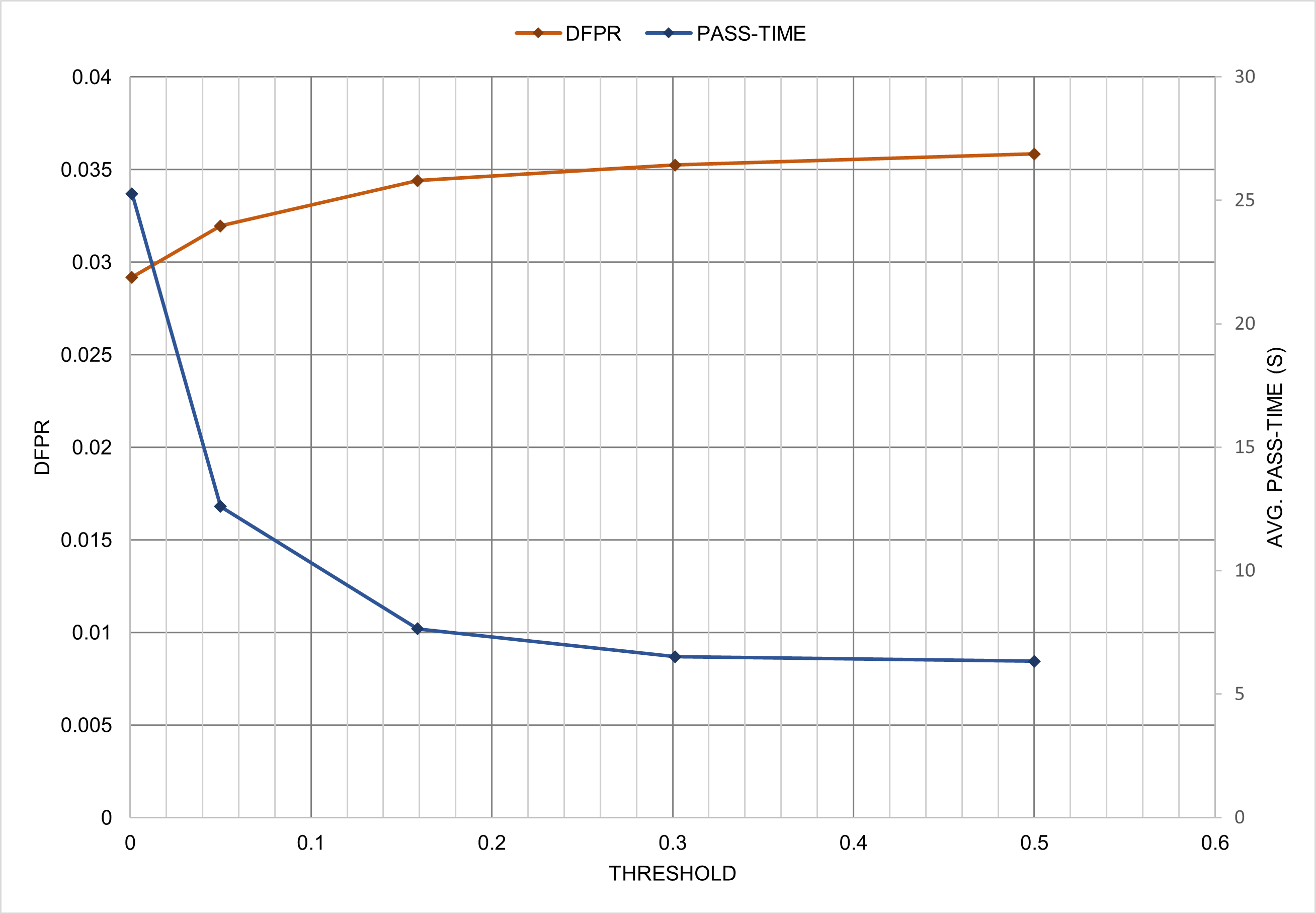}
\decoRule
\caption[]{DFPR and average pass-time for different gradient thresholds in adaptive subdivision. Lightmap resolution is $512\times512$, max-node size $8\times8$ and batch-size $64\times64$.} 
\label{AS_DFPR_PT}
\end{figure}

The gains made in pass-time, however, largely depend on the gradient threshold that is chosen (see fig. \ref{AS_DFPR_PT}).

Undersampling does not require the storage or retrieval of quad-tree metadata, and as such runs significantly faster in most cases.
Rudimentary undersampling is simpler and its parameters are scene-independent, which makes the results more predictable. With adaptive subdivision even minor changes in the chosen gradient threshold can manifest themselves in significant differences in pass-times.

It is noteworthy that our approach is entirely contained within a textures alpha-channel, which allows for a simple implementation and visualization, but may offer worse performance than a quad-tree contained in a memory buffer or a separately generated sampling texture.

\begin{figure}[H]
\centering
\includegraphics[scale=1.2]{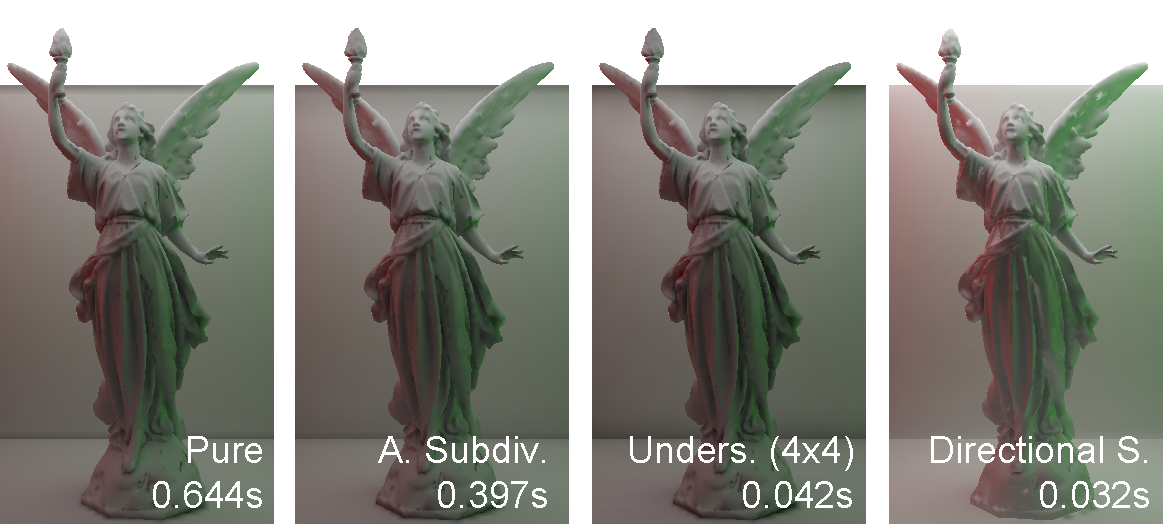}
\decoRule
\caption[]{Avg. pass-time and visual comparison between different sampling techniques on a $256\times256$ lightmap. Lighting quality is worse with directional sampling, but pass-time scale linearly. Undersampling can exaggerate ambient occlusion, but is generally faster than adaptive subdivision.} 
\label{general_comparison}
\end{figure}

In general we found that adaptive subdivision, although offering vastly reduced pass-times over pure radiosity, gets frequently outclassed by the swifter pass-times offered by Monte-Carlo undersampling (see fig. \ref{general_comparison}).

\subsection{Scene Complexity}

Measuring performance relative to a scene's triangle count is difficult to do accurately, as the cost incurred by BVH traversal can highly vary depending on the specific arrangement and concentration of these triangles.

Nevertheless, our results are roughly indicative of the logarithmic scaling that the underlying theory would predict.

\begin{figure}[th]
\centering
\includegraphics[scale=0.7]{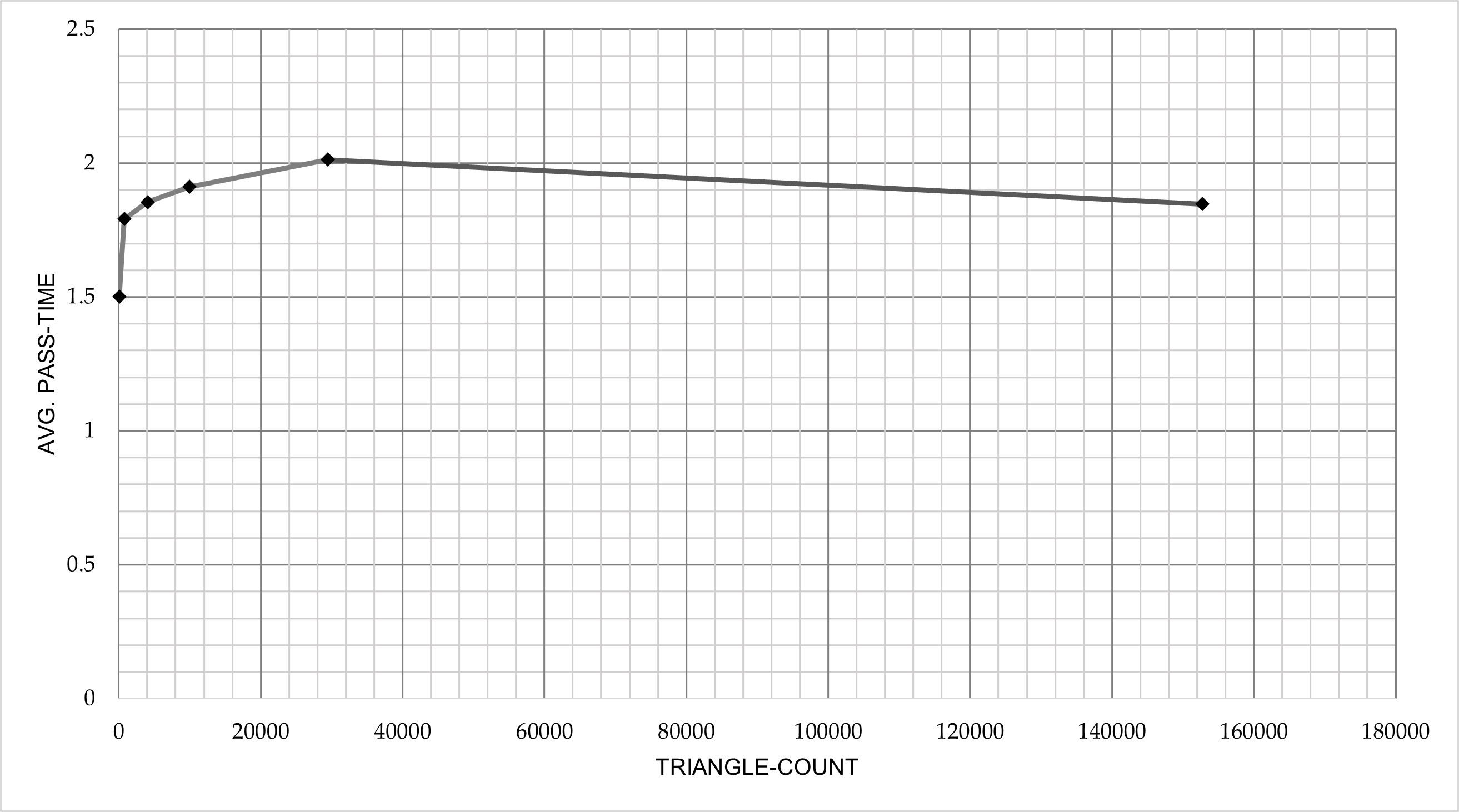}
\decoRule
\caption[]{Average pass-time (across five passes) for each of our testing scenes. Parameters: $256\times256$, pure radiosity, batch-size of $64\times64$.} 
\label{TriangleScaling}
\end{figure}

Fig. \ref{TriangleScaling} depicts the pass-time for each of our scenes under the same pass parameters. The curve generally follows a logarithmic pattern, apart from our largest scene (SponzaScene) being significantly faster than our second largest (CornellLucy).

The underlying cause for this anomaly is likely the specific arrangement of the geometry in these scenes. CornellLucy has the vast majority of its triangles at its center, with the surrounding walls consisting of a large number of radiosity patches. Correspondingly, most rays cross this center area and will likewise be traced against large portions of the BVH.

SponzaScene on the other hand has its triangles spread out and segregated into smaller sub-volumes. Given that our rays are only as long as they need to be, it stands to reason that such an arrangement would induce a swifter BVH traversal per ray. Furthermore, RTX's ray-grouping technique \cite{RTX_examination} may contribute to this phenomenon as well.

\section{Extensions}

Below we disclose the impact provided by the implemented extensions that go beyond the scope of progressive refinement radiosity, such as visibility caching, raymarching and directional sampling.

\subsection{Visibility Caching}\label{viscaching_analysis}

\begin{figure}[th]
\centering
\includegraphics[scale=0.7]{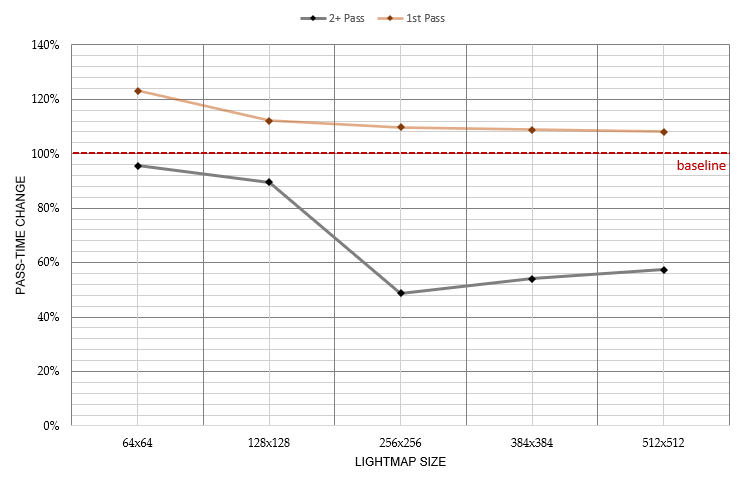}
\decoRule
\caption[]{Percentage change in pass-time after enabling visibility caching for the first and subsequent passes. A value of 50\% would equate to the pass running twice as fast, 200\% twice as slow etc. (CornellLucy scene with maximum batch-size).} 
\label{Viscache_Stats}
\end{figure}

Visibility data can be stored in its entirety for lightmaps smaller or equal to $512\times512$ pixels. Correspondingly, all lightmaps of this size demonstrate a significant speedup for all passes beyond the first (see fig. \ref{Viscache_Stats}). The first pass, in contrast, is slowed down by $10\%$ to $20\%$, due to the necessity to compute the cantor-index and store the visibility bit.

Smaller lightmaps benefit less from this increase in speed, but the maximum size remains capped at $512\times512$.

The overall computation speed is greatly improved for resolutions between $256\times256$ and $512\times512$. Unfortunately, higher resolutions cannot viably be covered by our method due to the respective buffer index exceeding the maximum value of an unsigned, 32bit memory pointer.

The unfortunate conclusion is thus that visibility caching is only \textit{possible} for lower resolutions, which least require it.
Although the approach may find itself useful for real-time lightmapping on small to medium sized lightmaps, the results demonstrate why GPU-based radiosity implementations generally refrain from caching visibility data: The available memory is unable to harbour the exponential amount demanded by larger lightmaps. 

\subsection{Voxel Raymarching}\label{VoxelRaymarch_Analysis}

\begin{figure}[H]
\centering
\makebox[\textwidth][c]{\includegraphics[scale=0.8]{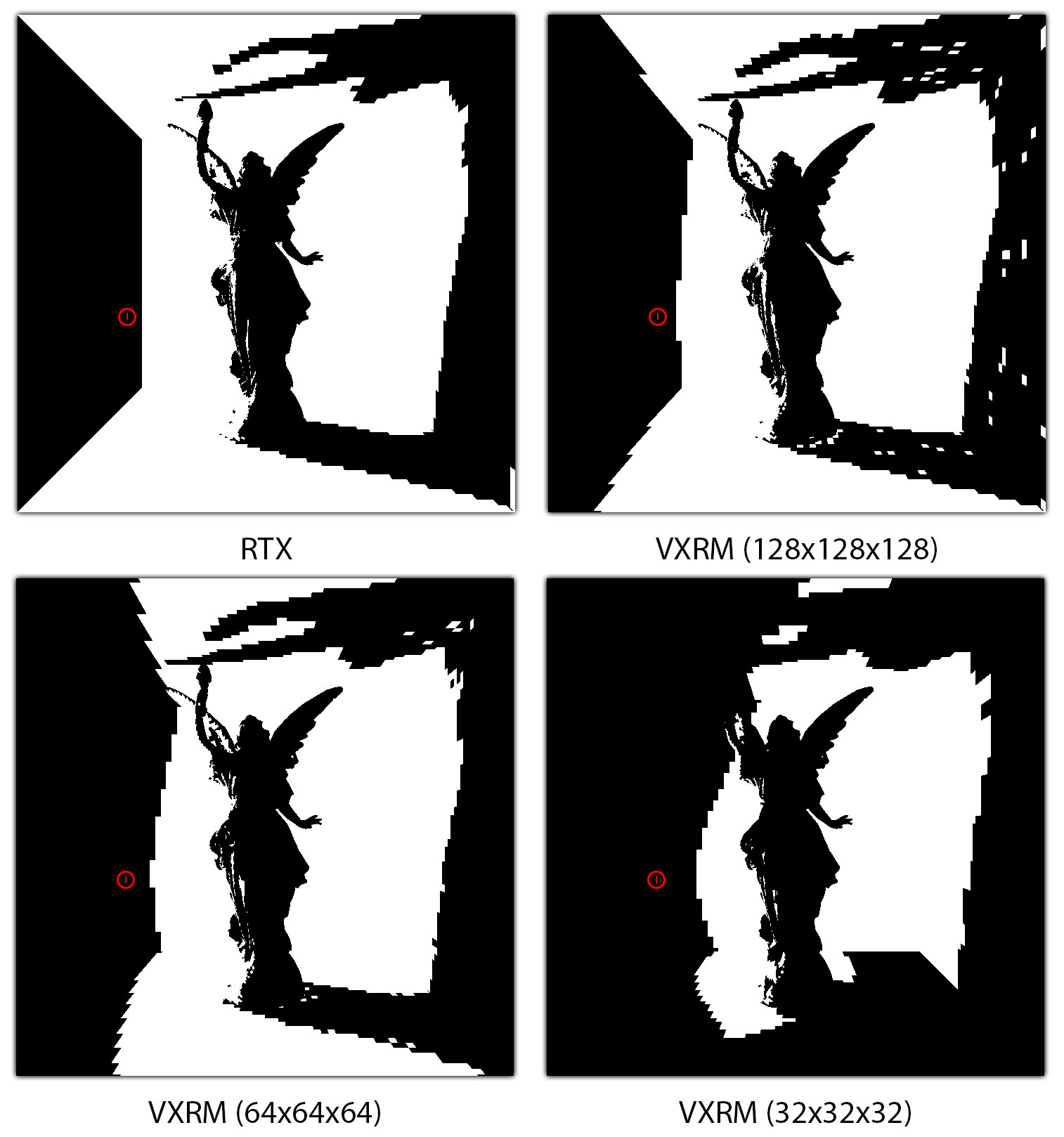}}
\decoRule
\caption[]{The red patch (highlighed by a circle) is visible to the white patches. Visibility accuracy of RTX compared with voxel raymarching on differently sized voxelmaps.} 
\label{VXRM_Accuracy}
\end{figure}

\begin{figure}[H]
\centering
\makebox[\textwidth][c]{\includegraphics[scale=0.8]{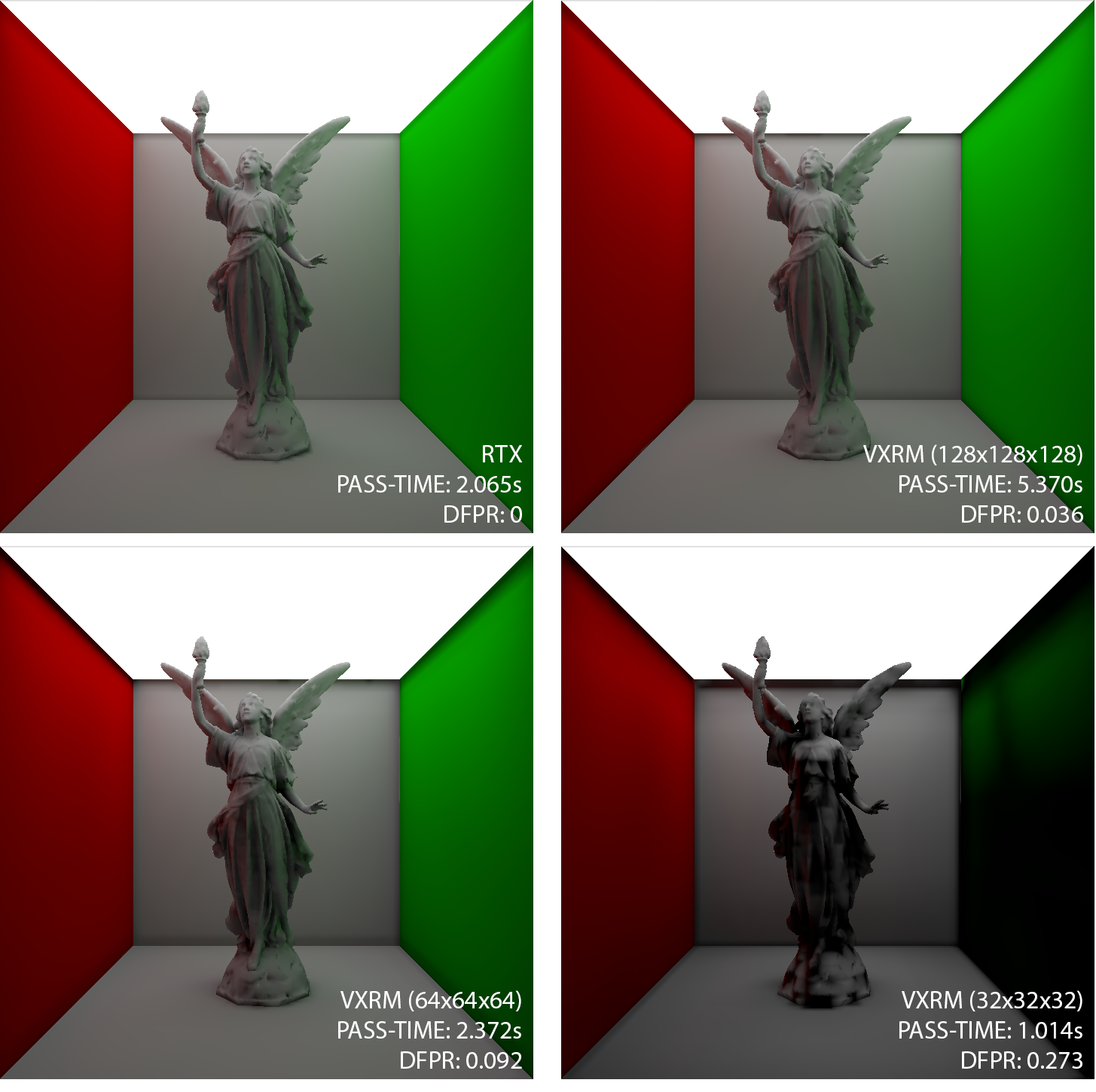}}
\decoRule
\caption[]{Pass-time and DFPR with regular RTX (left) and voxel raymarching on voxelmaps of different sizes. Pass-times only become worthwhile for very small voxelmaps, whereupon the quality degrades too much to remain viable.} 
\label{VXRM_Screenshots}
\end{figure}

Visibility estimation through voxel-raymarching is inherently less accurate than RTX. 
Fig. \ref{VXRM_Accuracy} shows the precision of visibility estimation for a single patch, which is \textit{favourable} to raymarching. Patches located in non-flat, detailed environments are likely to degrade accuracy even further, because all the detail within the vicinity gets simplified into a single voxel. This deterioration in quality can clearly be observed in fig. \ref{VXRM_Screenshots}, where complex geometry, or geometry miss-aligned with the voxelmap produces unrealistic shadows.

In addition, voxel raymarches do not run on dedicated hardware and, as such, will not benefit from RT-core parallelization.
In our testing, voxel raymarching was consistently slower than RTX-based raytracing, except for very small voxelmaps, which are less suited to complex scenes and provide unsatisfactory lighting quality.

In chapter \ref{Chapter5} we postulated the idea of complementing our RTX-based algorithm with regular voxel raymarches to relieve pressure on the limited amount of RT cores.

Unfortunately, this concept did not materialize and ended largely in failure.
The only instance for which interleaving ray-traces and raymarches demonstrated improved pass-times was when the voxel-maps were so small that raymarches were inherently faster.

However, using larger voxelmaps does provide accurate visibility and may serve as a fallback method for graphics cards that are not RTX compatible or in scenes that contain an exorbitant amount of triangles.

If nothing else, it is still serves as an indicative testament to RTX's impressive performance.

\subsection{Directional Sampling}

Employing a directional sampling approach as opposed to a rudimentary, patch-pair approach led to a dramatic improvement in performance for large lightmaps, even though the overall performance \textit{per ray} suffered, due to rays traversing further into the BVH than previously.

This method is of linear complexity $O(n)$ relative to the amount of patches, as the number of rays fired per patch is constant, which makes it is highly adequate for very large lightmaps. We were able to fully compute lightmaps of $2048^2$ pixels in under two seconds, with 1024 samples taken per patch.

\begin{figure}[h]
\centering
\includegraphics[scale=0.45]{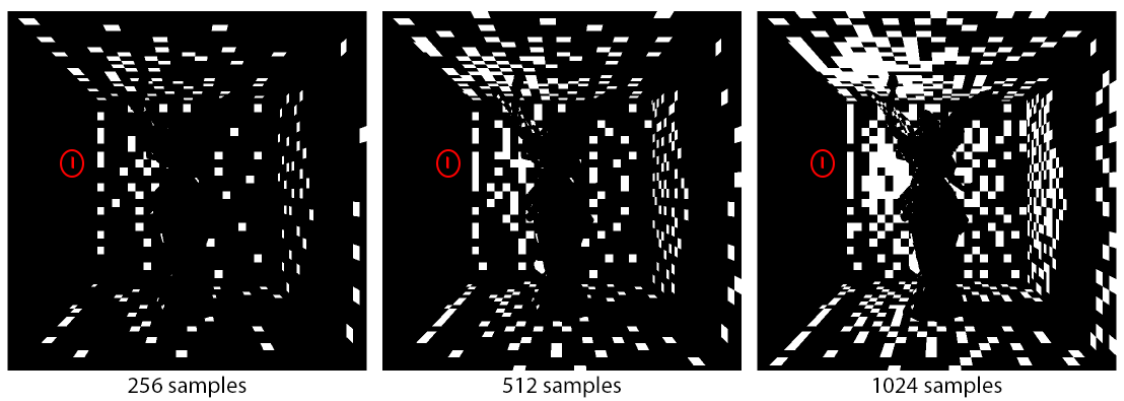}
\decoRule
\caption[]{Directional sampling with different amount of samples on a lightmap of $128\times128$ pixels. The white patches are sampled to compute the lighting value of the red patch (highlighted by a circle).} 
\label{directional_sampling_vis}
\end{figure}

\begin{figure}[h]
\centering
\includegraphics[scale=1.0]{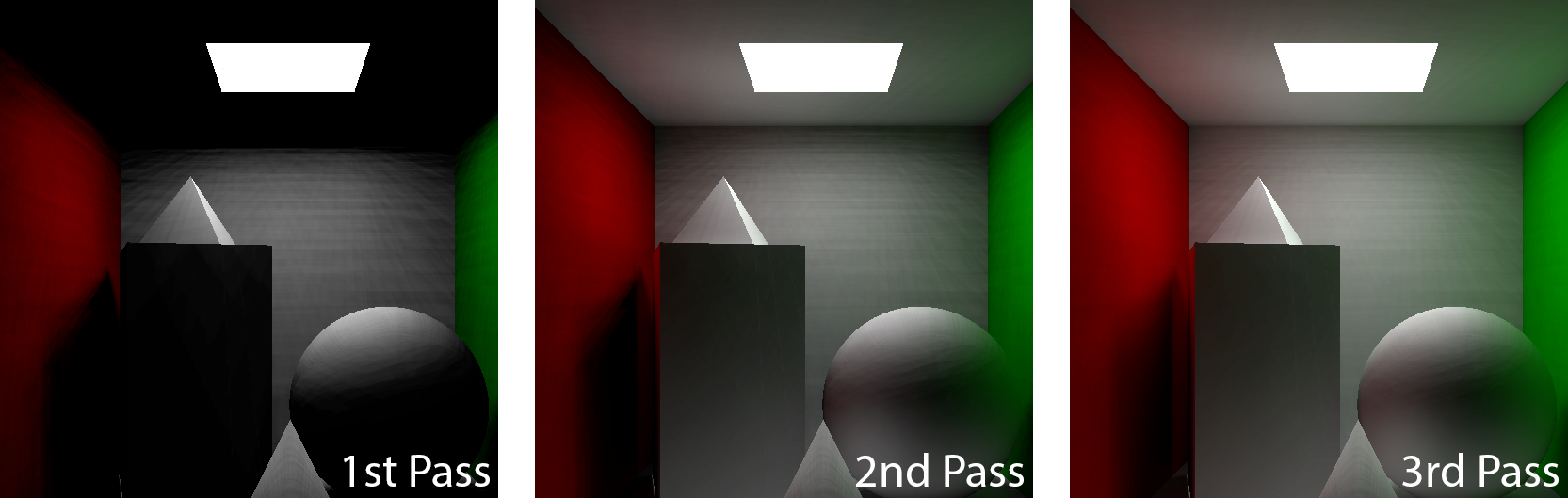}
\decoRule
\caption[]{Directional sampling in environments with small light-sources. Because the chance of a ray hitting the light-source is low, noisy shadows are produced on the walls. The noise becomes less pronounced with each subsequent pass.} 
\label{Directional_sampling_noise}
\end{figure}

Visual fidelity is comparable with that of undersampling techniques, but significantly deteriorates in scenes with small light-sources, as seen in fig. \ref{Directional_sampling_noise}.

Our implementation employs pre-computed directions that are hard-coded into the shader, which is very beneficial towards performance, but limits the maximum amount of samples to that of the pre-generated set. Examples of which patches our pre-generated sets sample can be seen in fig. \ref{directional_sampling_vis}.

We recommend using directional sampling for large lightmaps in scenes with large light-sources, but suggest falling back to Monte-Carlo undersampling in other instances. Alternatively, a hybrid approach that utilizes directional sampling in conjunction with discrete sampling of important patches (such as light-sources) may provide an ideal middle-ground.

\begin{figure}[h]
\centering
\includegraphics[scale=0.48]{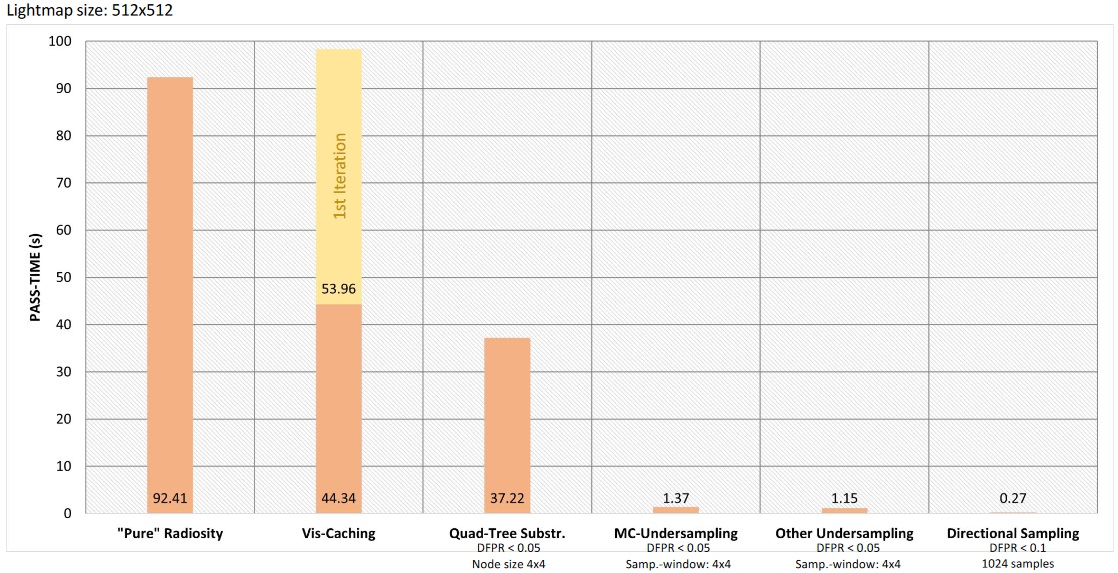}
\decoRule
\caption[]{Overall comparison of RTRad performance enhancements. Note that a different implementation of sub-structuring could yield results more comparable to undersampling.} 
\label{Overall_Comp}
\end{figure}

\section{Comparison}

In \textit{GPU Gems 2}\footnote{\textit{GPU Gems 2} was published in 2005. Given the advancements in hardware since then, it is safe to assume that their system would run significantly faster on modern hardware.}, Pharr et al. claim to achieve 2 frames per second on a scene with 10.000 patches using their GPU-based progressive refinement radiosity, with Coombe et al. achieving comparable results in their own implementation \cite{gpu_gems_2005, RadiosityOnGPUs_Coombe}. 

\begin{figure}[th]
\centering
\includegraphics[scale=0.85]{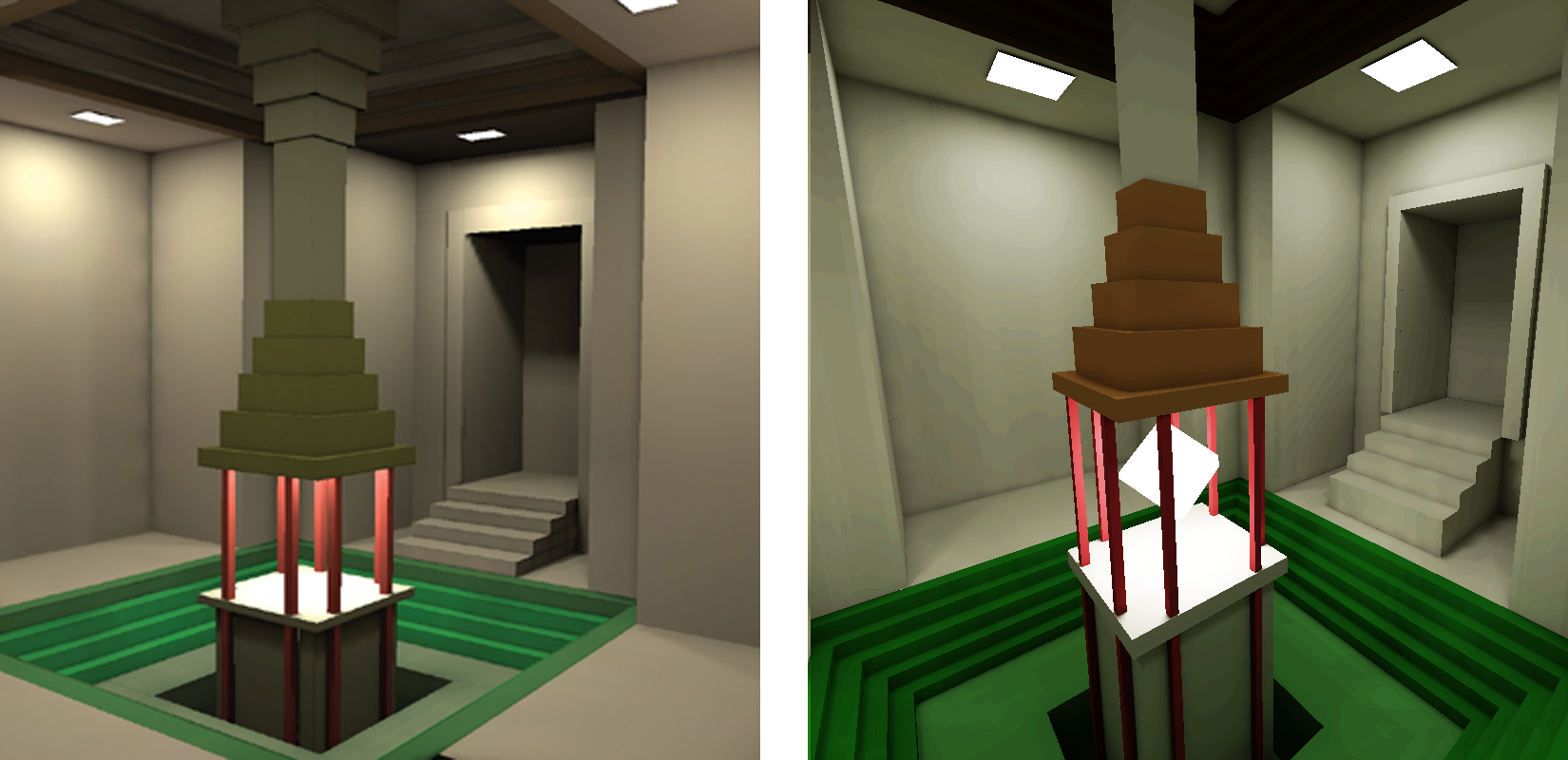}
\decoRule
\caption[]{Scene with one million elements as shown in \textit{GPU Gems 2} \cite{gpu_gems_2005} (left) and our reconstruction rendered with RTRad (right).} 
\label{nvidia_scene_comparison}
\end{figure}

In contrast, RTRad manages similar framerates with lightmaps of 60.000-100.000 patches. Fig. \ref{nvidia_scene_comparison} provides a side-by-side comparison for visual differences between the two programs.

\subsection{Unity and Unreal Engine}

Two of the most prolific rendering engines in the domains of video games, research and filmmaking, are \textit{Unity} and \textit{Unreal Engine} \cite{Render_Engines_Overview1, Render_Engines_Overview2}.

Both of these have their own built-in CPU lightmap generation system: Unity in its \textit{Progressive Lightmapper} and Unreal in its \textit{Lightmass} system. Either system can also be set up for GPU execution\footnote{Unity's Progressive GPU Lightmapper and Unreal's GPU Lightmass are both preview/beta features that, whilst usable, have not yet been officially released.}.

\begin{figure}[H]
\centering
\includegraphics[scale=0.6]{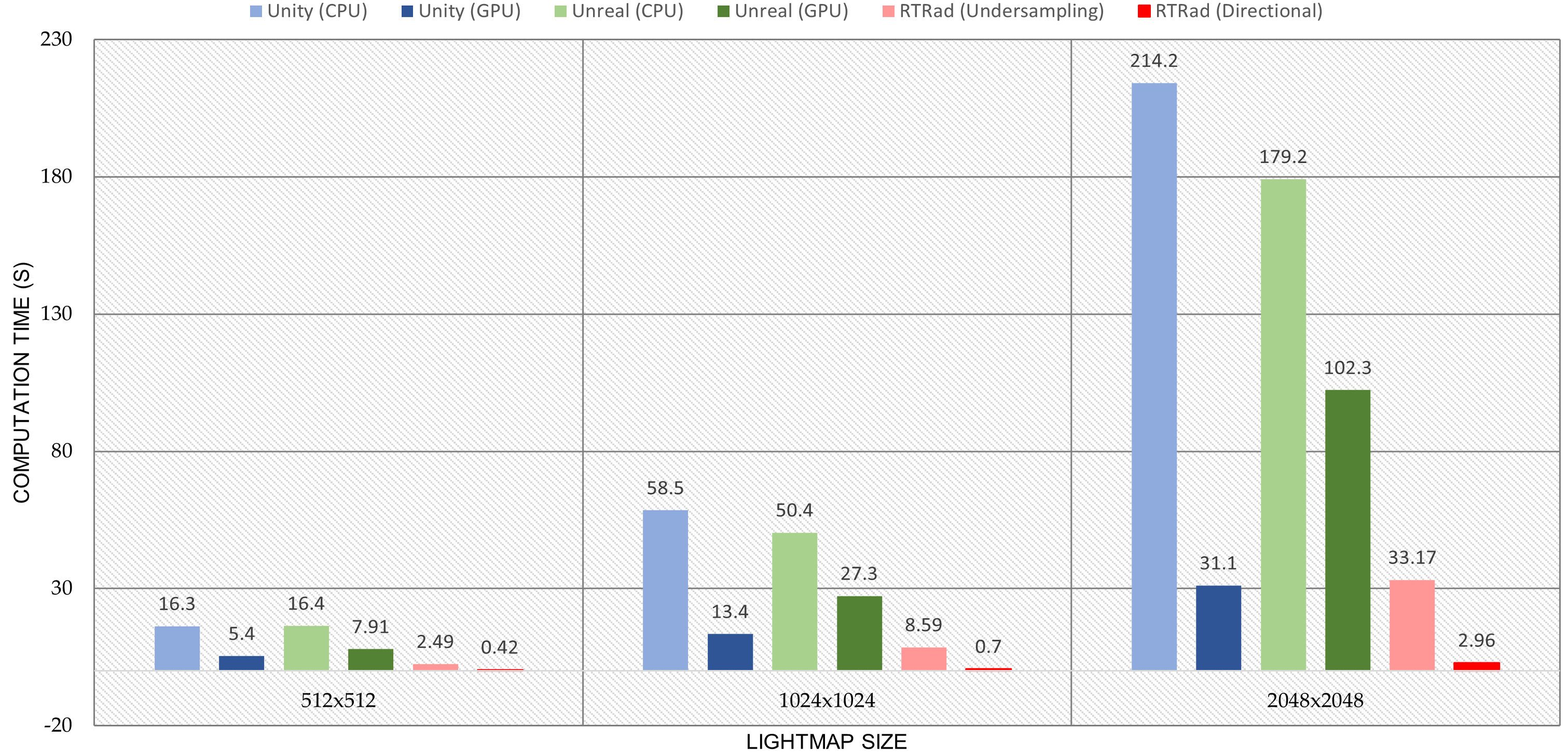}
\decoRule
\caption[]{Total bake time of the CornellLucy scene for two bounces of light with different applications and lightmap sizes.} 
\label{Comparison_graph}
\end{figure}

Fig. \ref{Comparison_graph} shows a performance comparison between each application
, with a detailed account of this data listed in table \ref{Comparison_Table}. Despite being a purely research-oriented application that was developed with limited time and resources, RTRad competes impressively well with these well-established industrial solutions.

\begin{figure}[H]
\centering
\includegraphics[scale=1.15]{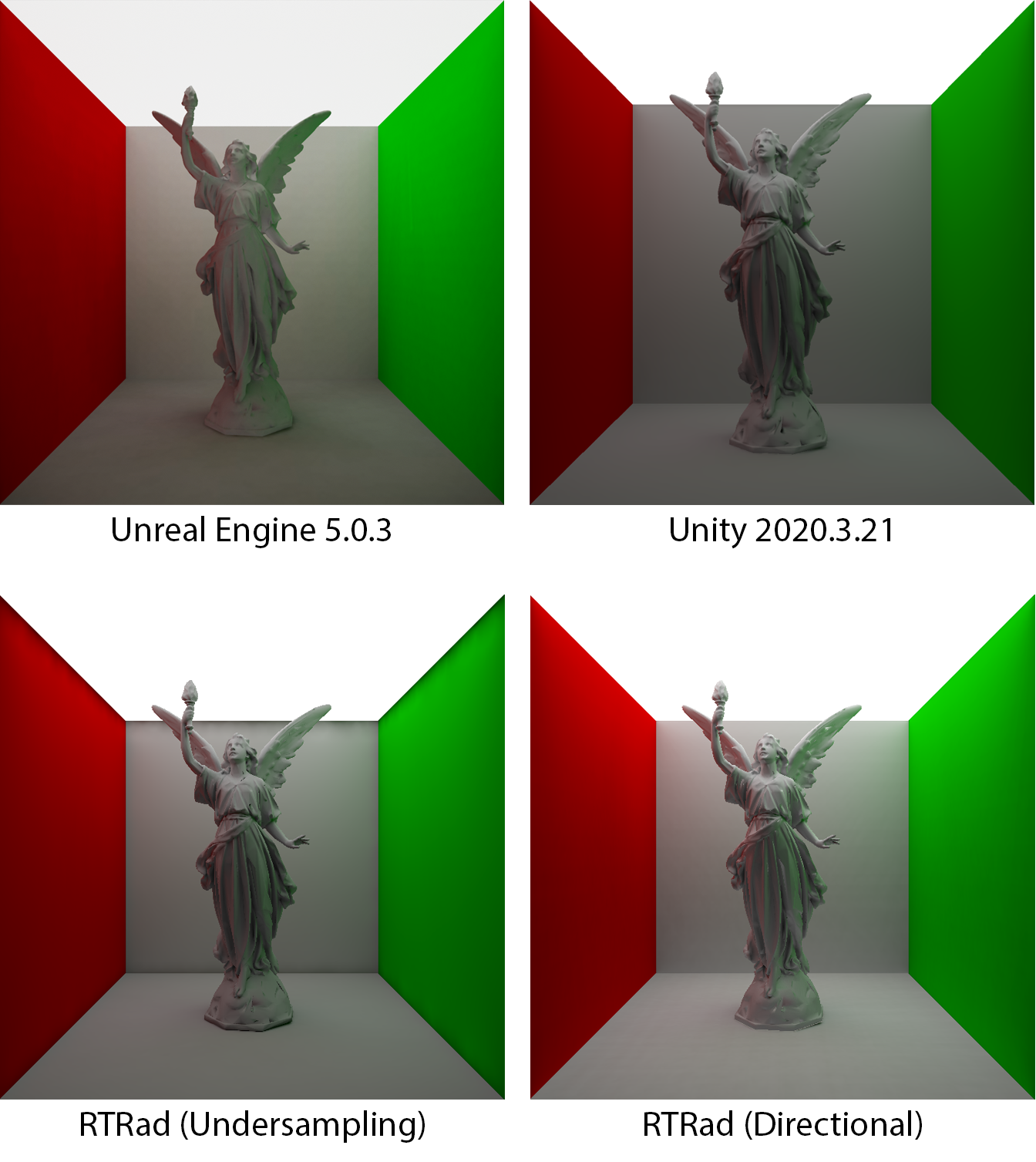}
\decoRule
\caption[]{Lightmap of $1024\times1024$ pixels for the same scene computed with different lightmapping tools.} 
\label{Comparison_side-by-side}
\end{figure}

\begin{table}[H]
    \centering
    \centerline{
    \begin{tcolorbox}[tablestyle_main,tabularx={l|l|X||r}]
        \rowcolor{gray!50!black}
        \textcolor{white}{Lightmap Size} & \textcolor{white}{Device} & \textcolor{white}{Engine} & \textcolor{white}{Bake Time} \\ \hline
        \hline
        \rowcolor[RGB]{236,236,245}
        $256^2$ & CPU & Unity 2020.3.21 & 4.1s \\ \hline
        \rowcolor[RGB]{226,226,245}
        $256^2$ & GPU & Unity 2020.3.21 & 4.2s \\ \hline
        \rowcolor[RGB]{236,245,236}
        $256^2$ & CPU & Unreal Engine 5.0.3 & 6.5s \\ \hline
        \rowcolor[RGB]{226,245,226}
        $256^2$ & GPU & Unreal Engine 5.0.3 & 3.36s \\ \hline
        \rowcolor[RGB]{245,236,236}
        $256^2$ & GPU & RTRad (Undersampling) & 0.5s \\ \hline
        \rowcolor[RGB]{245,226,226}
        $256^2$ & GPU & RTRad (Directional) & 0.12s \\ \hline
        \hline
        \rowcolor[RGB]{236,236,245}
        $512^2$ & CPU & Unity 2020.3.21 & 16.3s \\ \hline
        \rowcolor[RGB]{226,226,245}
        $512^2$ & GPU & Unity 2020.3.21 & 5.4s \\ \hline
        \rowcolor[RGB]{236,245,236}
        $512^2$ & CPU & Unreal Engine 5.0.3 & 16.4s \\ \hline
        \rowcolor[RGB]{226,245,226}
        $512^2$ & GPU & Unreal Engine 5.0.3 & 57.91s \\ \hline
        \rowcolor[RGB]{245,236,236}
        $512^2$ & GPU & RTRad (Undersampling) & 2.49s \\ \hline
        \rowcolor[RGB]{245,226,226}
        $512^2$ & GPU & RTRad (Directional) & 0.42s \\ \hline
        \hline
        \rowcolor[RGB]{236,236,245}
        $1024^2$ & CPU & Unity 2020.3.21 & 58.5s \\ \hline
        \rowcolor[RGB]{226,226,245}
        $1024^2$ & GPU & Unity 2020.3.21 & 13.4s \\ \hline
        \rowcolor[RGB]{236,245,236}
        $1024^2$ & CPU & Unreal Engine 5.0.3 & 50.4s \\ \hline
        \rowcolor[RGB]{226,245,226}
        $1024^2$ & GPU & Unreal Engine 5.0.3 & 27.3s \\ \hline
        \rowcolor[RGB]{245,236,236}
        $1024^2$ & GPU & RTRad (Undersampling) & 8.59s \\ \hline
        \rowcolor[RGB]{245,226,226}
        $1024^2$ & GPU & RTRad (Directional) & 0.70s \\ \hline
        \hline
        \rowcolor[RGB]{236,236,245}
        $2048^2$ & CPU & Unity 2020.3.21 & 214.2s \\ \hline
        \rowcolor[RGB]{226,226,245}
        $2048^2$ & GPU & Unity 2020.3.21 & 31.1s \\ \hline
        \rowcolor[RGB]{236,245,236}
        $2048^2$ & CPU & Unreal Engine 5.0.3 & 179.2s \\ \hline
        \rowcolor[RGB]{226,245,226}
        $2048^2$ & GPU & Unreal Engine 5.0.3 & 102.3s \\ \hline
        \rowcolor[RGB]{245,236,236}
        $2048^2$ & GPU & RTRad (Undersampling) & 33.17s \\ \hline
        \rowcolor[RGB]{245,226,226}
        $2048^2$ & GPU & RTRad (Directional) & 2.96s \\ \hline
    \end{tcolorbox}
    }
    \caption{Expanded data corresponding to fig. \ref{Comparison_graph}.}
    \label{Comparison_Table}
\end{table}

To create comparable results (see fig. \ref{Comparison_side-by-side}), each configuration was given its own custom settings, which we list below:

\begin{itemize}
    \item \textit{Unity Engine (CPU and GPU)}: Shadowmask, medium quality.
    \item \textit{Unreal Engine (CPU)}: High quality preset.
    \item \textit{Unreal Engine (GPU)}: 512 GI samples, no irradiance caching.
    \item \textit{RTRad (Undersampling)}: Monte-Carlo undersampling in line with our recommended settings from table \ref{recommended_stettings} and maximum batchsize.
    \item \textit{RTRad (Directional)}: 1024 samples with maximum batchsize.
\end{itemize}

The exhibited measurements are naturally not all-conclusive, as Unity and Unreal are both powerful game engines made for far more than just lightmapping \cite{UnrealDocu, unity_docs}. Each algorithm employs diverging techniques and operates on entirely different parameters in addition to potential differences in their approach for time measurement.

We can, however, establish a clear trend in that leveraging RTX for radiosity is not only competitive, but highly advantageous in most circumstances.

\section{Specular Reflections}

\begin{figure}[H]
\centering
\includegraphics[scale=1.15]{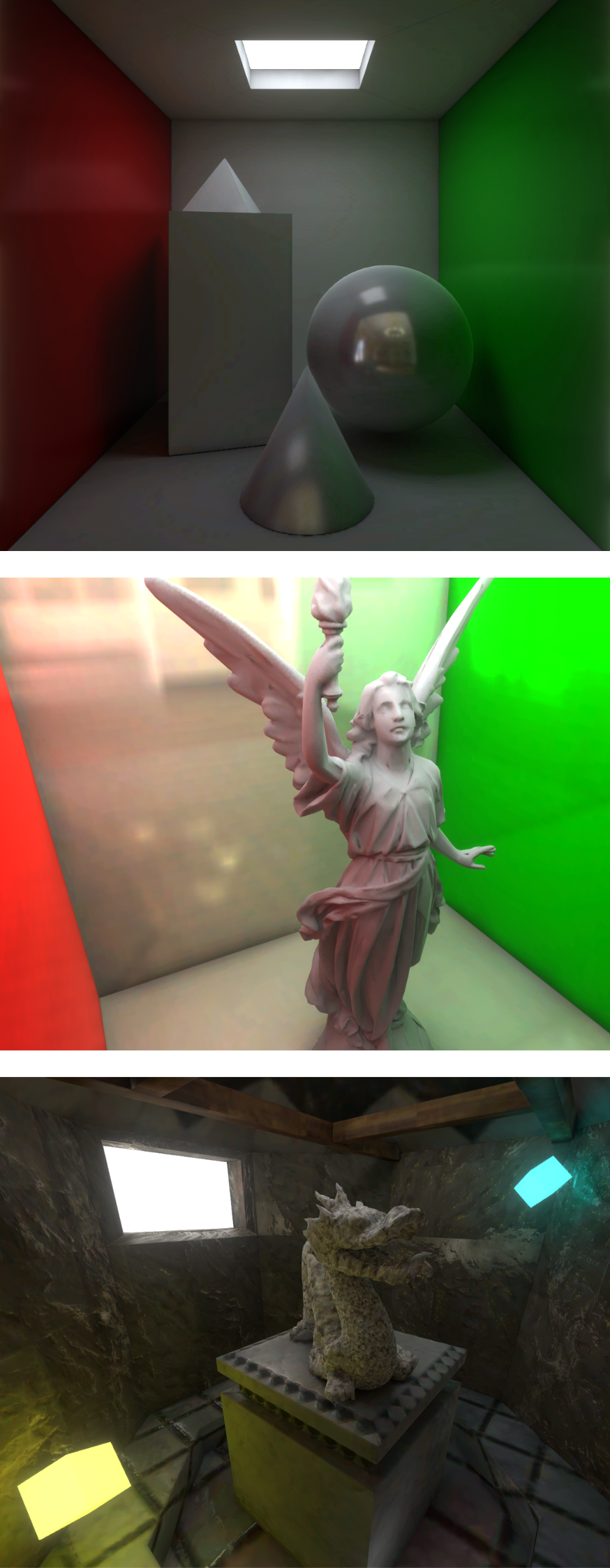}
\decoRule
\caption[]{RTRad-generated lightmaps rendered in real-time using Unity's built-in rendering pipeline with specular reflection probes and some rudimentary post-processing effects.} 
\label{Gallery}
\end{figure}

A limitation inherent to radiosity as a whole is that only diffuse reflections are accounted for. RTRad does contain any non-diffuse functionality itself, but allows generated lightmaps to be exported as textures. 
These can then be brought into a different rendering pipeline to be complemented with specular reflections, as we did in fig. \ref{Gallery} with Unity.

\chapter{Verdict} 

\label{Chapter7} 

In this chapter we sum up our primary conclusions in relation to our initial goals.

Additionally, we list some of the limitations encountered as well as questions that require further research to answer.

\section{Summary}

Radiosity has long proven itself an ideal solution for real-time global illumination in static scenes. Unfortunately, drawn-out periods of pre-computation can hinder productivity in designing 3D environments with realistic lighting.

Attempts to speed up this process have been centered around either

\begin{itemize}
\item the general reduction of computational complexity, as is done in instant radiosity, or
\item the exploitation of its parallelizable nature.
\end{itemize}

Both of these techniques manifest themselves in \textit{progressive refinement} radiosity, where patches are organized in quad-trees and updated simultaneously. 
Yet, few instances of this algorithm are implemented for GPU execution, and those that are typically rely on a z-buffered hemicube approximation for visibility, which comes at a significant expense whilst providing less realistic results.

Nvidia's Turing GPUs come equipped with dedicated raytracing hardware intended to speed up real-time raytracing. In chapter \ref{Chapter2} we demonstrated how raytracing and radiosity both share the same underlying principles by deriving each method from the rendering equation, implying that the performance gains enabled by RTX ought not only to speed up raytracing, but also radiosity.

In chapter \ref{Chapter6} we successfully demonstrated that this idea is both viable and competitive. Through our implementation we put a significant number of different methods and configurations to the test as well as examining additional extensions that serve as performance enhancements.

\section{Limitations}

In our evaluation we described the limitations encountered concerning visibility caching and voxel-raymarching in particular. 

Since these do not present an obstacle to the core idea of leveraging RTX for progressive refinement radiosity, holistically, the concept appears viable.

\newpage

Below we list our encountered limitations that are specific to our implementation of RTRad:

\begin{itemize}
\item Unless lightmap resolutions are sufficiently low or directional sampling is used, the performance remains marginally outside the domain of real-time.

\item Directional samples either require large light-sources or a vast amount of rays to produce serviceable results.

\item Our implementation of adaptive subdivision is subject to significant speed constraints because all pixels need to be tested for their alpha value. This problem may be ameliorated by creating an entirely separate texture consisting only of the pixels with an alpha value larger than zero.

\item Our implementation only uses a single UV channel both for color textures and lightmaps. This approach is unsuitable for a PBR material system, but is easily expandable by including an additional channel.

\item All light-sources must be reflected as a patch in the texturegroup. Point-lights or directional lights, as they are traditionally used in the phong illumination model, do not work.
\end{itemize}

\section{Conclusions}

Our analysis in chapter \ref{Chapter6} comes with several implications. In regards to our initial goals, we arrived at the following overarching conclusions:

\medskip
\textbf{Conclusion 1:} \textit{On the GPU, simpler radiosity variants perform better.}

Performance-wise, our findings were generally more favourable to simpler techniques over adaptive subdivision.
In our assessment, Monte-Carlo undersampling using the recommended settings listed in section \ref{Undersampling_Analysis} generally provides the best balance between speed, visual fidelity and reliability. For very large lightmaps and well-lit scenes, directional sampling is fastest.

Visibility-caching and voxel-raymarching are generally discouraged, but can be advantageous in specific use-cases.

We expect that a potential hybrid solution utilizing a conjoined set of patches sampled directionally and based on a quad-tree to provide ideal results.

\medskip
\textbf{Conclusion 2:} \textit{RTX can significantly improve GPU radiosity performance.}

As a whole, we were successful in demonstrating that RTX has the capability to speed up the offline lightmap generation process and, as more applications begin to move their heavy workloads onto the GPU, we expect radiosity to follow suit.

According to the \textit{Steam Hardware Survey} of July 2022, approximately 30\% of gaming-oriented PCs are equipped with a Turing graphics card \cite{SteamHardwareSurvey}. If this number continues to grow, we anticipate engines utilizing GPU lightmapping (such as Unity, Blender or Unreal) to begin leveraging this power.

\medskip
\textbf{Conclusion 3:} \textit{RTX can quickly compute vast amounts of accurate visibility data for arbitrary point-pairs.}

Extensively, RTX appears to be a great potential asset to \textit{any} application that requires large sets of highly accurate visibility data. This precedent is not strictly tied to just the domains of graphics or lighting, but may be expanded to computer vision, physics simulations and photogrammetry.

\section{Future Work}

RTX may prove useful for a variety of purposes depending on how widespread its adoption becomes. Our demonstration of its viability in lightmap generation opens up several new paths that would benefit from further research:

\begin{itemize}
\item We believe that the directional sampling approach in particular, which this thesis only touched on superficially, is the way forward for progressive radiosity on a very large scale. The potential that a hybrid approach may provide requires further examination. We expect that introducing multiple-importance sampling by prioritizing light-sources in combination with a number of directional samples for indirect light to function as a reasonable foundation for this approach.
\item The utilization of \textit{proxy geometry} has proved itself useful in computing global illumination with RTX \cite{ProxyGeometry}. It thus stands to reason that radiosity's performance may benefit from this as well.
\item Whilst our implementation performs all calculations on the GPU, hybrid methods that combine both GPU and CPU have demonstrated potential in the past \cite{ComplexGPURadiosity}. This opens the door to expanding a batch-wise view-factors computation on the GPU with RTX, whilst the CPU performs the steps related to light-transfer.
\item Elias' radiosity implementation \cite{HugoElias_Radiosity} utilizes a static texture that is multiplied with the samples of a hemicube. The brightness of this texture is proportional to the cosine of the angle between the surface normal, and the direction of the light which saves a significant amount of work on view factor calculation. Using a directional sampling approach with a hemicube and storing this value in the ray payload may prove to be faster and more flexible.
\item Diffuse indirect light scatters uniformly and usually does not require the same resolution as direct light. A common practice in applications handling global illumination is to isolate direct and indirect light by computing them separately \cite{gpu_gems_2005}. Using a model like Phong to compute direct light may allow a far lower lightmap resolution to be generated in RTRad to exclusively include the diffuse indirect component, enabling significantly shorter pass-times.
\end{itemize}

\titleformat{\chapter}[display]{\normalfont\bfseries}{}{0pt}{\Huge}

\chapter{Bibliography}\label{bibliography}

\begingroup
\renewcommand{\cleardoublepage}{}
\renewcommand{\clearpage}{}

\printbibliography[heading=subbibliography,title={Books},type=book]

\defbibfilter{theses}{
  type=masterthesis or
  type=phdthesis
}
\printbibliography[heading=subbibliography,title={Theses and Dissertations},filter=theses]

\defbibfilter{articlesandproceedings}{
  type=article or
  type=inproceedings
}
\printbibliography[heading=subbibliography,title={Articles and Proceedings}, filter=articlesandproceedings]

\printbibliography[heading=subbibliography,title={Manuals and Documentation},type=manual]

\printbibliography[heading=subbibliography,title={Repositories and Databases},type=online]

\printbibliography[heading=subbibliography,title={Lectures and Slides},type=custom]

\printbibliography[heading=subbibliography,title={Other},
nottype=book,
nottype=article,
nottype=manual,
nottype=online,
nottype=article,
nottype=inproceedings,
nottype=masterthesis,
nottype=phdthesis,
nottype=custom,
]

\endgroup






\nocite{c1}
\nocite{c3}
\nocite{c4}
\nocite{c5}
\nocite{c7}
\nocite{RTX_visibility_koch}
\nocite{Hybrid_RTX_Mader}
\nocite{FermiArchitecture}
\nocite{TensorCores}
\nocite{Turing_InDepth}
\nocite{RTX_Intro}
\nocite{RTX_Tuto}
\nocite{fu_cg}
\nocite{blender}
\nocite{unity_docs}
\nocite{Crassin_preview}
\nocite{unreal_uv_unwrap}
\nocite{takeshige}
\nocite{opengl_46_spec}

\end{document}